\documentclass[nohyper]{JHEP3}

\usepackage{amsmath}
\usepackage{amssymb}
\usepackage{graphicx}
\usepackage{latexsym,epsfig,cite}
\usepackage{ulem}
\usepackage{lineno}

\def\beq{\begin{equation}}
\def\eeq{\end{equation}}
\def\bea{\begin{eqnarray}}
\def\eea{\end{eqnarray}}
\def\barr{\begin{array}}
\def\earr{\end{array}}

\def\ee{\bar{\epsilon}}

\def \tj {\tau_\text{jet}}
\def \bj {b_\text{jet}}

\def\lsim{\mathrel{\rlap{\raise 2.5pt \hbox{$<$}}\lower 2.5pt\hbox{$\sim$}}}
\def\gsim{\mathrel{\rlap{\raise 2.5pt \hbox{$>$}}\lower 2.5pt\hbox{$\sim$}}}

\newcommand{\half}{{\textstyle\frac{1}{2}}}

\allowdisplaybreaks 

\title{Flavour Structure of R-violating Neutralino Decays at the LHC}
\author{Nils-Erik Bomark,\\
Department of Physics,
University of Bergen, Postboks 7803, N-5020 Bergen, Norway\\
E-mail: \email{Nils-Erik.Bomark@ift.uib.no}}
\author{Debajyoti Choudhury,\\
Dept.\ of Physics and Astrophysics, University of
Delhi, Delhi 110007, India \\
E-mail: \email{debajyoti.choudhury@gmail.com}}
\author{Smaragda Lola,\\
TH Division, Physics Department, CERN, 1211 Geneva 23, Switzerland and \\
Department of Physics, University of Patras, GR-26500 Patras, Greece\\
E-mail: \email{Magda.Lola@cern.ch}}
\author{Per Osland\\
Department of Physics,
University of Bergen, Postboks 7803, N-5020 Bergen, Norway\\
E-mail: \email{Per.Osland@ift.uib.no}}

\abstract{
We study signatures of R-parity violation in the production of supersymmetric
particles at the LHC, and the subsequent decay of the lightest neutralino
being the end product of a supersymmetric cascade decay. In doing so, we pay
particular attention to the possible flavour structure of the operators, and
how one may discriminate between different possibilities. A neutralino LSP
would couple to all quarks and leptons and a comparative study of its decays
provides an optimal channel for the simultaneous study of all 45 R-violating
operators. By studying the expected signals from all these operators, we
demonstrate the ability to understand whether more than one coupling
dominates, and to map the experimental signatures to operator hierarchies that
can then be compared against theoretical models of flavour.  Detailed
comparisons with backgrounds, including those from MSSM cascade decays are
made, using the PYTHIA event simulator.}


\preprint{}

\begin{document}

\section{Introduction}

Over the last decades, there has been increasing interest in
R-violating supersymmetry, as a plausible alternative to the
Minimal Supersymmetric Standard Model (MSSM). Supersymmetrizing the Standard
Model (SM) does allow for additional terms in the superpotential
over and above those corresponding to the Higgs
potential or  the Yukawa terms within the SM. The
supersymmetric versions of the latter read
\begin{equation}
\mu \, H_1 \, H_2 + m_e L_{i}\bar{E}_{j} H_1
+ m_d  Q_{i}\bar{D}_{j} H_1
+ m_u  Q_{i}\bar{E}_{j} H_2,
\label{mssm}
\end{equation}
where $H_{1,2}$ are the Higgs superfields, $L$ $(Q)$ the
left-handed lepton (quark) doublet superfields, and ${\bar{E}}$
(${\bar{D}},{\bar{U}}$) the corresponding left-handed singlet
fields. Gauge symmetry as well as supersymmetry do allow
for additional bilinear terms of the
form
\beq
\mu_i \, L_i \, H_2
\label{bilin}
\eeq
as well as trilinear ones, namely
\begin{equation}
\lambda_{ijk} L_{i}L_{j}{\bar{E}}_{k}
+\lambda ^{\prime}_{ijk}L_{i}Q_{j}{\bar{D}_{k}}
+\lambda ^{\prime \prime }_{ijk}{\bar{U}_{i}}{\bar{D}_{j}}{\bar{D}_{k}}.
\label{Rviol}
\end{equation}
Each of the terms in eqs.~(\ref{bilin}) \& (\ref{Rviol})
violate R-parity. In particular, the existence of
any of  $\mu_i, \lambda_{ijk}, \lambda^{\prime}_{ijk}$ implies
lepton number violation  while  a non-zero $\lambda''_{ijk}$
violates baryon number.
SU(2) and SU(3) invariance imply that there are 48 R-violating
couplings in all (3 $\mu_i$'s, 27 $\lambda'_{ijk}$s and 9 each of
$\lambda_{ijk}$ and $\lambda''_{ijk}$).

The strictest bounds on lepton and baryon-number violating operators
come from proton stability. The assumption of a conserved R-parity
automatically rules out all of the terms of eqs.~(\ref{bilin}) \&
(\ref{Rviol}), rendering the proton stable\cite{fayet}, modulo higher
dimensional terms endemic to the MSSM.  However, alternative
symmetries, namely baryon or lepton parities \cite{IR,LR} can also
exclude the simultaneous presence of dangerous $LQ\bar{D}$ and
$\bar{U}\bar{D}\bar{D}$ couplings~\cite{SMVIS}.  Experimental
constraints from the non-observation of modifications to Standard
Model rates, or from possible exotic
processes~\cite{constraints_baryo, constraints_lepto, Bhattacharyya:1995pr,constraints_rev,Allanach:1999ic}
also impose additional limits.  Overall, the phenomenology to be
expected out of such theories is very rich, since the LSP (Lightest
Supersymmetric Particle) is no longer stable and the missing-energy
signatures of the MSSM \cite{HabKan} are substituted by multi-lepton
or multi-jet events \cite{Rpar,Hall:1983id,barb}; single superparticle
productions\cite{single_prod,berger_harris_sullivan} are also possible.

In addition to the consequences for collider searches, R-violation
implies that gravitinos (which may have been thermally produced after a
period of inflation) are also unstable. However, gravitino dark matter
in the framework of R-violating supersymmetry is plausible
\cite{BM,LOR,LOR2}, provided that the gravitino decays
slowly enough for its lifetime to be larger than the age of the universe
\cite{TY,CM}.  This is an exciting possibility that allows for
supersymmetric dark matter, even if the R-violating couplings
are sufficiently large to lead to observable signatures at colliders
\cite{LOR,LOR2}. Moreover, it was found that
the branching ratios for gravitino decays are
very sensitive to the flavour structure of R-violating operators
\cite{LOR,LOR2}.

In general, the flavour structure of R-violating couplings
is of particular relevance in defining the nature
of the signals to be expected and any information
on it would be crucial for understanding the flavour structure of the
fundamental theory. Indeed, one may try to relate hierarchies amongst
R-violating couplings
to those in fermion masses \cite{MODELS,ELR}, using models with family
symmetries. For example, a large class of such models
allow only the third generation
fermions to be massive, while the remaining masses are generated
by the spontaneous breaking of the family symmetry.  In such a scheme,
if R
parity is violated, couplings with different family charges are likely to
appear with different powers of the family symmetry-breaking
parameter, and thus with different magnitudes.

Testing the above models against observations is very hard, since by
allowing the most general family structure of couplings one ends up with a
large number of possibilities. Single superparticle productions can lead to
spectacular signatures, but they are very dependent on the flavour structure
and one must ensure that the initial scattering states couple to the operators
of interest in a specific model. If this is not the case, the respective
single superparticle production mechanisms will be invisible.

From this point of view, pair production of superpartners and their
subsequent cascade decays via an unstable neutralino LSP have a great
advantage, since the latter (by coupling to all quarks and leptons)
could decay via any of the 45 trilinear operators, thereby allowing
a {\it comparative} study.  Through a detailed, correlated study of
these decay chains, one may also investigate whether more than one
R-violating couplings are of substantial size, ``map'' their
magnitudes and hierarchies, and compare against theoretical
models. This is the aim of the present paper, which is structured as
follows: in Section~\ref{sect:framework}, we summarise the generic
framework of the analysis. In Section~\ref{sect:lle}, we study energy
and invariant mass distributions, for $LL\bar{E}$ operators, as well
as leptonic R-violating branching fractions. In
Sections~\ref{sect:lqd} and \ref{sect:udd} we discuss energy and
invariant mass distributions for
$LQ\bar{D}$ and $\bar{U}\bar{D}\bar{D}$ operators, respectively. In
section~\ref{sect:models} we focus on the correlations
between different couplings
and possible mappings between the kinematic distributions
 and operator flavour hierarchies. In section~\ref{sect:summary} we present
our conclusions. Finally, in Appendices A and B we present
spectra from three-body decays and theoretical invariant mass
distributions, respectively, integrating over the unobserved neutrino energy.

\section{Framework for the Analysis}
\label{sect:framework}
\setcounter{equation}{0}

We study possible effects of R-parity violation (RPV) in the
end stages of cascade decays of supersymmetric particles. In other words,
we assume that the RPV-couplings are small enough not to materially
change either the production processes (which continue to be gauge-interaction
driven), nor in the cascade decays down to the neutralino LSP. The LSP, though,
decays, within the detector, on account of a non-zero RPV-coupling.
A neutralino LSP is motivated even in the case
of R-violating supersymmetry, since, due to the lack of electromagnetic or
colour interactions, its mass after renormalisation group runs tends to be
smaller than those of other sparticles for a very broad region of the
supersymmetric parameter space. A sneutrino or a stau LSP (due to the large
Yukawa coupling, staus can be anomalously light) is still plausible,
especially if $m_0$ is sufficiently small compared to $m_{1/2}$. Those scenarios
are beyond the scope of this work.

Depending on the scenario, a variety of supersymmetric particles could
be produced in proton-proton collisions, mostly squarks and gluinos,
\begin{equation}
q q \to \tilde q \tilde q, \quad
q g \to \tilde q \tilde g, \quad
g g \to \tilde g \tilde g, \quad
g g \to \tilde q \tilde{\bar q}, \quad
q \bar q \to \tilde q \tilde {\bar q},
\end{equation}
but also charginos and neutralinos. The squarks and gluinos will typically
decay to quarks and leptons, one important channel being
\begin{align} \label{Eq:chains}
\tilde q &\to q \tilde\chi_2^0 \to q \ell^+ \tilde\ell^-
\to q \ell^+ \ell^- \tilde\chi_1^0, \nonumber \\
\tilde g &\to \bar q \tilde q \to \bar q q \tilde\chi_2^0
\to \bar q q \ell^+ \tilde\ell^-
\to \bar q q \ell^+ \ell^- \tilde\chi_1^0.
\end{align}

The importance of this is due to the presence of leptons, which allows
easier extraction from the background. However, when confronted with
leptonic or semi-leptonic R-violating neutralino decay, such leptons
will constitute a background for our signals.

We shall, in turn, allow the LSP to decay via $LL\bar E$, $LQ\bar D$ and $\bar
U\bar D\bar D$ couplings, as depicted in Fig.~\ref{fig-Feyn}.
\FIGURE[ht]{
\let\picnaturalsize=N
\def\picsize{15cm}
\ifx\nopictures Y\else{
\let\epsfloaded=Y
\centerline{\hspace{4mm}{\ifx\picnaturalsize N\epsfxsize \picsize\fi
\epsfbox{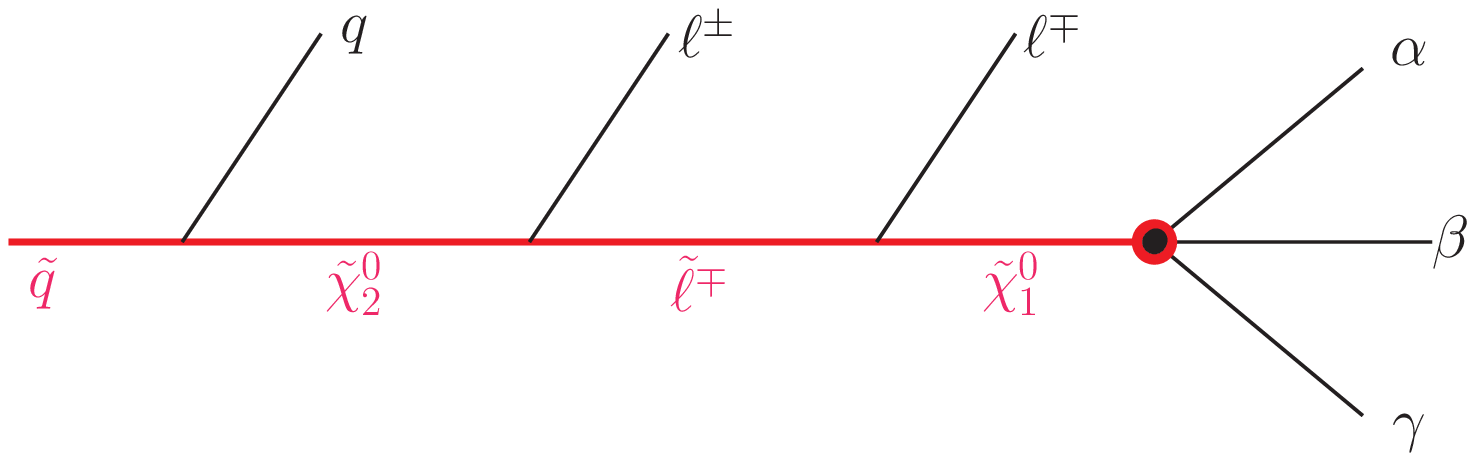}} }
}\fi
\vspace{-6mm}
\caption{Feynman diagram describing one important supersymmetric decay chain
  followed by an R-parity violating neutralino decay.
\label{fig-Feyn} } }

Our analysis has been performed for some of the SPS points
\cite{Allanach:2002nj}.
These points have been identified so as to satisfy the WMAP
constraint on dark matter \cite{Bennett:2003bz}. While we do not require stable dark matter,
these benchmark points are, nevertheless, convenient
reference points in the multidimensional parameter space and allow for direct
comparisons between the MSSM results in the literature and the expectations
in the presence of R-violating operators.
Several of the results we obtain are similar for all these points
and therefore, for many of the distributions to be discussed, we focus on
the SPS1a case. However, for  those  predictions that are more
sensitive to the SUSY parameter space, we present more global results,
elaborating on the differences to be expected in each case.

In the above cascade decay, the actual strength of the R-violating coupling
does not lead to qualitative differences in the predictions, as long as:
\begin{itemize}
\item[(i)] it is strong enough for the neutralino decays
to be inside the detector ($\geq 10^{-6}$ for 100~GeV sparticles and
scaled accordingly for higher masses \cite{DR}, in the case that we do not
have any additional effects from phase space suppressions);
\item[(ii)] it is not sufficiently large for sparticles to
decay directly via the R-violating operator instead
of reaching the end of the chain and then
decaying via neutralinos.
\end{itemize}
For couplings closer to $10^{-6}$ there might be displaced vertices
that can improve the extraction of the signal \cite{deCampos:2007bn,deCampos:2005ri}. This, however, has not
been studied in this work.

The numerical value of the upper limit of couplings does depend on the
SUSY parameter space and the flavour of the R-violating operator under
consideration. In Section~\ref{subsect:CascDom}, we show that the
cascade decay actually dominates for a large part of parameter-space
as long as the R-violating coupling is kept
within experimental
bounds.  Especially when more than one coupling is large, the strong
bounds imposed on products of couplings from flavour violating
processes \cite{constraints_fcnc} will assure cascade dominance.

To analyse the above scenario, we simulate proton-proton collisions at
a center of mass energy of 14 TeV using the
PYTHIA~8.1~\cite{Sjostrand:2007gs} Monte Carlo event generator. In order
to correctly describe the kinematical distributions, we have modified
PYTHIA to use the energy distributions of appendix~A for the neutralino decay.
For the SUSY input we use the SUSY Les Houches Accord
(SLHA)~\cite{Allanach:2008qq} spectra from
SoftSUSY~\cite{Allanach:2001kg}, together with SLHA decay tables
calculated in PYTHIA~6.4~\cite{Sjostrand:2006za}.

The jet analysis is done using FastJet~2.4~\cite{Cacciari:2005hq} with
a $kt$-algorithm \cite{k_t} where the parameter $R_{\rm jet}$,
denoting the upper limit on $\Delta R
\equiv\sqrt{(\Delta\eta)^2+(\Delta\phi)^2}$ for a merge to be allowed,
is set to 0.4 unless stated otherwise. Included in the analysis are
all detectable particles, i.e.\ all particles except neutrinos (and stable neutralinos and gravitinos if applicable).

For analyses including taus and b-quarks, the definition of tau (b-)
jets is that at least $60\%$ of the $p_T$ of the jet should come from
decay products of the tau (b-quark). This fraction is known in the
Monte Carlo but for a more realistic study one would need to resort to
proper tau and b tagging. For the rest of the paper we shall denote
the tau-jets, $\tj$ and the b-jets, $\bj$, except in figures where we
use $\tau$ and $b$ in order to save space.

A lepton is considered isolated if there is no jet with $p_T > 10$ GeV
within a cone of $\Delta R < 0.4$ and the total additional
energy within $\Delta R < 0.2$ is less than 10
GeV. Note that leptons are included in the jet clustering, but jets
that, after the clustering, contain essentially nothing but a lepton,
are not included in the further calculations involving jets.

We also impose acceptance cuts, requiring $p_T > 5$ GeV for leptons and
$p_T > 10$ GeV for $\tj$'s and $\bj$'s. Where light quark jets are used, they
are required to have $p_T > 20$ GeV. The reason for the harder cut on light
jets as compared to $\tj$'s and $\bj$'s, is to remove some background jets from
beam remnants, bremsstrahlung etc.

All data presented is based on simulations of one million events for each
operator/SPS point combination. The backgrounds, $t\bar t$, $W,Z$ and
QCD (with total $p_T>300$ GeV) are studied simulating 10 million events
for each in order to account for their larger cross-sections.

\FIGURE[ht]{
\let\picnaturalsize=N
\def\picsize{6cm}
\ifx\nopictures Y\else{
\let\epsfloaded=Y
\centerline{\hspace{4mm}{\ifx\picnaturalsize N\epsfxsize \picsize\fi
\epsfbox{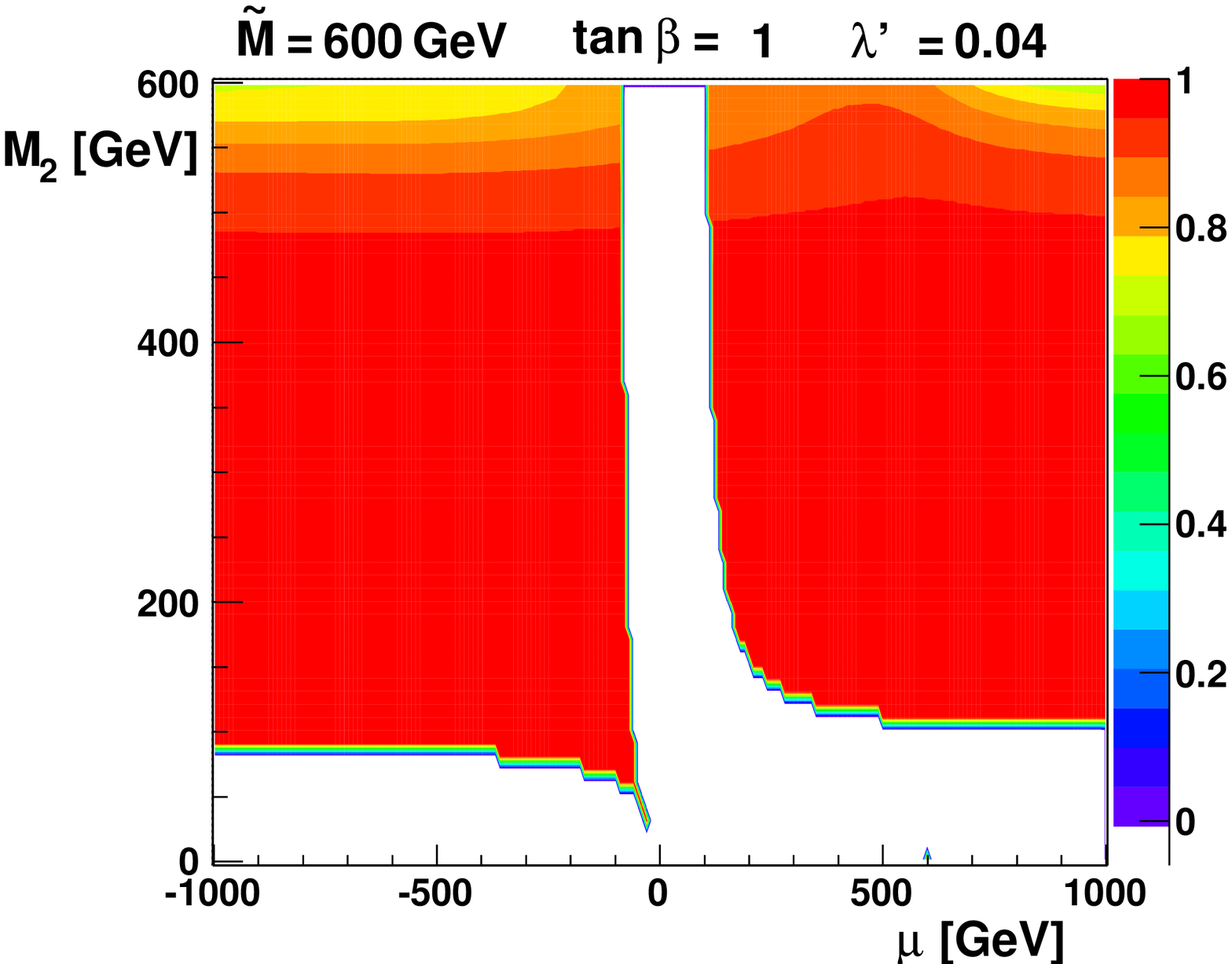}
\hspace{4mm}{\ifx\picnaturalsize N\epsfxsize \picsize\fi
\epsfbox{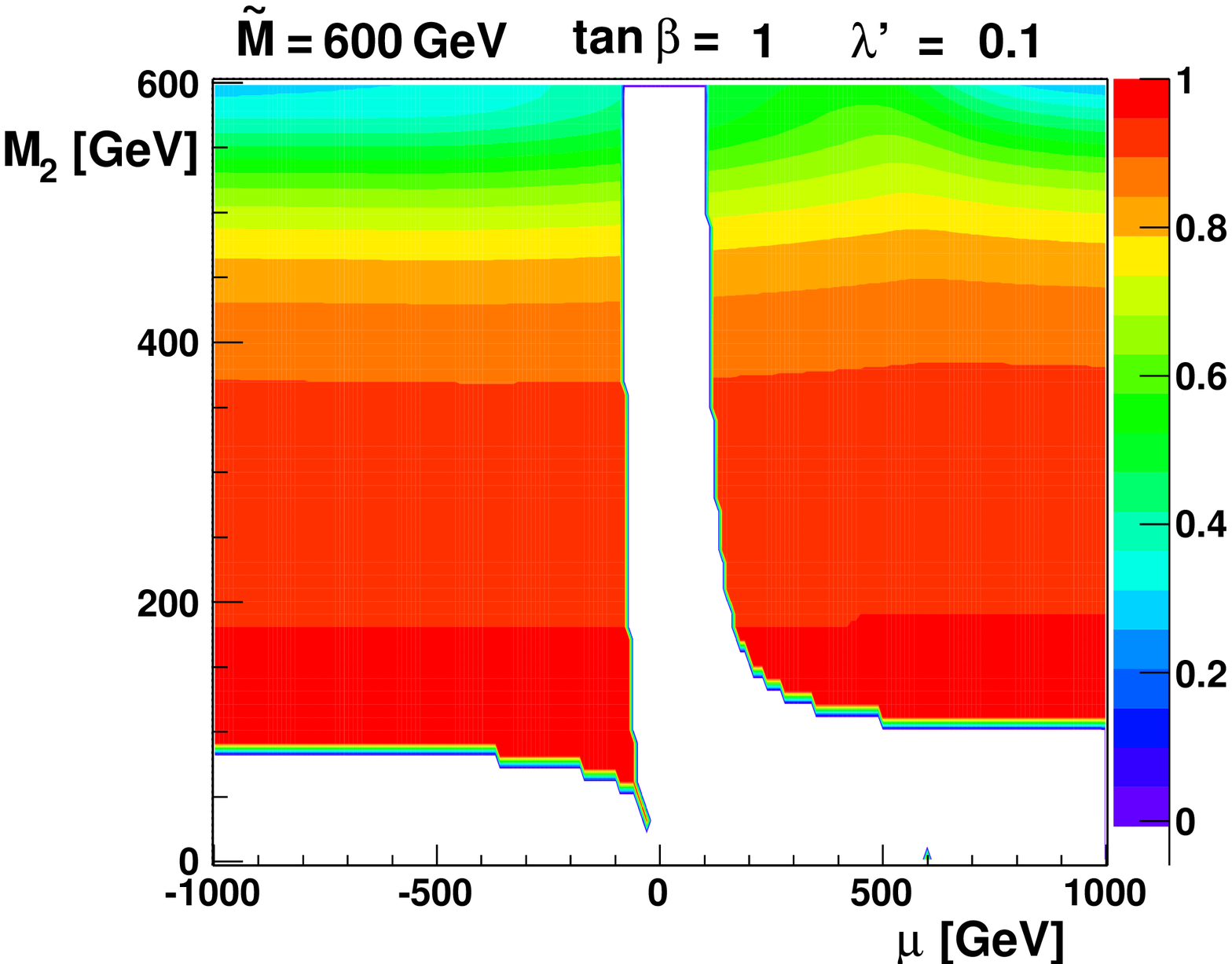}} } }
\centerline{\hspace{4mm}{\ifx\picnaturalsize N\epsfxsize \picsize\fi
\epsfbox{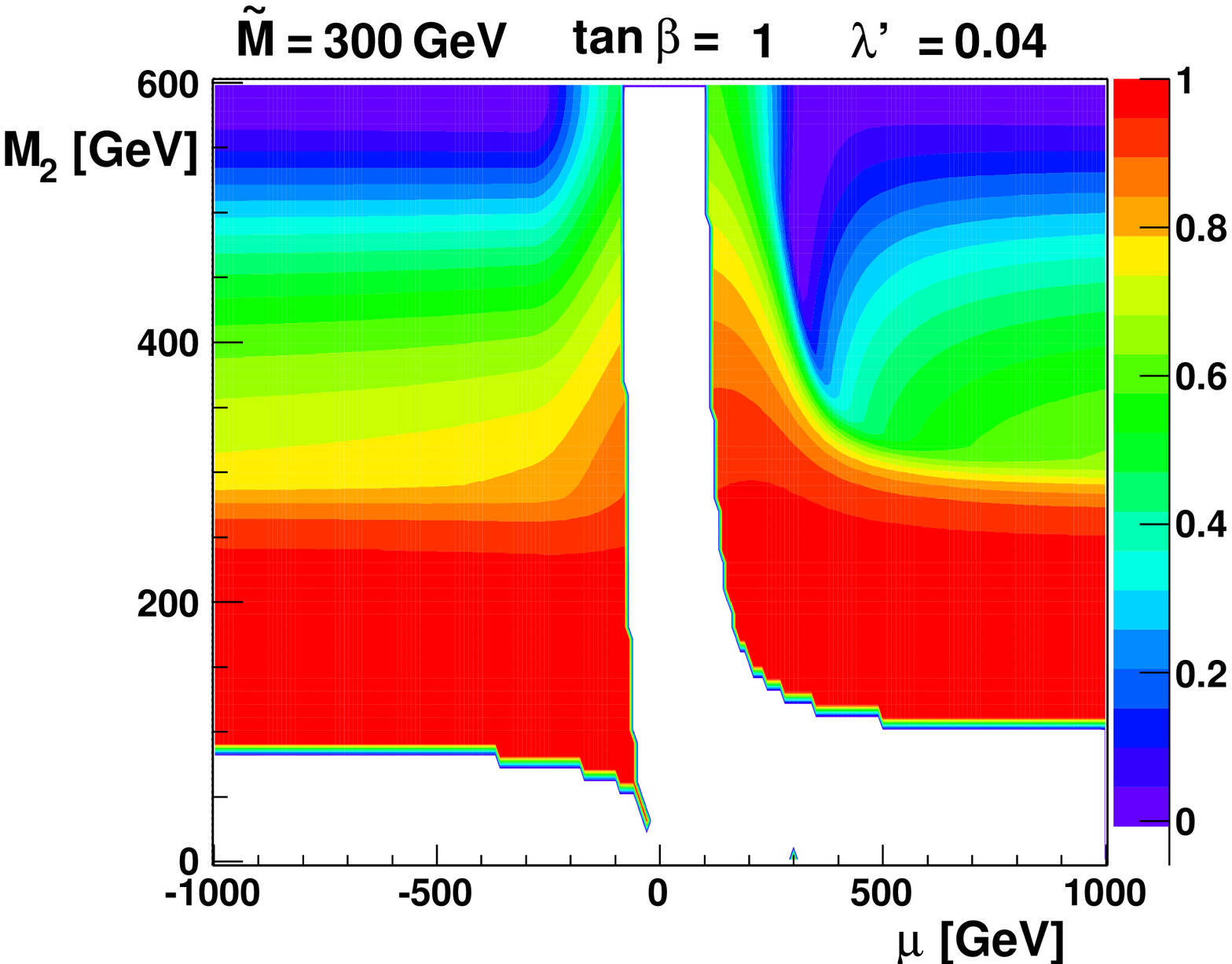}
\hspace{4mm}{\ifx\picnaturalsize N\epsfxsize \picsize\fi
\epsfbox{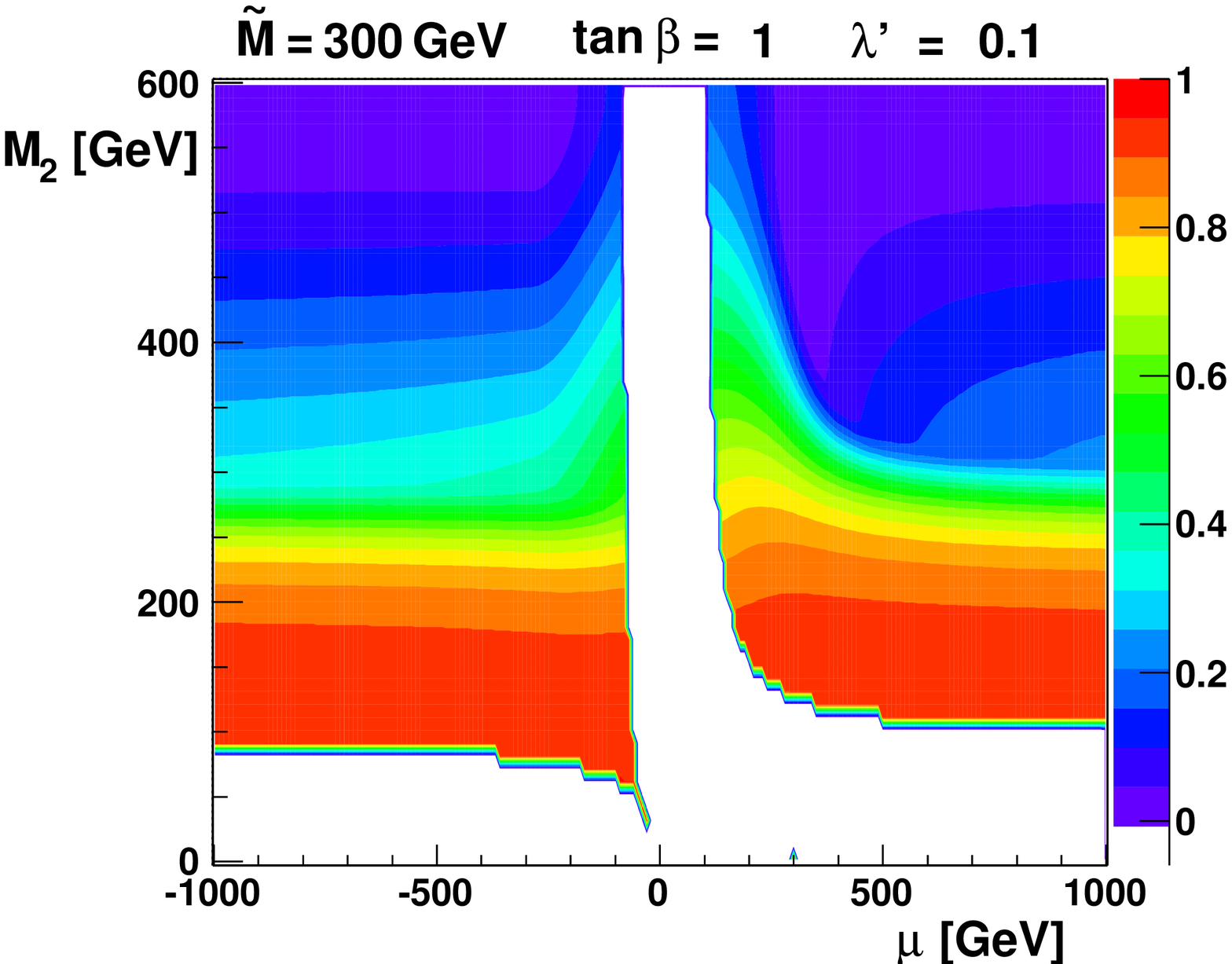}} } }
\centerline{\hspace{4mm}{\ifx\picnaturalsize N\epsfxsize \picsize\fi
\epsfbox{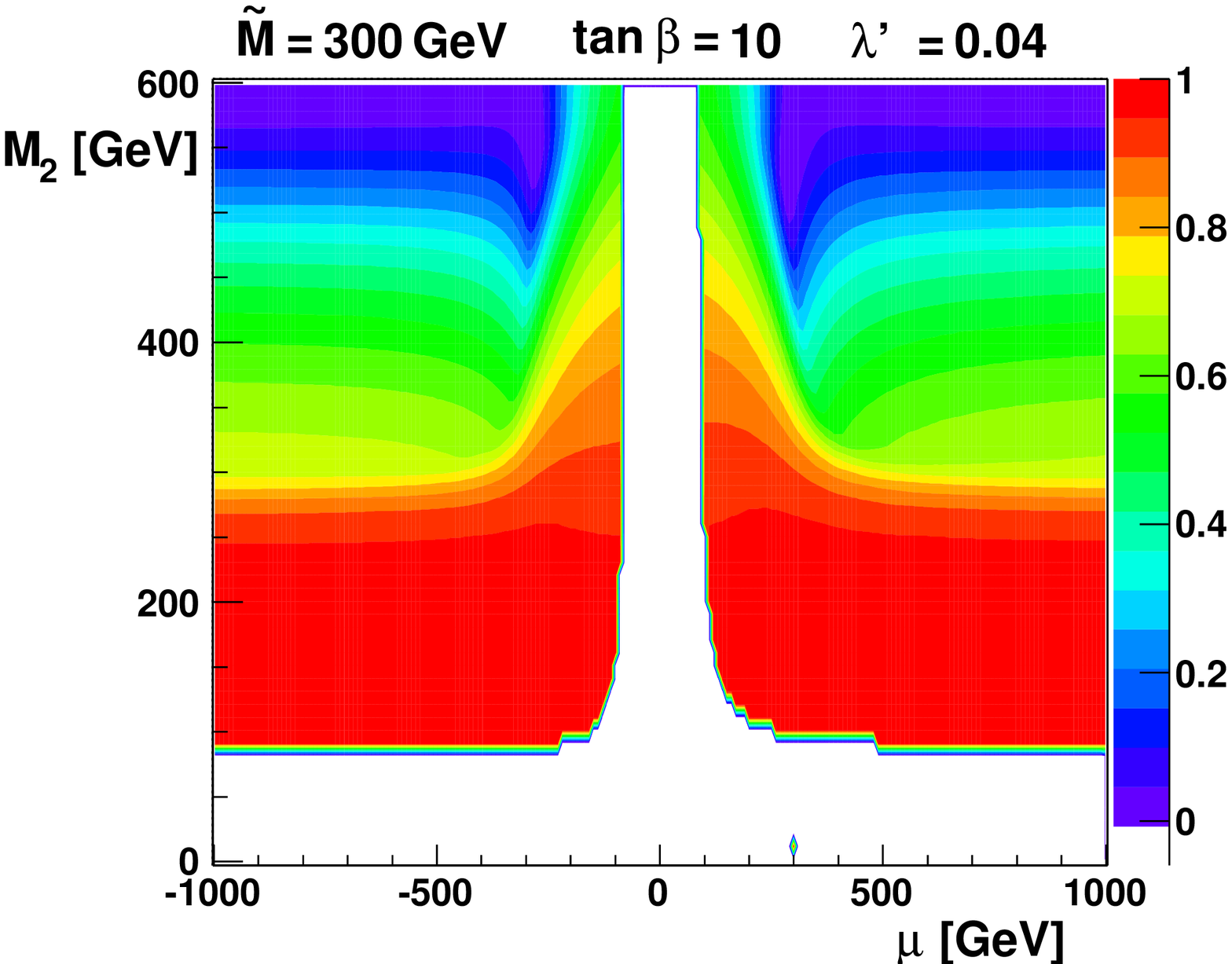}
\hspace{4mm}{\ifx\picnaturalsize N\epsfxsize \picsize\fi
\epsfbox{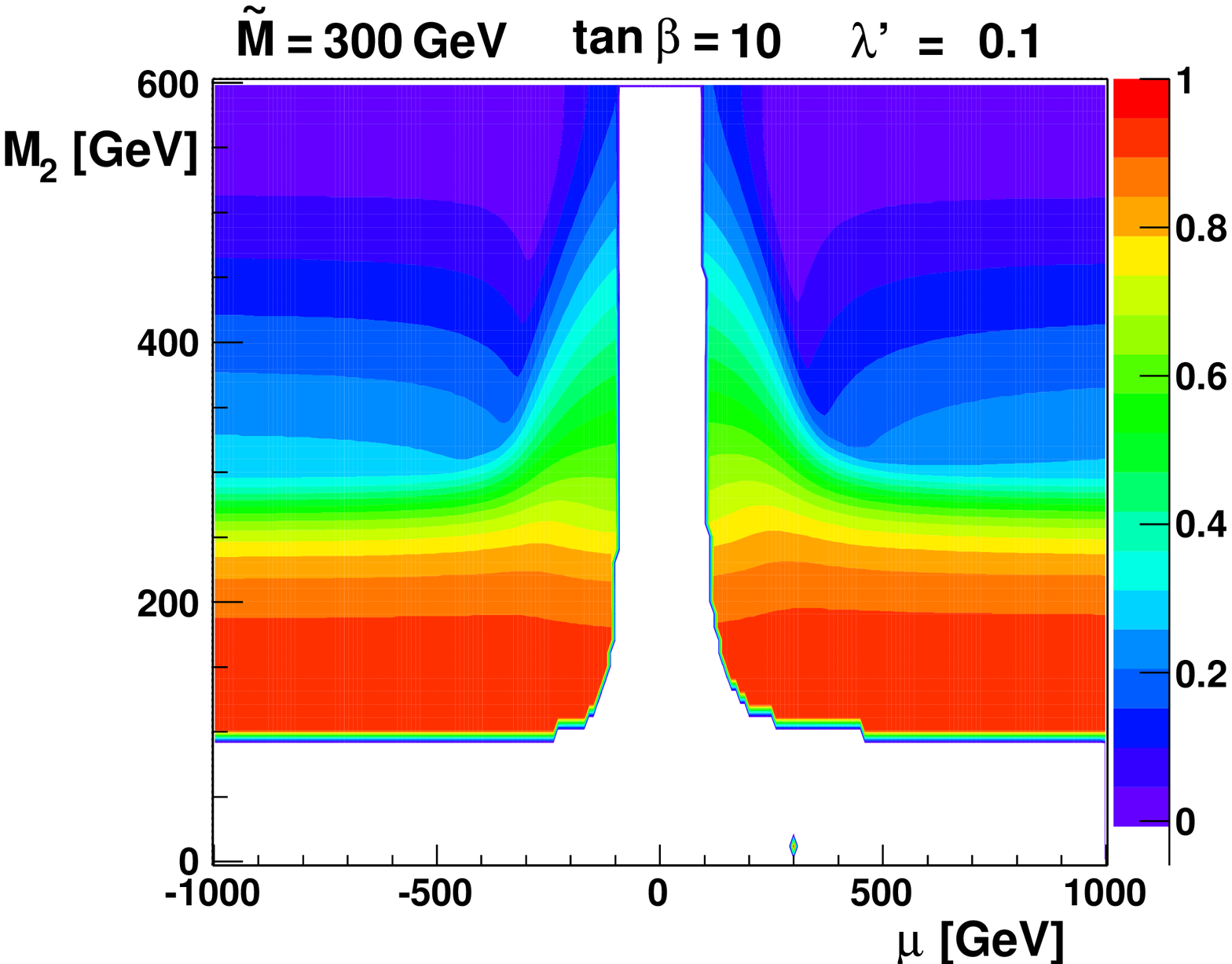}} } }
}\fi
\vspace{-6mm}
\caption{Dominance of MSSM decay of squark vs its R-parity-violating decay, shown for two squark masses, and two values each of $\tan\beta$ and $\lambda'$.
\label{br_ratio} } }

\subsection{Cascade dominance}\label{subsect:CascDom}
In order for the cascade to proceed to the end, giving an (unstable)
LSP, the competing R-violating decay of the squark cannot be too
strong. The decay of the squark to a quark and a neutralino (or
chargino) proceeds via a combination of gauge and Yukawa couplings, in
competition with an R-violating decay to a quark and a lepton with a
strength proportional to some $\lambda'$
squared\footnote{Analogous statements hold for decays through
$\lambda''$, and in case of slepton/sneutrino decays
for $\lambda$-couplings.}. The relative rates
of these channels will depend on the squark mass, as well as on the
neutralino and chargino spectrum, and the coupling $\lambda'$. In
order to illustrate this competition, we determine neutralino and
chargino spectra in terms of $M_2$, $\mu$ and
$\tan\beta$ (holding $M_1=(5/3)\tan^2\theta_W M_2$) and plot,
 in Fig.~\ref{br_ratio}, the ratio
\begin{equation}
\xi \equiv\frac{\sum\Gamma(\tilde q\to q\tilde\chi_i^0)+\sum\Gamma(\tilde q\to q'\tilde\chi_i^\pm)}{\Gamma(\tilde q\to \text{all})}.
\end{equation}
Where this ratio is high (represented in red and yellow), the decay of
the squark will proceed via a cascade to the lightest neutralino,
which then will decay via the R-parity-violating interaction.
Note that we do not need any assumption here about where in the decay chain
the studied squark is produced and hence a high $\xi$ for all sparticles
present in the coupling, will ensure
that the cascade continues down to the
lightest neutralino.

The
white regions in these panels are excluded by the lightest chargino
mass being in conflict with the LEP bound,
$M_{\tilde\chi_1^\pm}\geq94~\text{GeV}$ \cite{Nakamura:2010zzi}.

For some representative combinations of the squark mass $\tilde M$,
$\tan\beta$ and $\lambda'$, we see in Fig.~\ref{br_ratio}
that even for rather substantial R-violating
couplings, the cascade decay dominates for large regions of
parameter space.
One should also remember that typical decay chains may involve particles other than those present
in the dominant R-violating coupling and thus have no open channels for
direct R-violating decay. Since these cascades will always proceed to the
lightest neutralino, the probability of interrupting the cascade with a
direct decay, is further reduced compared to the above discussion.

\section{$LL\bar E$ couplings}
\label{sect:lle}
\setcounter{equation}{0}

We shall first consider the couplings $L_iL_j\bar E_k$, which, typically,
lead to
decays of the LSP to two charged leptons and a neutrino, for example
\begin{equation} \label{Eq:LLE121}
\tilde\chi_1^0\underset{L_1L_2\bar E_1}{\longrightarrow}
e^+\mu^-\nu_e, \quad
e^-\mu^+\bar\nu_e, \quad
e^+e^-\nu_\mu, \quad
e^+e^-\bar\nu_\mu.
\end{equation}
While isolated leptons in the final state would also arise in
R-conserving scenarios from decays higher up in the cascade chain, we
will see that differences in kinematic distributions allow us to
statistically disentangle the composition of the final state.

We assume that only one $LL\bar E$ operator is non-vanishing and set
the corresponding coupling to be
\begin{equation}
\lambda_{ijk}=10^{-4}.
\end{equation}
It should be noted that, even if two (or more) couplings of
such a size are introduced, the constraints from
FCNC~\cite{constraints_fcnc} are satisfied almost
trivially. There do exist a few combinations for which even
such sizes can be of concern, but these we take care
of specifically. Note also that many such constraints
actually apply in combinations like $ \sum_i \lambda_{i j k}
\lambda^*_{i \ell n} / m^2_{i}$, and given that the
couplings $\lambda$ can assume either sign (and, in
principle, any phase), larger couplings could, in principle,
be accommodated if one were to admit cancellations. Collider
signals, on the other hand, would be insensitive to such
cancellations. We, however, desist from considering
such possibilities. Identical observations would hold for
$\lambda'$ couplings as well, while, for $\lambda''$
couplings, the bounds are, in general, even weaker.
However, we will return to the issue of a comparative
study of the magnitude of the operators, if several of them
are large---a main aim of the paper---in a subsequent
section.

For the analysis of the $LL\bar E$ operators, the following selection cuts have been used \cite{AtlasTDR}:
\begin{itemize}
\item
    at least three isolated leptons with $p_\text{T}>20~\text{GeV}$;
\item
    the leading lepton has $p_\text{T}>70~\text{GeV}$;
\item
    $p_\text{T}^\text{miss}>100~\text{GeV}$.
\end{itemize}

These cuts are sufficient to reduce $t\bar t$ events by a factor
$\approx 3\cdot 10^{-6}$ and $W$ and $Z^0$ events (including all
production of weak bosons) by a factor $\approx 6\cdot 10^{-5}$. The
RPV events, on the other hand, are reduced to approximately $1-40\%$
of the total, with the suppression factor depending on the point in
the SUSY parameter space as well as the nature of the RPV
interaction. For operators with a lot of tau flavour, the reduction is
significantly stronger than for operators involving only light
flavours.

This means that in the case of a SUSY parameter point with low cross
section and low lepton production in combination with a tau-rich
$LL\bar E$ operator, we might get some $W$ and $Z^0$ events in the
sample, but detailed
simulations show that this contamination in the studied
distributions is not a significant problem.

Overall, in the presence of R-violation via $LL\bar{E}$ operators, we
can see spectacular signatures (final-state flavours, like-sign leptons, etc.)
\cite{barb}; these become more pronounced for the
$L_2L_3\bar{E}_1$ and $L_1L_3\bar{E}_2$ couplings, which mix all flavours.

\subsection{Final-state flavour correlations}

Since we cannot detect the neutrinos or their flavours, the
characteristics of various decay channels have to be described in
terms of the charged leptons in the decay. From each neutralino we get
one pair of charged leptons and their identities are correlated with
the flavours in the operator.

\begin{TABLE}{
  \centering
\begin{tabular}{|c|c|c|c|c|c|c|}
  \hline
 $ijk$ &  $ee$ & $\mu\mu$ & $\tau\tau$ &  $e\mu$ &  $e\tau$ & $\mu\tau$  \\\hline
  121 & $\checkmark$ &  &  & $\checkmark$ &  &  \\\hline
  122 &  & $\checkmark$ &  & $\checkmark$ &  &  \\\hline
  123 &  &  &  &  & $\checkmark$ & $\checkmark$ \\\hline
  131 & $\checkmark$ &  &  &  & $\checkmark$ &  \\\hline
  132 &  &  &  & $\checkmark$ &  & $\checkmark$ \\\hline
  133 &  &  & $\checkmark$ &  & $\checkmark$ &  \\\hline
  231 &  &  &  & $\checkmark$ & $\checkmark$ &  \\\hline
  232 &  & $\checkmark$ &  &  &  & $\checkmark$ \\\hline
  233 &  &  & $\checkmark$ &  &  & $\checkmark$ \\\hline
\end{tabular}
  \caption{Possible final states for $LL\bar E$ couplings.}\label{tab:LLEChannels} }
\end{TABLE}

It is important to remember that every operator gives rise to two distinct
channels, a consequence of the fact that one of the
charged leptons is associated with the $\bar E$ superfield and the other
with the $L$ superfield.  Since $SU(2)$ invariance requires
the $L$-type fields to have different flavours, this leads to
exactly two different channels, modulo charge conjugation. The
operator $L_1L_2\bar E_1$, for example, will give both $e^+e^-$ and
$e\mu$ pairs with equal frequency.

A further complication arises when tau fields are involved, then
light-lepton pairs may arise from leptonic tau decays. And since our
final states require isolated light leptons, the leptonic tau decays
will be favoured over hadronic ones. Similarly, decay channels wherein
the superparticles decay directly into light leptons would have a
higher detection efficiency compared to those involving intermediate
tau states. Thus, the composition of final states measured is not the same
as given directly by the branching fractions but depends nontrivially on the supersymmetric
spectrum as well as the flavour structure of the coupling.

The open channels for each coupling are given in Table~\ref{tab:LLEChannels}.

\subsection{$p_T$ distributions}

We first look at the $p_T$ spectra of the electrons (henceforth,
``electrons'' will refer collectively to both electrons and positrons)
and muons as well as hadronic taus.  We shall assume that hadronic
taus ($\tj$) can be identified (with some efficiency). Note that in the
$\tj$s the neutrino is not included.

\FIGURE[ht]{
\let\picnaturalsize=N
\def\picsize{15cm}
\ifx\nopictures Y\else{
\let\epsfloaded=Y
\centerline{\hspace{4mm}{\ifx\picnaturalsize N\epsfxsize \picsize\fi
\epsfbox{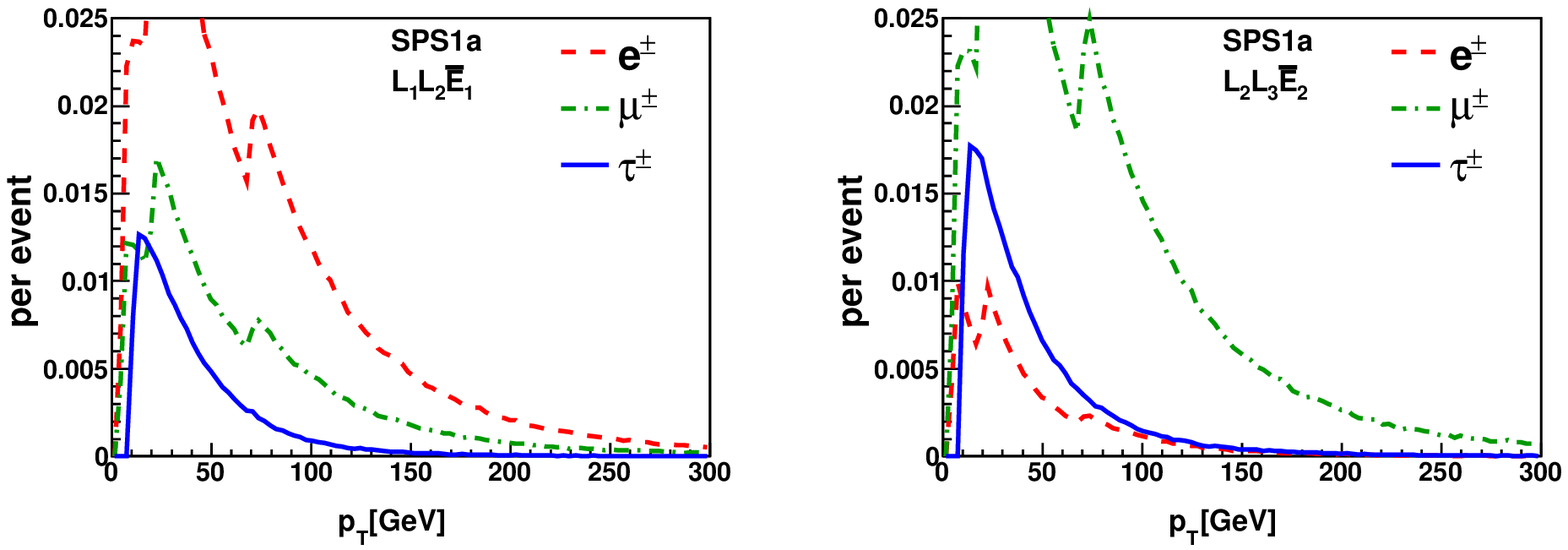}} }
}\fi
\vspace{-6mm}
\caption{Electron, muon and $\tj$ spectra for $L_1L_2\bar E_1$ (left panel) and
  $L_2L_3\bar E_2$ (right panel) at SPS1a. Notice the dips in the spectra at 20
  GeV; this comes from the fact that a lepton with  $p_T<20~\text{GeV}$
  can only pass the cuts if there are at least 3 other high
  $p_T$ leptons in the event. One can also see features at 70 GeV
  due to the cut.
\label{fig-eSPS1aLLE-121} } }

The electron, muon and $\tj$ spectra for $L_1L_2\bar E_1$ and
$L_2L_3\bar E_2$ at SPS1a are shown in Fig.~\ref{fig-eSPS1aLLE-121}.
The two panels look the same apart from different
relative normalizations of individual curves. These different
normalizations can give insight into the flavour composition of the
operator. However, since we do not know how much the R-conserving
decay chain contributes and also since the overall
number of leptons is more or less dictated by the cuts, the most
interesting information here might be the difference in normalization
between the electron and the muon spectra. Since the R-conserving
chain is not expected to give any difference here, it might give some
hint on the dominant flavour in the coupling.

\subsection{Invariant mass distributions}

One of the best tools for studying leptonic decays of the neutralino
are invariant mass distributions. The charged leptons from the decay
will give rise to invariant mass distributions as calculated in
Appendix~B. Since the flavours of the involved leptons are determined
by the flavours in the coupling, the channels in which these
distributions are to be expected, can be deduced from
Table~\ref{tab:LLEChannels}.

Let us illustrate how this works: $L_2L_3\bar E_1$ gives
rise to $e^\pm\mu^\mp$ and $e^\pm\tau^\mp$
final states (see Table~\ref{tab:LLEChannels}).
Therefore, we expect a prompt contribution of the type
(\ref{Eq:InvMass}) in the $e\mu$ invariant mass distribution as well
as contributions of the type (\ref{Eq:Mad}) (from leptonic tau decays)
in the $e\mu$ as well as $ee$ distributions.  In addition, the
$e^\pm\tau^\mp$ final state should give a contribution of type
(\ref{Eq:Mtaul}) to the $e\tj$ distribution.

Since our goal is to determine the flavours of the coupling, we want
to study all dilepton distributions available, as well as
distributions including $\tj$'s. We then want to use the presence of
the various signals together with Table~\ref{tab:LLEChannels} to
determine the operator flavours.

Since we cannot tell which leptons come from the same vertex, we will
have a problem with unrelated lepton pairs, i.e., pairs of leptons
from different parts of the event.  This combinatorial background
will, in some cases, make it difficult to see the signal. However,
since these pairs are not related, they will equally often have the
same charge as the opposite, and since we do not expect any signals in
the same-sign distributions, subtracting those distributions turns out
to be a very effective way of removing these backgrounds.

The effect of the same-sign subtraction can be seen in
Fig.~\ref{fig-InvMemusps1aLLE-231} where we show the di-lepton
distributions for $L_2L_3\bar E_1$ at SPS1a both without and with
same-sign (SS) subtraction.  From this we can see that the same-sign
subtraction is especially important in order to extract the smaller
signals where leptonically decaying taus are involved.  Also for the
distributions involving $\tj$'s it is crucial to apply same-sign subtraction
in order to see the signals.

\subsection{Backgrounds and uncertainties}
The interpretation of any detected excess as a signal is possible
only if we understand well the (backgrounds from) R-conserving decay
chains and also the effect of the cuts on the shapes of the
distributions.
The latter issue is expected to affect the accuracy of the fit of the
theoretical distributions to the data, which makes, e.g., a
reconstruction of the neutralino mass more challenging. The first
issue, on the other hand, depends on the SUSY model and is especially
important for the distributions including $\tj$'s where the
contributions from the R-conserving
chains look very similar to the contributions
from the neutralino decay.  This can be seen in
Fig.~\ref{fig-InvMemusps1aLLE-231} where both the $\mu\tj$ and the
$\tj\tj$ distributions show unexpected features, closely resembling
the expected signals. On the other hand, the corresponding triangular
distributions in the $ee$ and $\mu\mu$ distributions are more distinct
and can therefore be removed from the analysis (practically invisible in
Fig.~\ref{fig-InvMemusps1aLLE-231}).

There are, of course, backgrounds from standard-model processes too, but
these are well handled by the event selection criteria. Other experimental
issues include photon conversion as well as photon and lepton
misidentification. The necessity of same-sign subtraction also implies a
sensitivity to the ability to correctly measure the sign of the
leptons, especially the corresponding charge of $\tj$'s.

For some choices of SUSY parameters,
a real $Z^0$ may be produced in the decay chain and
this would show up in the same-flavour (opposite-sign)
dilepton invariant mass
distributions. However, since this only gives a peak at the $Z^0$
mass, this effect can be easily accounted for.

\FIGURE[ht]{
\let\picnaturalsize=N
\def\picsize{15cm}
\ifx\nopictures Y\else{
\let\epsfloaded=Y
\centerline{\hspace{4mm}{\ifx\picnaturalsize N\epsfxsize \picsize\fi
\epsfbox{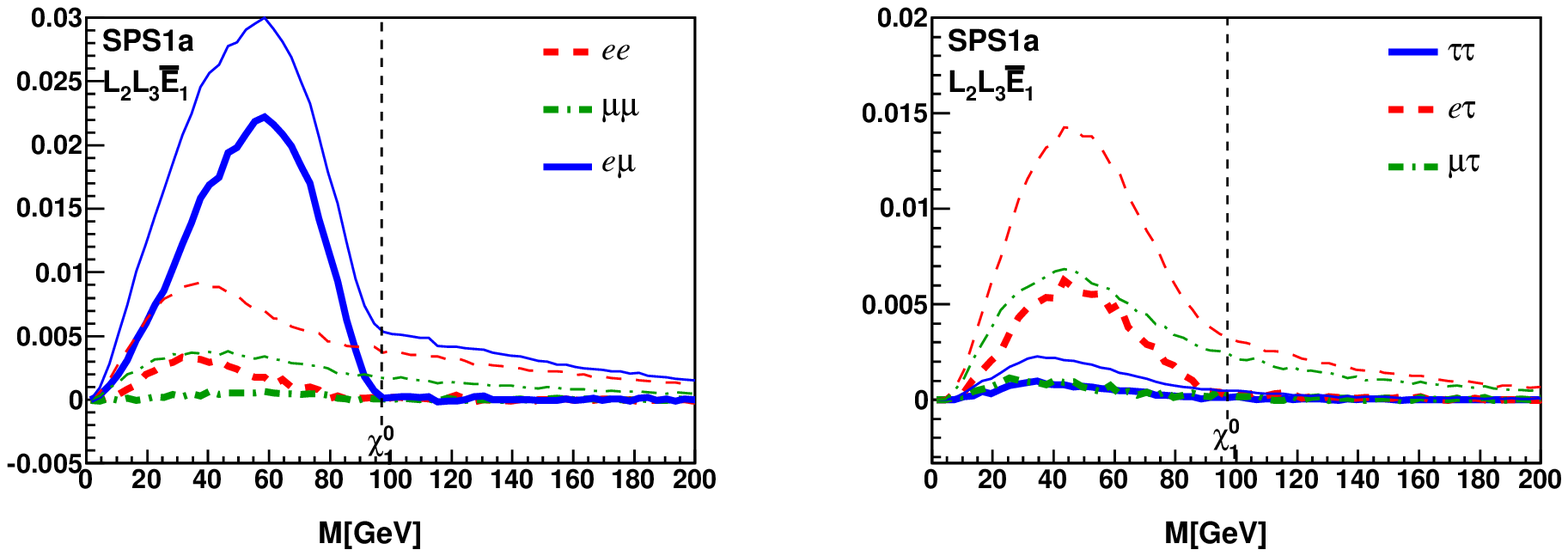}} }
}\fi
\vspace{-6mm}
\caption{Di-lepton invariant masses for $L_2L_3\bar E_1$ at SPS1a, where the neutralino mass is 97.0~GeV. Thin curves: without same-sign subtraction. Heavy curves: $M(\ell\ell^\prime)$ distributions after same-sign subtraction. Here, ``$\tau$'' refers to a hadronic tau.
\label{fig-InvMemusps1aLLE-231} } }

\subsection{Comparison of Monte Carlo results with theory}

We now compare Monte Carlo simulations with the theoretical
distributions derived in Appendix~B.  In what follows, the neutralino
mass and the normalizations of the spectra are fitted to the Monte
Carlo distributions, thereby allowing for a determination of the
neutralino mass.

As can be seen in e.g.\ Fig.~\ref{fig-MabTheory} in appendix~B, most of the expected
distributions look very similar, the main differences being
different kinematical cutoffs. This means we have to be careful with the fitting
procedure to avoid too large degeneracies in the fitted parameters; e.g.\ a
distribution of type (\ref{Eq:InvMass}) may look very similar to a distribution of
type (\ref{Eq:Mad}) with a larger neutralino mass.

To avoid problems with this, each distribution has only been fitted using the
distributions that are relevant due to the assumed operator flavours. For the relevant distributions we
also include a triangular distribution where the cutoff is assumed
known (only for SUSY models where such a contribution from the chain
exists), and a peak at the $Z^0$ mass. The parameters to be fitted are
then the neutralino mass as well as the normalizations of the various
components.

In a realistic scenario where the operator is not known but to be
determined, the relevant distributions can be determined by a
comparison with Monte Carlo data. In cases where this is
difficult one can exploit the expected correlations between the
distributions; if one distribution has a clear cutoff, the others
cannot have neutralino masses significantly lower than that cutoff.
Note that at least one of the distributions will always have a cutoff
close to the neutralino mass.

Assuming $L_2L_3\bar E_1$ to be the dominant R-violating term, various
dilepton mass distributions corresponding to the benchmark points
SPS1a, SPS1b and SPS6 are shown in Fig~\ref{fig-InvM-LLE231}. The left
panels show the pure dilepton invariant mass distributions while the
right panels show the distributions involving $\tj$'s.  As one can
see, the fits are in all cases quite good, but there are some
deviations at the lower end, especially in the $\tj$
distributions. This is due to the cuts imposed and the consequent
deterioration of the signal profile. Especially in the case of SPS6 we
find that the inclusion of the triangular contribution from the chain
into the fit is crucial for a correct fit to the smaller signals.

The situation changes somewhat when the operators
give large numbers of taus in the decay. However,
as Fig.~\ref{fig-InvM-taurich}  shows, the fit is still rather
good. Of special interest might be the case of $L_1L_3\bar E_3$ at
SPS6 where there is no tau from the cascade decay chain and we can therefore
see all the distributions expected, including the softer $\mu\tj$
signal where the $\mu$ comes from a decaying tau.  When comparing the
various couplings at SPS1b it is clear that one cannot tell whether a
signal in the $\tj\tj$ distribution is from the neutralino decay or
the decay chain, the expected distributions are much too similar.

\FIGURE[ht]{
\let\picnaturalsize=N
\def\picsize{15cm}
\ifx\nopictures Y\else{
\let\epsfloaded=Y
\centerline{\hspace{4mm}{\ifx\picnaturalsize N\epsfxsize \picsize\fi
\epsfbox{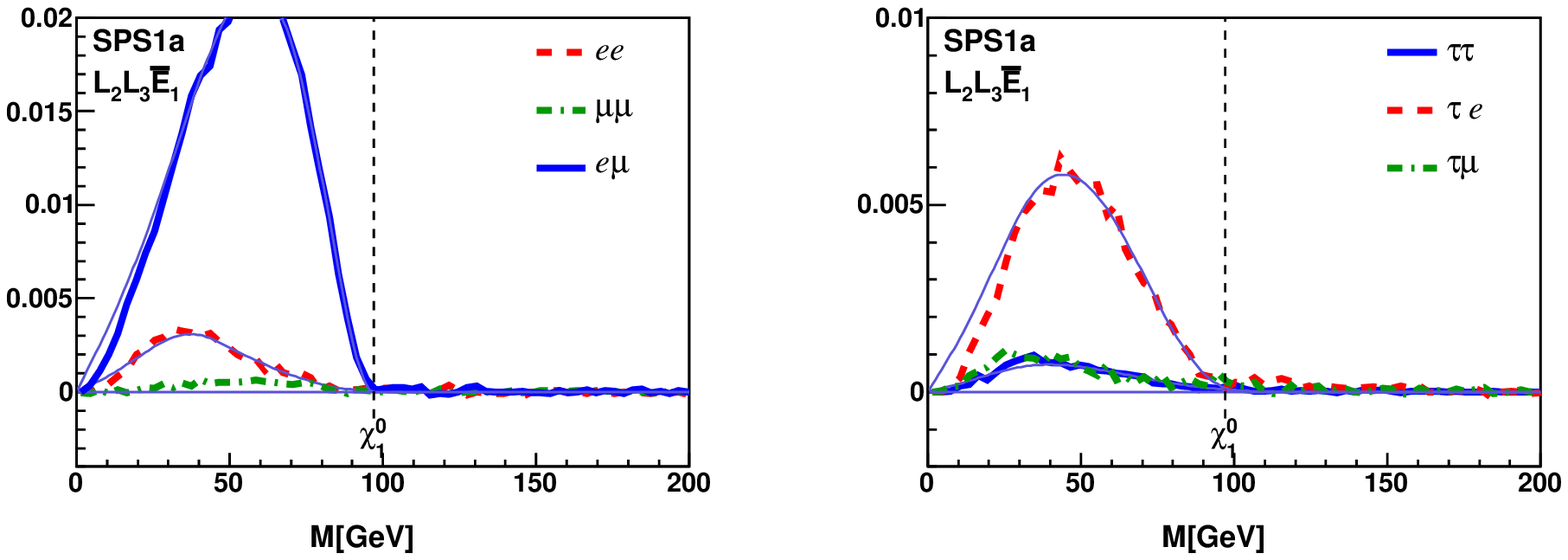}} }
\centerline{\hspace{4mm}{\ifx\picnaturalsize N\epsfxsize \picsize\fi
\epsfbox{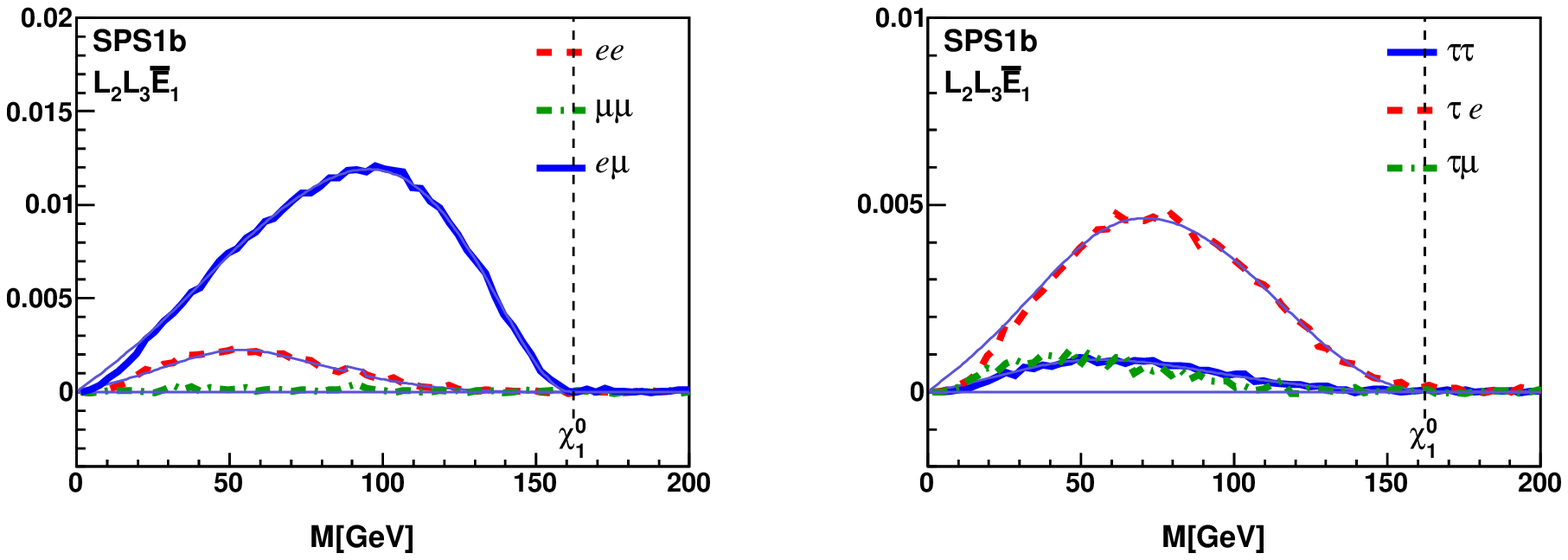}} }
\centerline{\hspace{4mm}{\ifx\picnaturalsize N\epsfxsize \picsize\fi
\epsfbox{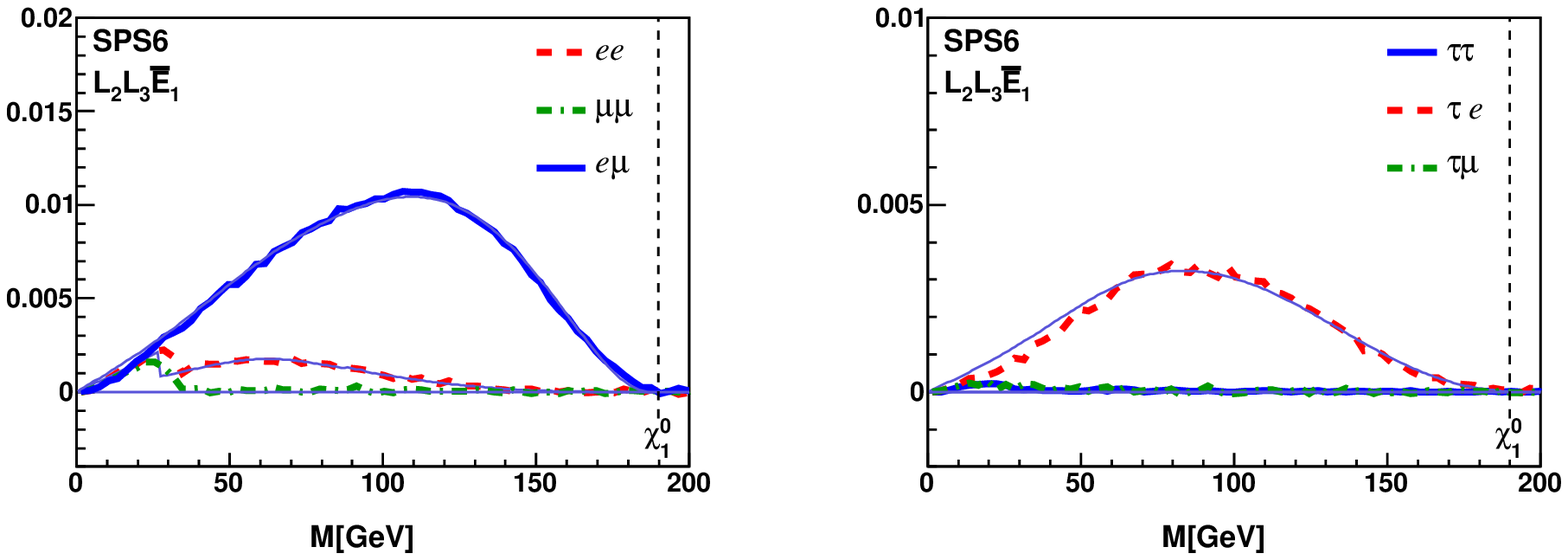}} }
}\fi
\vspace{-6mm}
\caption{Comparison between fitted theoretical distributions (thin curves in violet)
   and Monte Carlo results (heavy curves). The results for $L_2L_3\bar E_1$ are shown for three different SPS points.
\label{fig-InvM-LLE231} } }

\FIGURE[ht]{
\let\picnaturalsize=N
\def\picsize{15cm}
\ifx\nopictures Y\else{
\let\epsfloaded=Y
\centerline{\hspace{4mm}{\ifx\picnaturalsize N\epsfxsize \picsize\fi
\epsfbox{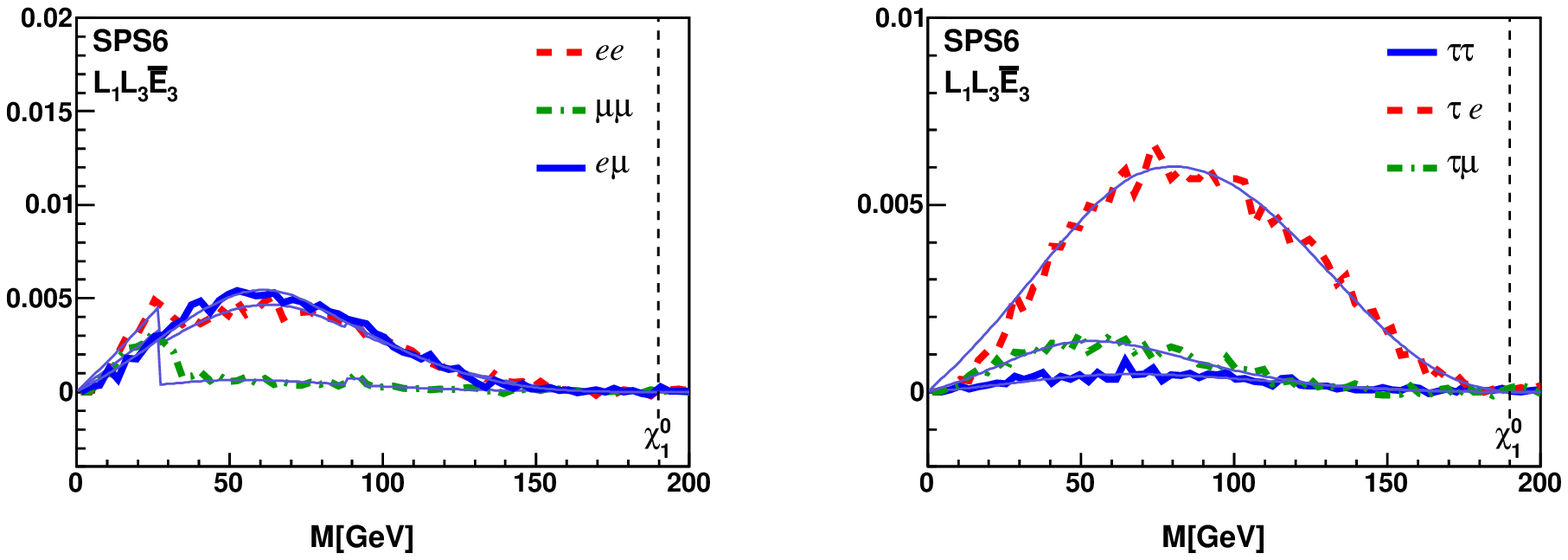}} }
\centerline{\hspace{4mm}{\ifx\picnaturalsize N\epsfxsize \picsize\fi
\epsfbox{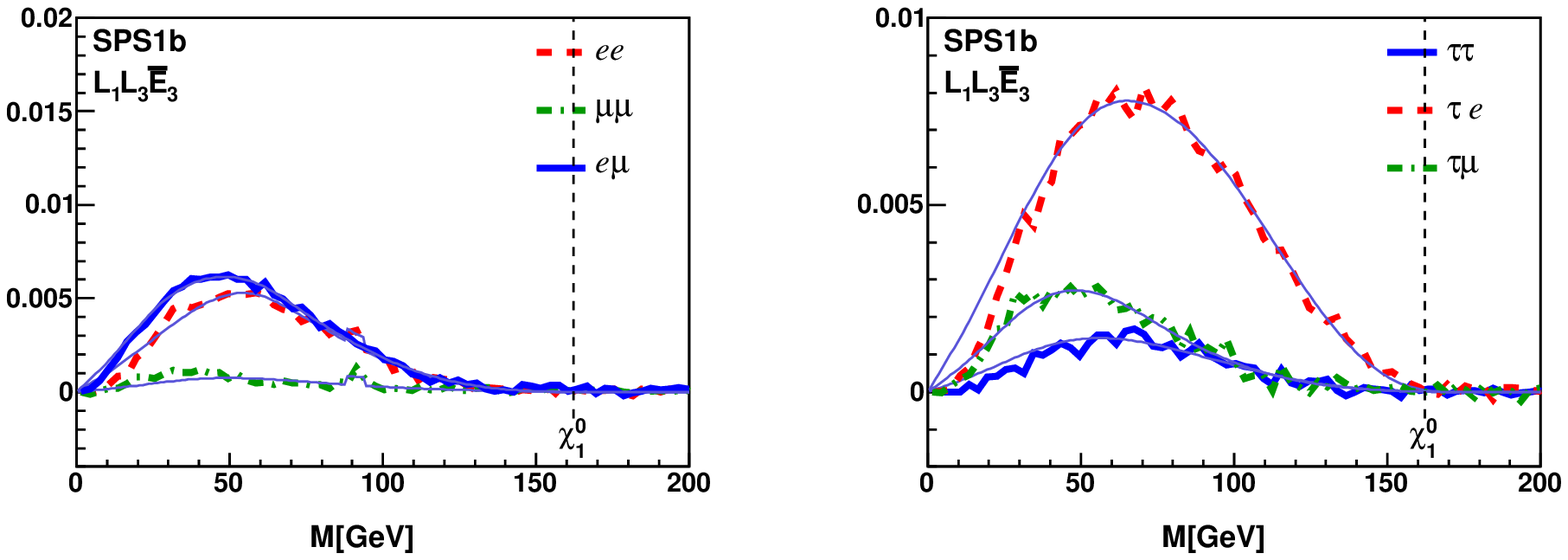}} }
\centerline{\hspace{4mm}{\ifx\picnaturalsize N\epsfxsize \picsize\fi
\epsfbox{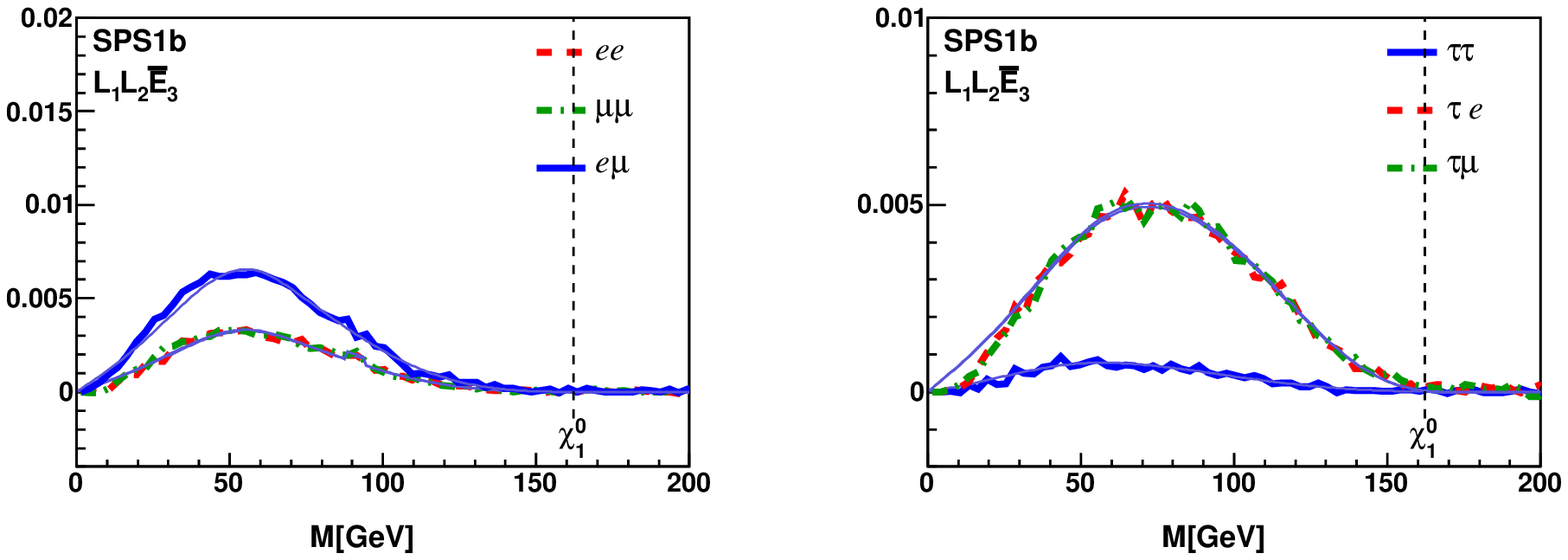}} }
}\fi
\vspace{-6mm}
\caption{The same as Fig.~\ref{fig-InvM-LLE231} for some tau-rich operators at SPS1b and SPS6.
\label{fig-InvM-taurich} } }

\begin{TABLE}{
  \centering
  \caption{Fitted neutralino mass (all values are in GeV) from the various distributions.}
  \label{Tab:chi0Mass}
  \vspace*{8pt}
{\small
  \begin{tabular}{|l|r|r|r|r|r|r|r|}
    \hline
\multicolumn{1}{|c|}{model} & 
\multicolumn{1}{c|}{$M_{\chi_1^0}$} & 
\multicolumn{1}{c|}{$ee$} & 
\multicolumn{1}{c|}{$\mu\mu$} & 
\multicolumn{1}{c|}{$e\mu$} & 
\multicolumn{1}{c|}{$e\tau$} & 
\multicolumn{1}{c|}{$\mu\tau$} & 
\multicolumn{1}{c|}{$\tau\tau$} \\\hline
    $L_2L_3\bar E_1$, SPS1a & 97.0  & 116 $\pm$ 0.3 & --- & 96.8 $\pm$ 0.1 & 106 $\pm$ 0.5 & --- & --- \\
    $L_2L_3\bar E_1$, SPS1b & 162.2 & 167 $\pm$ 0.5 & --- & 171 $\pm$ 0.1  & 171 $\pm$ 0.4 & --- & --- \\
    $L_2L_3\bar E_1$, SPS6  & 189.8 & 195 $\pm$ 1  & --- & 187 $\pm$ 0.1   & 200 $\pm$ 0.5 & --- & --- \\
    $L_1L_3\bar E_3$, SPS6  & 189.8 & 205 $\pm$ 0.2 & 295 $\pm$ 6 & 195 $\pm$ 1  & 198 $\pm$ 0.8 & 227 $\pm$ 2 & 224 $\pm$ 3\\
    $L_1L_3\bar E_3$, SPS1b & 162.2 & 172 $\pm$ 2  & 272 $\pm$ 13  & 176 $\pm$ 0.4 & 172 $\pm$ 0.7 & 204 $\pm$ 3 & 187 $\pm$ 3\\
    $L_1L_2\bar E_3$, SPS1b & 162.2 & 173 $\pm$ 1  & 172 $\pm$ 1 & 172 $\pm$ 0.5 & 172 $\pm$ 0.4 & 172 $\pm$ 0.4& ---\\
    \hline
  \end{tabular}  }
}
\end{TABLE}

Table~\ref{Tab:chi0Mass} shows the neutralino masses estimated from
the fits shown in Figs.~\ref{fig-InvM-LLE231} and
\ref{fig-InvM-taurich}. Included are also statistical errors on the fitted neutralino masses.
In addition we expect systematic uncertainties from the theory, especially concerning
the inclusion of the event selection. Therefore, the uncertainties given in Table~\ref{Tab:chi0Mass}
should not be taken too seriously but rather as lower bounds on the uncertainty.

It seems that the measurement of the
neutralino mass from the $e \tj$ mass always gives an
overestimate. The same is true for the di-lepton mass distributions
when one of the leptons comes from a decaying tau, although, in the
latter case, the overestimate is less severe. In the relevant cases, the
distribution when both leptons come from decaying taus are included but
the impact on the resulting fit is not very large; we see from the
$\mu\mu$ distribution for the $L_1L_3\bar E_3$ operator, where both muons
have to come from decaying taus, that the signal is too small to get a good fit
and the resulting neutralino mass estimate is completely useless. This is reflected in the 
very large statistical uncertainty. In general the
possible improvement by including this last distribution is smaller than
the uncertainties from e.g.\ the effects of the event selection.

When interpreting these neutralino mass estimates, one has to keep in
mind the large uncertainty associated with the effects of the event
selection on the distribution. In the theoretical curves, these
effects are included through the lower limit on the mass ratio
entering in Eq.~(\ref{Eq:Mad}). The value of this limit has great
impact on the resulting distribution and therefore on the estimated
neutralino mass. It is therefore clear that a better understanding of
the effect of the cuts is very important in order to get the best
possible neutralino mass estimate.

In order to improve these fits one could make use of various expected correlations;
e.g.\ as mentioned before, the neutralino mass of course has to be the same for all
distributions and one could use the most suitable distribution, presumably the one
with a cutoff at the neutralino mass, to constrain the fits to the smaller distributions
with cutoffs significantly below the neutralino mass. Ideally one would want to use the
fitted parameters to determine the decay channels for the neutralino, but due to the
aforementioned degeneracy problems it is hard to achieve the required accuracy in the
parameters, this could be improved if one uses the fact that e.g.\ an $e\tau$ decay channel
for the neutralino gives correlated signals in both the $e\tj$, $ee$ and $e\mu$ distributions,
where the relative sizes of the signals are basically given by the branching fractions of the tau
(together with tau tagging efficiency and detector acceptances).

\subsection{Some counting observables}
\label{subsect:counting}

We now consider observables that can bring
additional insight into the operator structure. One obvious candidate
is the determination of the branching fractions of the neutralino into
various channels. This can be achieved by studying the leptonic
structure of the final states.

To get a clean sample to work with, we look at the subset of all
events where we have exactly four leptons of which two are positively
charged and two are negatively charged. This restriction
increases the probability that the leptons have all come from
decaying neutralinos.

Let $P_{2e^-2e^+}$ denote the fraction of these
events that have exactly two electrons and two positrons:
\begin{equation}\label{Eq:P2e2eDef}
P_{2e^-2e^+}
=
\frac{N[(2e^-)\&(2e^+)]}
{N_{4\ell}}
\end{equation}
(and similarly for $P_{2\mu^-2\mu^+}$), with
\begin{equation}
N_{4\ell}=N[(4e)]+N[(3e)\&(1\mu)]+N[(2e)\&(2\mu)]+N[(1e)\&(3\mu)]+N[(4\mu)],
\end{equation}
subject to the constraint that the charges should sum to zero.
Further, let $P_{2e2\mu}$ denote the
fraction that has two same-sign electrons and two same-sign muons,
i.e., $2e^-2\mu^+$ or $2e^+2\mu^-$:
\begin{equation}\label{Eq:P2e2muDef}
P_{2e2\mu}
=\frac{N[(2e^-)\&(2\mu^+)]+N[(2e^+)\&(2\mu^-)]}
{N_{4\ell}}.
\end{equation}

In the event that {\it all four leptons come from decaying
neutralinos}, $P_{e^-e^+}$ is the probability that a
neutralino decays to an electron-positron pair. Since under these
assumptions, the only possibility to achieve $2e^-2e^+$ in one event
is to have two neutralinos that both decay to electron-positron pairs,
we get
\begin{equation}
P_{2e^-2e^+}=P_{e^-e^+}^2.
\end{equation}
Note that $P_{e^-e^+}$ is a probability\footnote{This is the
  probability that a randomly picked neutralino in an event with
  exactly four leptons, is decaying through a certain channel. In
  other words; it is a branching fraction, but depends strongly on the
  cuts used to calculate it and it should not be confused with a
  ``true'' branching fraction.}  referring to {\it one} neutralino,
whereas $P_{2e^-2e^+}$ refers to the whole event.

Similarly, $P_{2\mu^-2\mu^+}=P_{\mu^-\mu^+}^2$ and $P_{2e2\mu}=\half
P_{e\mu}^2$ where $P_{e\mu}=P_{e^-\mu^+}+P_{e^+\mu^-}$ and
$P_{\mu^-\mu^+},P_{e^-\mu^+}$ and $P_{e^+\mu^-}$ are defined in
accordance with $P_{e^-e^+}$. The factor $\half$
in the second of these
equations comes from the requirement that the same-flavour pairs
also have the same charge.
From this we can calculate the values $P_{e^-e^+},P_{\mu^-\mu^+}$ and
$P_{e\mu}$.

A problem with the above calculation is that it does not tell us
anything about channels including taus.  Since we require exactly four
leptons, events with taus are only included when the taus decay
leptonically. This also means that decay channels with light leptons
are favoured by the cuts over those with taus and therefore care has
to be taken when interpreting the numbers. However, keeping this in
mind, the probabilities $P_{e^-e^+},P_{\mu^-\mu^+}$ and $P_{e\mu}$ do
tell us which decay channels are important.

In order to get some information about decay channels involving taus,
we need to also study events with three rather than four leptons. Let
us study events with one pair of same-sign, same-flavour leptons
($\ell^\pm\ell^\pm$) and one extra lepton of opposite charge and
different flavour ($\ell^{'\mp}$).  Such events must all contain one
different-flavour, opposite-sign lepton pair ($\ell^\pm\ell^{'\mp}$)
and one extra lepton ($\ell^\pm$), the extra lepton stemming from a
neutralino that decayed either to a $\tau$ plus a lepton where the $\tau$
decayed hadronically, or to a lepton pair where one of the leptons was
lost to the detector.

Since counting observables depend not only on production rate and
detector efficiencies, but also on {\it a priori\ } unknown factors
such as selection efficiencies, it is useful to study, instead, ratios
such as
\begin{equation}\label{Eq:3l-event-frac}
    \frac{P_{2e\mu}}{P_{2\mu e}}
\equiv\frac{N_{2e^-\mu^+}+N_{2e^+\mu^-}}{N_{2\mu^-e^+}+N_{2\mu^+e^-}}
=\frac{N_{2e^-\mu^+}}{N_{2\mu^-e^+}},
\end{equation}
where the numbers $N$ refer to the number of events with the specified
number of leptons. The last equality above is based on the assumption that
both charges are produced with the same probability and detected with
the same efficiency.

Let us now consider a particular case wherein
we start with four leptons ($e, \mu$) stemming directly from the decays
of two neutralinos with one of the four (randomly picked) getting
lost (absorbed in a jet,  too little $p_T$, too large a rapidity etc.).
The flavour distribution of the four-lepton state is then given by the probabilities
$P_{\mu^-\mu^+},P_{e\mu}$ and $P_{e^-e^+}$. Assuming that the efficiencies
for muons are the same as for electrons\footnote{While this is not exactly
true, the differences are marginal, and, at the level of sophistication
of this study, can be ignored.}, we have
\begin{equation}\label{Eq:3l-frac-est}
    \frac{P_{2e\mu}}{P_{2\mu e}}=\frac{P_{e\mu}(2P_{e^-e^+}+P_{e\mu})}{P_{e\mu}(2P_{\mu^-\mu^+}+P_{e\mu})} = \frac{2P_{e^-e^+}+P_{e\mu}}{2P_{\mu^-\mu^+}+P_{e\mu}},
\end{equation}
where the factor 2 is a combinatorial factor arising from the
fact that the individual leptons could have
arisen from the two different neutralinos.

Since the prediction of Eq.~(\ref{Eq:3l-frac-est}) does not take into
account the possibility of the fourth missing particle being a
hadronically decaying tau, it will, occasionally, give an inaccurate
answer. This apparent lacuna could, however, be used
to our advantage:
a comparison of the prediction of (\ref{Eq:3l-frac-est}) with a direct
measurement of the ratio (\ref{Eq:3l-event-frac}) would give us
a quantitative measure of the tau decay channel. If
the decay channel $\chi_1^0\rightarrow
e+\tau+\nu$ is a significant one,
the measured fraction will be much larger than the
prediction (\ref{Eq:3l-frac-est}), while if the channel
$\chi_1^0\rightarrow \mu+\tau+\nu$ is present,
Eq.~(\ref{Eq:3l-frac-est}) will give an overestimate.

Independent measures of the branching fractions to various final states
are given by the integrals over invariant mass, namely
\begin{equation}\label{Eq:InvMIntegral}
    \int \mathcal{N}_{ll'} \equiv
\int_0^\infty \frac{d\mathcal{N}_{ll'}(M_{ll'})}{dM_{ll'}} dM_{ll'},
\end{equation}
where $l,l'\in \{e,\mu\}$ and the distributions are evaluated after same-sign
subtraction. Whereas the $P$'s above refer to the four-lepton events where the charges sum to zero, these integrals refer to {\it all} events that pass the cuts.

Both the integrals over the invariant mass distributions and the
probabilities $P_{e^-e^+},P_{\mu^-\mu^+}$ and $P_{e\mu}$ must be
interpreted with the used cuts in mind; in both cases leptons are
favoured over taus. The resulting branching fractions include
leptonically decaying taus in addition to prompt leptons
but the former
contribution is strongly dependent on the cuts and therefore hard to
estimate analytically.
Due to the requirement of exactly four leptons, the
probabilities $P_{e^-e^+},P_{\mu^-\mu^+}$ and $P_{e\mu}$ essentially
include only events where there are no taus in the decay or all taus
decay leptonically. The invariant mass distributions, on the other
hand, are expected to contain a larger amount of lepton pairs from
sources other than neutralino decay. This can, however, be handled by the same-sign
subtraction as well as by comparing the distributions with the
theoretical expectations. The probabilities
$P_{e^-e^+},P_{\mu^-\mu^+}$ and $P_{e\mu}$ are more sensitive to
contamination by single leptons from other sources and are therefore
more dependent on a clean event sample.

\begin{TABLE}{
  \centering
    \begin{tabular}{|l|r|r|r|r|r|r|r|r|}
    \hline
    model & $P_{e^-e^+}$ & $P_{e\mu}$ & $P_{\mu^-\mu^+}$ & (\ref{Eq:3l-event-frac}) & (\ref{Eq:3l-frac-est}) & $\int \mathcal{N}_{ee}$ & $\int \mathcal{N}_{e\mu}$ & $\int \mathcal{N}_{\mu\mu}$ \\\hline
    $L_1L_2\bar E_1$, SPS1a & 0.48 & 0.51 & 0.02 & 2.5 & 2.69 & 0.48 & 0.51 & 0.01\\
    $L_1L_2\bar E_2$, SPS1a & 0.02 & 0.49 & 0.50 & 0.37 & 0.35 & 0.01 & 0.49 & 0.50\\
    $L_1L_2\bar E_3$, SPS1a & 0.26 & 0.48 & 0.27 & 0.93 & 0.98 & 0.27 & 0.47 & 0.26\\
    $L_1L_3\bar E_1$, SPS1a & 0.81 & 0.16 & 0.01 & 19.3 & 10.1 & 0.86 & 0.12 & 0.03\\
    $L_1L_3\bar E_2$, SPS1a & 0.02 & 0.81 & 0.17 & 0.26 & 0.75 & 0.02 & 0.87 & 0.11\\
    $L_1L_3\bar E_3$, SPS1a & 0.45 & 0.45 & 0.09 & 3.77 & 2.15 & 0.43 & 0.44 & 0.13\\
    $L_2L_3\bar E_1$, SPS1a & 0.17 & 0.81 & 0.03 & 3.26 & 1.33 & 0.11 & 0.87 & 0.02\\
    $L_2L_3\bar E_2$, SPS1a & 0.01 & 0.16 & 0.82 & 0.05 & 0.10 & 0.02 & 0.11 & 0.86\\
    $L_2L_3\bar E_3$, SPS1a & 0.09 & 0.45 & 0.45 & 0.27 & 0.47 & 0.12 & 0.44 & 0.44\\
    \hline
    \end{tabular}
  \caption{The three first columns give calculated values of $P_{e^-e^+},P_{\mu^-\mu^+}$ and
$P_{e\mu}$ for all the different operators at SPS1a. We also give the Monte Carlo  value  of the fraction (\ref{Eq:3l-event-frac})
as well as the prediction from Eq.~(\ref{Eq:3l-frac-est}). Finally, we tabulate the integrals of the invariant
    mass distributions, normalized so that the sum is one. This should
    provide an independent estimate of the branching fractions of the
    first three columns. }\label{tab:Counting} }
\end{TABLE}

\begin{TABLE}{
  \centering
    \begin{tabular}{|l|r|r|r|r|r|r|r|r|}
    \hline
    model & $P_{e^-e^+}$ & $P_{e\mu}$ & $P_{\mu^-\mu^+}$ & (\ref{Eq:3l-event-frac}) & (\ref{Eq:3l-frac-est}) & $\int \mathcal{N}_{ee}$ & $\int \mathcal{N}_{e\mu}$ & $\int \mathcal{N}_{\mu\mu}$ \\\hline
    $L_2L_3\bar E_1$, SPS1a & 0.17 & 0.81 & 0.03 & 3.26 & 1.33 & 0.11 & 0.87 & 0.02\\
    $L_2L_3\bar E_1$, SPS1b & 0.18 & 0.80 & 0.02 & 3.72 & 1.38 & 0.12 & 0.87 & 0.01\\
    $L_2L_3\bar E_1$, SPS6  & 0.17 & 0.81 & 0.03 & 3.54 & 1.33 & 0.12 & 0.85 & 0.03\\\hline
    $L_1L_2\bar E_3$, SPS1a & 0.26 & 0.48 & 0.27 & 0.93 & 0.98 & 0.27 & 0.47 & 0.26\\
    $L_1L_2\bar E_3$, SPS1b & 0.25 & 0.50 & 0.25 & 1.00 & 0.99 & 0.25 & 0.50 & 0.26\\\hline
    $L_1L_3\bar E_3$, SPS1a & 0.45 & 0.45 & 0.09 & 3.77 & 2.15 & 0.43 & 0.44 & 0.13\\
    $L_1L_3\bar E_3$, SPS1b & 0.43 & 0.49 & 0.08 & 3.37 & 2.08 & 0.41 & 0.50 & 0.08\\
    \hline
    \end{tabular}
  \caption{Similar to table~\ref{tab:Counting} for $L_2L_3\bar E_1$ at SPS1a, SPS1b and SPS6, as well as $L_1L_2\bar E_3$ and $L_1L_3\bar E_3$ at SPS1a and SPS1b.}\label{tab:Counting2} }
\end{TABLE}

All the observables discussed in this section for all $LL\bar E$
couplings at SPS1a are shown in Table~\ref{tab:Counting}. When
comparing the probabilities $P_{e^-e^+},P_{\mu^-\mu^+}$ and $P_{e\mu}$
and the integrals over invariant mass distributions, we see that they
show similar structures, implying
that they both are good
indications for the relevant decay channels.

In table~\ref{tab:Counting2}, we compare these observables for the same
operator ($L_2L_3\bar E_1$) but at three different parameter points:
SPS1a, SPS1b and SPS6. As the values are similar at all three points,
we may conclude that they are largely independent of the underlying
SUSY model.

However, also shown in table~\ref{tab:Counting2} are $L_1L_2\bar E_3$
and $L_1L_3\bar E_3$ at SPS1a and SPS1b and here some
noticeable differences arise.  First of all, for $L_1L_2\bar
E_3$ we naively expect to get $0.25$, $0.5$ and $0.25$ for
the branching fractions to $e^+e^-$, $e\mu$ and $\mu^+\mu^-$
respectively, a consequence of the fact that the tau will go
as often to a $\mu$ as an $e$.
As can be seen in table~\ref{tab:Counting2}
this is what we get at SPS1b while SPS1a gives slightly larger values
for the same-flavour ($ee$ and $\mu\mu$) channels.  This difference is
consistent with contamination by lepton pairs from the cascade decay
chain, which is important for SPS1a but not for SPS1b.

Since this kind of contamination will be most important for tau-rich
operators where there are fewer leptons from the neutralino decay, we
also expect to see it in $L_1L_3\bar E_3$, which, in fact, we do. For
a $L_1L_3\bar E_3$ coupling, we expect the same amount of $e^+e^-$ and
$e\mu$ from the $e\tau$ channel and from the $\tau\tau$ channel we
expect contributions to $e^+e^-$, $e\mu$ and $\mu^+\mu^-$ in the
ratios $1:2:1$ (again due to the two possibilities for $e\mu$). The
result should be that the branching fraction to $e\mu$ is larger than
the fraction to $e^+e^-$ by about the same amount as the total
branching fraction to $\mu^+\mu^-$ and this is rather close to what we
see for SPS1b, but at SPS1a the $e^+e^-$ and $e\mu$ values are of the same size,
indicating some contamination in the $e^+e^-$ (and $\mu^+\mu^-$)
channel from the cascade chain.

Four out of the nine $LL\bar E$ operators are such that they allow one
channel with prompt light leptons and one channel with a light lepton
and a tau. These operators are $L_1L_3\bar E_1$, $L_1L_3\bar E_2$,
$L_2L_3\bar E_1$ and $L_2L_3\bar E_2$.  Let us look at $L_2L_3\bar
E_1$ as an example (the same logic applies to all of them). In this
case, the prompt lepton channel is $e\mu$ and, in addition, we get
decays to $e\tau$. Since the tau will decay equally often to electrons
and muons, we expect a large signal in the $e\mu$ channel both from
the prompt leptons and the $e\tau$ channel. In addition we expect some
contribution to the $e^+e^-$ channel from the $e\tau$ decays, while
$\mu^+\mu^-$ should give nothing. As can be seen from
tables~\ref{tab:Counting} and \ref{tab:Counting2} this is, to a good
approximation, indeed the case.  We also notice that the dominance of
the largest channel is quite similar for all the SPS points considered
and all the four couplings. However, there is a clear difference
between the two ways of measuring the branching fractions; for
$L_2L_3\bar E_1$ we have $P_{e\mu}\approx0.81$ while $\int
\mathcal{N}_{e\mu}\approx0.87$ and this difference is nearly independent
of the choices for the SPS point and the R-violating coupling as well.

The estimate of the branching fraction of the dominant channel will
depend strongly on which cuts are used as well as how many leptons not
originating from neutralino decay we have. Therefore it is very
difficult to make any precise predictions, but the difference seen
between the two measures is expected due to the differences in the
cuts used.

If we now compare the predictions from Eq.~(\ref{Eq:3l-frac-est}) with
the direct measurement of the fraction~(\ref{Eq:3l-event-frac}), we
see that operators with an $e\tau$ channel give a measured fraction
about twice as large as the prediction~(\ref{Eq:3l-frac-est}) (see
$L_1L_3\bar E_1$, $L_1L_3\bar E_3$ and $L_2L_3\bar E_1$ in
table~\ref{tab:Counting}), while for operators with a $\mu\tau$
channel the prediction~(\ref{Eq:3l-frac-est}) overestimates the
fraction by about a factor two (see $L_1L_3\bar E_2$, $L_2L_3\bar E_2$
and $L_2L_3\bar E_3$ in table~\ref{tab:Counting}).  This is in
accordance with our expectation and illustrates that we can infer the
presence of tau in the operator without having to rely on tau tagging.

\subsection{Redefining taus}
Another way of getting a grip on the decay channels containing taus is
to include hadronic taus in the analysis; in this section leptons also
include $\tj$'s and the only cuts imposed on
the leptons are the acceptance cuts of $p_T > 5$ GeV (10 GeV for $\tj$) and isolation.

The event selection used here is {\it exactly} four leptons (including $\tj$'s),
two positive and two negative.
In addition, we require $p_T^\text{miss}>100$ GeV.

\begin{TABLE}{
  \centering
    \begin{tabular}{|l|l|l|l|l|l|l|}
    \hline
    model & $P_{e^-e^+}$ & $P_{\mu^-\mu^+}$ & $P_{\tj^-\tj^+}$ & $P_{e\mu}$ & $P_{e\tj}$ & $P_{\mu\tj}$ \\\hline
    $L_1L_2\bar E_1$, SPS1a & 0.38 & 0.02 & 0.004 & 0.41 & 0.08 & 0.03\\
    $L_1L_2\bar E_2$, SPS1a & 0.02 & 0.40 & 0.004 & 0.40 & 0.03 & 0.07\\
    $L_1L_2\bar E_3$, SPS1a & 0.08 & 0.09 & 0.06 & 0.16 & 0.30 & 0.31\\
    $L_1L_3\bar E_1$, SPS1a & 0.44 & 0.003 & 0.03 & 0.10 & 0.36 & 0.02\\
    $L_1L_3\bar E_2$, SPS1a & 0.02 & 0.09 & 0.03 & 0.44 & 0.04 & 0.33\\
    $L_1L_3\bar E_3$, SPS1a & 0.09 & 0.02 & 0.26 & 0.10 & 0.40 & 0.13\\
    $L_2L_3\bar E_1$, SPS1a & 0.09 & 0.02 & 0.03 & 0.44 & 0.32 & 0.05\\
    $L_2L_3\bar E_2$, SPS1a & 0.005 & 0.45 & 0.03 & 0.10 & 0.02 & 0.35\\
    $L_2L_3\bar E_3$, SPS1a & 0.02 & 0.09 & 0.26 & 0.10 & 0.13 & 0.39\\
    \hline
    $L_2L_3\bar E_1$, SPS1a & 0.09 & 0.02 & 0.03 & 0.44 & 0.32 & 0.05\\
    $L_2L_3\bar E_1$, SPS1b & 0.09 & 0.01 & 0.04 & 0.44 & 0.35 & 0.04\\
    $L_2L_3\bar E_1$, SPS6 & 0.10 & 0.02 & 0.01 & 0.54 & 0.32 & 0.02\\
    \hline
    \end{tabular}
  \caption{Branching fractions for different final states, estimated from events with 4 leptons (including $\tj$) with a total charge-sum of zero.}\label{tab:Counting3} }
\end{TABLE}

We can then expand the definitions (\ref{Eq:P2e2eDef}) and
(\ref{Eq:P2e2muDef}) to include $\tj$'s.  Using these probabilities we
can estimate all branching fractions of neutralinos decaying to lepton
pairs.  The result for all couplings at SPS1a is shown in
table~\ref{tab:Counting3} along with the result for $L_2L_3\bar E_1$
at SPS1a, SPS1b and SPS6. It is clear from table~\ref{tab:Counting3}
that this method does pick out the correct channels as the dominant
ones; for all couplings, the two
dominant ones are the same as those given in
table~\ref{tab:LLEChannels}.  It should be noted
that channels with tau will give rise to several different final
states depending on the decay of the tau.
If we look at $L_2L_3\bar E_1$ at
SPS6 the above points become clear; the two dominant final states are
$e\mu$ and $e\tj$ and
the $e\tau$ channel contributes
about $0.05-0.10$ to both $P_{e^+e^-}$ and $P_{e\mu}$ and about
$0.3$ to $P_{e\tj}$ which is consistent with the branching ratios of
taus. We also have some contamination of around $0.02$ in all
channels.

Compared to the cases of table~\ref{tab:Counting}, there are
now more leptons that do not arise from neutralino decays. For example,
for the SPS1a point, each of the couplings considered results in more
$\tj$'s than expected
from the neutralino decays alone. And when comparing the values for
$L_2L_3\bar E_1$ at SPS1a, SPS1b and SPS6 we notice some differences
which are consistent with the differences in the contributions from
the cascade decay chain; e.g.\ we see that SPS6 gives fewer $\tj$'s
than SPS1a and SPS1b. The conclusion is that the numbers are rather
sensitive to the parameter point.

These issues make it hard to interpret the exact values of the
branching fractions. It is also important to remember that tau tagging
efficiencies should be added on top of this and that will further
suppress the $\tj$'s in comparison with the light leptons.

\subsection{Combinations of couplings--Case study}
To illustrate the usefulness of the observables discussed in the
previous sections, let us study two cases where two operators are
comparable, namely,
\begin{equation}
  \begin{array}{ll}
    \text{mix1}:\qquad & \hbox{$L_1L_2\bar E_1=L_2L_3\bar E_2=10^{-4}$;} \\
    \text{mix2}:\qquad & \hbox{$L_1L_2\bar E_2=L_2L_3\bar E_1=10^{-4}$.}
  \end{array}
\end{equation}

\FIGURE[ht]{
\let\picnaturalsize=N
\def\picsize{15cm}
\ifx\nopictures Y\else{
\let\epsfloaded=Y
\centerline{\hspace{4mm}{\ifx\picnaturalsize N\epsfxsize \picsize\fi
\epsfbox{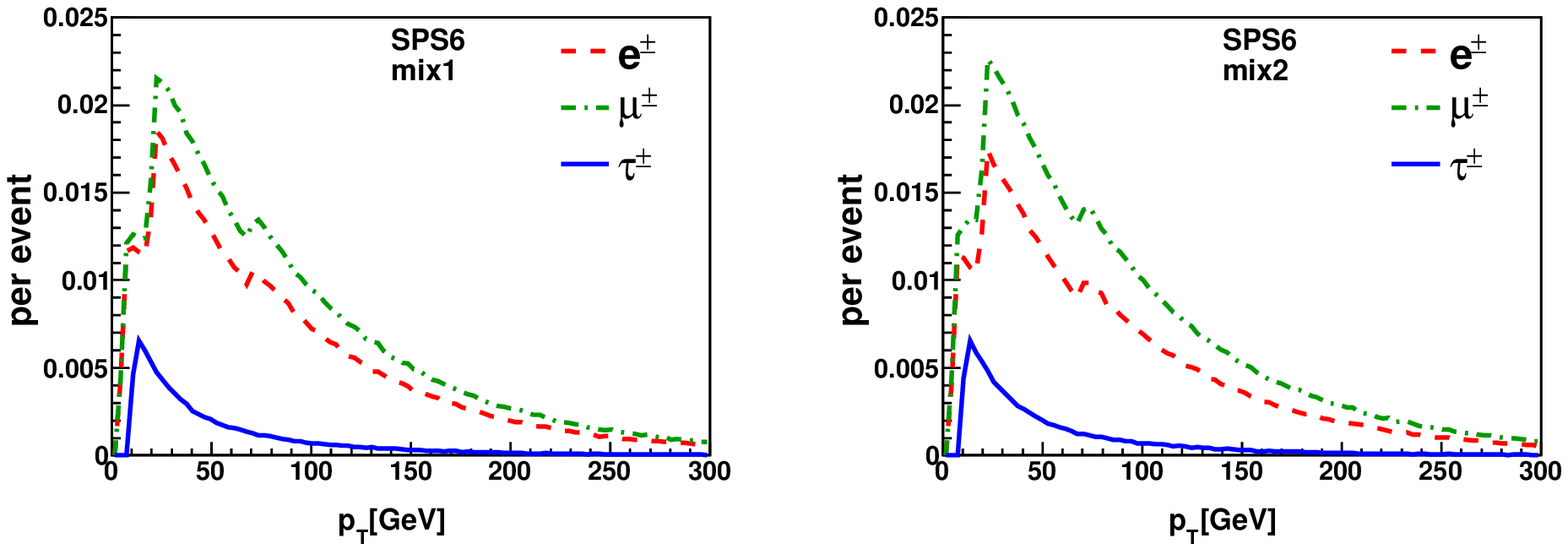}} }
}\fi
\vspace{-6mm}
\caption{Electron, muon and $\tj$ spectra for mix1 (left) and mix2 (right) at SPS6.
\label{fig-MixSpec} } }

\FIGURE[ht]{
\let\picnaturalsize=N
\def\picsize{15cm}
\ifx\nopictures Y\else{
\let\epsfloaded=Y
\centerline{\hspace{4mm}{\ifx\picnaturalsize N\epsfxsize \picsize\fi
\epsfbox{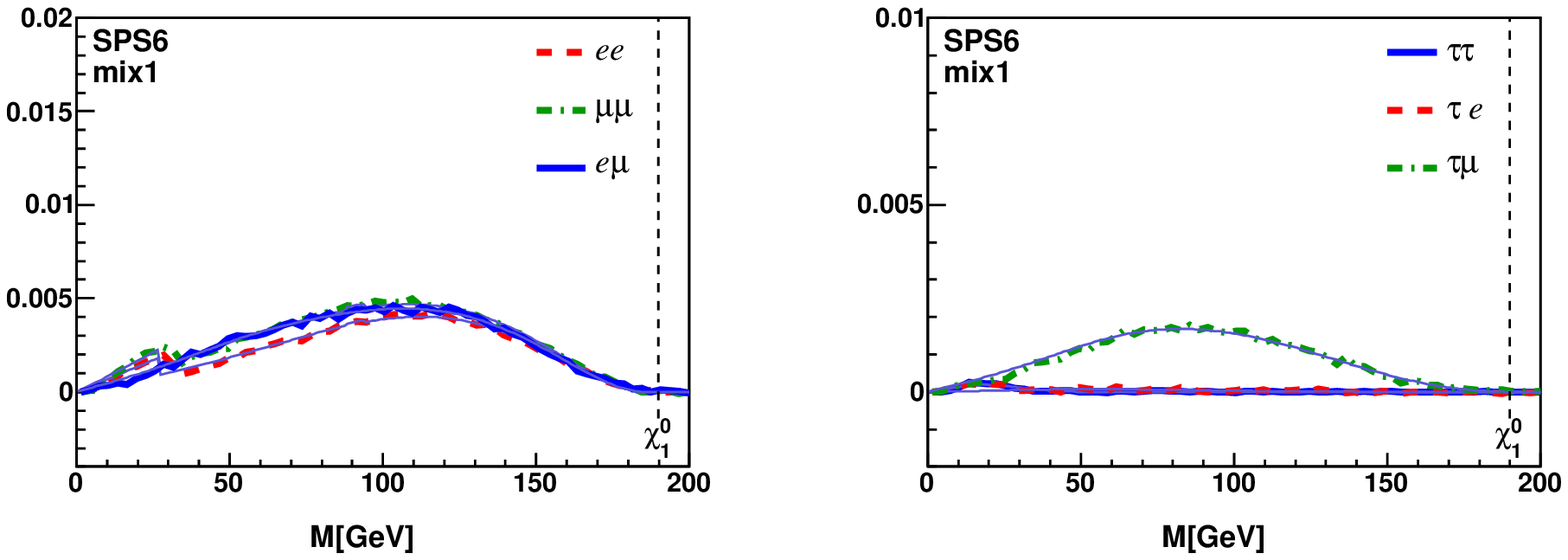}} }
\centerline{\hspace{4mm}{\ifx\picnaturalsize N\epsfxsize \picsize\fi
\epsfbox{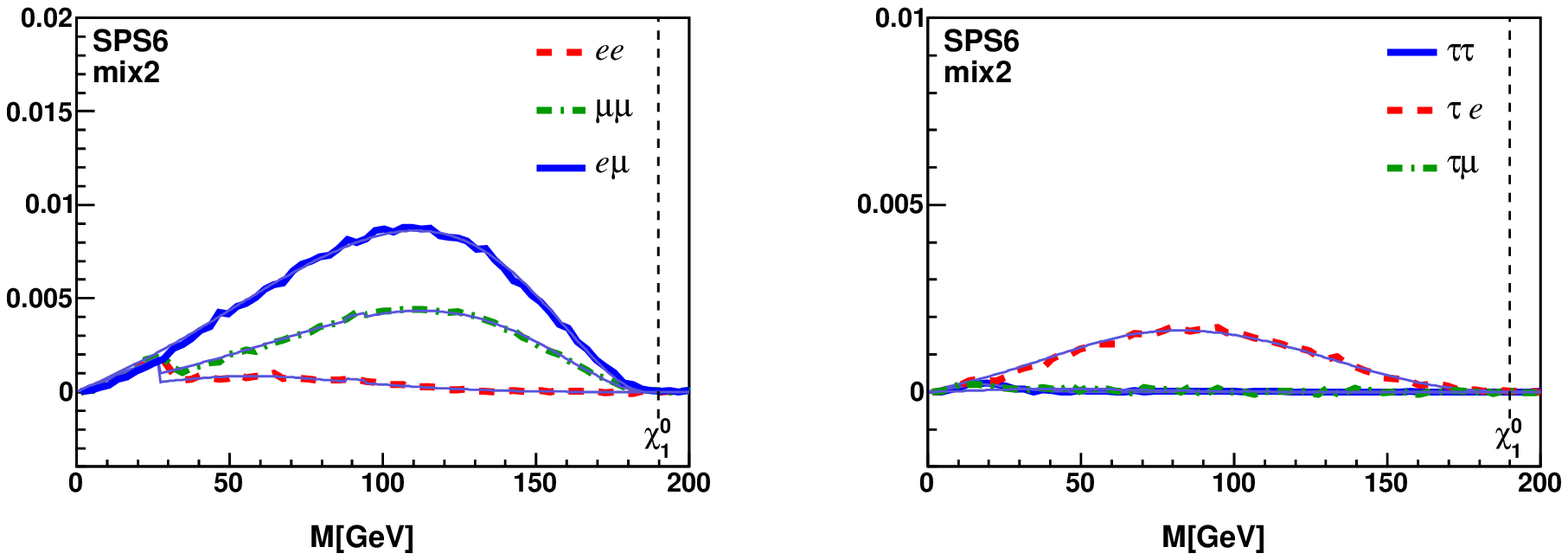}} }
}\fi
\vspace{-6mm}
\caption{Di-lepton invariant masses for mix1 (upper panels) and mix2
  (lower panels) at SPS6.
\label{fig-InvMmixSPS6} } }

These combinations are chosen such that the total flavour content of
mix1 and mix2 are the same at the operator level.  However, the effect
of the cuts will be somewhat different and that will induce some
differences. For example, the $p_T$ spectra for the two mixtures at SPS6 (as shown
in Fig.~\ref{fig-MixSpec}) do look similar,
but we can see some small differences caused by the fact that the
light lepton flavour accompanying the tau gets suppressed by the cuts.

The corresponding invariant mass distributions
(Fig.~\ref{fig-InvMmixSPS6}) are clearly different.
For mix1 we see prompt di-lepton
(i.e. Eq.~(\ref{Eq:InvMass})) signals in each of
the $ee$, $e\mu$ and $\mu\mu$ channels.
We also see that at masses below some 100 GeV, the latter two
distributions are slightly larger than the $ee$ one
(we ignore the triangular contributions from the cascade chain).
A fit confirms that the difference is well explained by an
additional contribution of type (\ref{Eq:Mad}), which in turn suggests
a $\mu\tau$ channel. It should be noted that the ability to detect a
$\mu\tau$ channel in the light di-lepton distributions depends on the
SUSY model so it cannot always be done, but we see that the $\mu\tau$
channel clearly shows up in the $\mu\tj$ distribution as well.

Looking at mix2 in the lower panels of Fig.~\ref{fig-InvMmixSPS6}, we
see a strong prompt $e\mu$ signal as well as a smaller prompt $\mu\mu$
signal. This is difficult to explain without assuming at least two
couplings where both give $e\mu$ and only one gives $\mu\mu$. We also
see a small signal in the $ee$ distribution, consistent with
Eq.~(\ref{Eq:Mad}) and the fit also prefers such a contribution to the
$e\mu$ channel. This suggests an $e\tau$ channel, which is confirmed
by the $e\tj$ distribution.

\begin{TABLE}{
  \centering
    \begin{tabular}{|l|r|r|r|r|r|r|r|r|}
    \hline
    model & $P_{e^-e^+}$ & $P_{e\mu}$ & $P_{\mu^-\mu^+}$ & (\ref{Eq:3l-event-frac}) & (\ref{Eq:3l-frac-est}) & $\int \mathcal{N}_{ee}$ & $\int \mathcal{N}_{e\mu}$ & $\int \mathcal{N}_{\mu\mu}$ \\\hline
    mix1, SPS1a & 0.27 & 0.37 & 0.36 & 0.48 & 0.84 & 0.29 & 0.35 & 0.36\\
    mix1, SPS1b & 0.27 & 0.37 & 0.37 & 0.44 & 0.82 & 0.29 & 0.36 & 0.36\\
    mix1, SPS6 & 0.28 & 0.37 & 0.36 & 0.48 & 0.85 & 0.30 & 0.34 & 0.37\\\hline
    mix2, SPS1a & 0.07 & 0.62 & 0.30 & 1.31 & 0.63 & 0.05 & 0.63 & 0.31\\
    mix2, SPS1b & 0.08 & 0.62 & 0.29 & 1.47 & 0.65 & 0.05 & 0.64 & 0.31\\
    mix2, SPS6 & 0.08 & 0.62 & 0.31 & 1.35 & 0.63 & 0.06 & 0.62 & 0.32\\
    \hline
    \end{tabular}
  \caption{Similar to table~\ref{tab:Counting} for mix1 and mix2 at SPS1a, SPS1b and SPS6.}\label{tab:mixtab1} }
\end{TABLE}

\begin{TABLE}{
  \centering
    \begin{tabular}{|l|l|l|l|l|l|l|}
    \hline
    model & $P_{e^-e^+}$ & $P_{\mu^-\mu^+}$ & $P_{\tj^-\tj^+}$ & $P_{e\mu}$ & $P_{e\tj}$ & $P_{\mu\tj}$ \\\hline
    mix1, SPS1a & 0.19 & 0.24 & 0.02 & 0.25 & 0.05 & 0.21\\
    mix1, SPS6 & 0.23 & 0.29 & 0.003 & 0.29 & 0.01 & 0.17\\\hline
    mix2, SPS1a & 0.06 & 0.21 & 0.02 & 0.42 & 0.19 & 0.06\\
    mix2, SPS6 & 0.06 & 0.25 & 0.006 & 0.51 & 0.16 & 0.02\\
    \hline
    \end{tabular}
  \caption{Similar to table~\ref{tab:Counting3} for mix1 and mix2 at SPS1a and SPS6.}\label{tab:mixtab2} }
\end{TABLE}

Let us now look at the counting observables from the last
section. They are all listed in tables~\ref{tab:mixtab1} and
\ref{tab:mixtab2}. If we first look at table~\ref{tab:mixtab1}, we
again see that these numbers seem largely independent of the SPS
point. When looking at mix1 we see that we have significant branching
fractions to $ee$, $e\mu$ and $\mu\mu$ and that
the last two are a little larger than the first.
 We also see that the
measurement of the fraction~(\ref{Eq:3l-event-frac}) is about half the
prediction from (\ref{Eq:3l-frac-est}), suggesting a $\mu\tau$ channel
which would also explain the higher values for $e\mu$ and $\mu\mu$
compared to $ee$.

If we instead look at mix2 we see a large branching fraction to $e\mu$
and also a significant $\mu\mu$ channel while the $ee$ channel is
small but non-zero. We also notice a measured
fraction~(\ref{Eq:3l-event-frac}) that is twice that of the
prediction~(\ref{Eq:3l-frac-est}), indicating an $e\tau$ channel,
consistent with the small $ee$ contribution.

If we finally turn our attention to table~\ref{tab:mixtab2}, we see a
larger variation with SPS point but the structure of the numbers is
in agreement with our previous conclusions.

To summarize, the invariant mass distributions and the counting
observables from the last section can independently help identify the
relevant decay channels which then leads us to the correct
combinations of operators. The comparison between the measurement of
the fraction~(\ref{Eq:3l-event-frac}) and the
prediction~(\ref{Eq:3l-frac-est}) is the most reliable way of
detecting an $e\tau$ or $\mu\tau$ channel without relying on tau
tagging.

\section{$LQ\bar D$ coupling}
\label{sect:lqd}
\setcounter{equation}{0}

$LQ\bar D$ couplings can induce two types of neutralino decays.  The
neutralino can either go to one up-type and one down-type (anti)quark
and a charged lepton or to a down-type quark-antiquark pair and a
neutrino.  In the case of $L_iQ_3\bar{D}_k$ the latter decay dominates
since otherwise a top would have to be produced, which is either
strongly suppressed by phase space or requires an off-shell top;
hence, in these cases, there are no isolated charged leptons from the
decay and the lepton flavor of the $\lambda'$ operator cannot be
determined.

Table~\ref{Tab:LQD} summarizes possible scenarios for the various
flavour combinations. Operators easier to study are of the type
$L_{1,2}Q_{1,2}\bar D_{3}$, where lepton identification combined with
b-tagging might give some interesting signals.  The reason the $\ell
j$ signal is mentioned in Table~\ref{Tab:LQD} is that in many cases
the two jets can be so close in the detector that they merge into one
jet and the peak expected in the $\ell jj$ distribution instead shows
up in $\ell j$. This is also partially true for $\ell\bj$; however, in
that case a more realistic algorithm for b-tagging is needed to know
how well the b-tagging works if the jets are so close that they might
merge.

\begin{TABLE}{
  \centering
  \begin{tabular}{|l|l|l|}
    \hline
    Coupling & Decay products & Signals \\\hline
    $L_{1,2}Q_{1,2}\bar D_{1,2}$ & $\ell q\bar q$, $\nu q\bar q$ & $\ell j$, $\ell jj$ \\
    $L_{1,2}Q_{1,2}\bar D_{3}$ & $\ell q\bar b$, $\nu q\bar b$ & $\ell \bj$, $\ell \bj j$ \\
    $L_{3}Q_{1,2}\bar D_{1,2}$ & $\tau q\bar q$, $\nu q\bar q$ & $\tj j$, $\tj jj$ \\
    $L_{3}Q_{1,2}\bar D_{3}$ & $\tau q\bar b$, $\nu q\bar b$ & $\tj \bj$, $\tj \bj j$ \\
    $L_{1,2,3}Q_{3}\bar D_{1,2}$ & $\nu b\bar q$ & $\bj j$ \\
    $L_{1,2,3}Q_{3}\bar D_{3}$ & $\nu b\bar b$ & $\bj \bj $ \\
    \hline
  \end{tabular}
  \caption{The various possible scenarios with $LQ\bar D$ operators.
  It is here assumed that the neutralino is lighter than the top.
  Notation: $\tj$, $\bj$ and $j$
  denote a tau jet, a b-jet and a light-flavor jet, respectively.
  }\label{Tab:LQD}  }
\end{TABLE}

When the $\lambda'$ operator has $\tau$ flavour the situation is
rendered more complicated on account of the loss of information
involving the neutrinos in the decay of the tau. The most problematic
situation, however, appears when we have a $Q_3$ operator, as
discussed above we only expect decays to a neutrino and two jets and
therefore this scenario might be hard to identify.

\subsection{Event selection}
The task of suppressing backgrounds is a bit more difficult with
$LQ\bar D$ couplings as compared to $LL\bar E$ couplings, especially
when it comes to suppressing $t\bar t$ events.  However, the lepton
multiplicity is useful in this respect.

To cover most possibilities we use three different cuts in this
analysis. All three require:
\begin{itemize}
  \item Transverse sphericity $> 0.2$,
  \item $E_{T}=\Sigma_{leptons, jets}|p_T| > 1000$ GeV,
  \item Hardest jet $p_T>300$ GeV.
\end{itemize}

We define a ``hard cut'' to additionally
require at least two same-sign
isolated leptons with $p_T>20$ GeV. The same-sign requirement
suppresses the $t\bar t $ background by several orders of magnitude,
whereas the signal is reduced to only about $1\%$ of its precut value.

An alternative is to
use a ``loose cut'' which passes a lot more of the signal, at
the cost of a significant background from $t\bar t$ events. This
requires only one isolated lepton with $p_T>20$ GeV, and at least 7
jets with $p_T >20$ GeV.

The third cut is meant for the cases where we only get neutrinos and
no leptons. Therefore it requires instead of leptons,
$p_T^\text{miss}>200$ GeV, and at least 7 jets with $p_T >20$ GeV.
We name this the ``pTmiss cut''.

\begin{TABLE}{
  \centering
    \begin{tabular}{|l|l|l|l|}
        \hline
        model & hard cut & loose cut & pTmiss cut\\\hline
        $L_1Q_1\bar D_2$, SPS1a & 0.011 & 0.082 & 0.043\\
        $L_1Q_1\bar D_2$, SPS1b & 0.043 & 0.24 & 0.16 \\
        $L_1Q_1\bar D_2$, SPS3 &  0.06 & 0.25 & 0.16 \\
        $L_1Q_1\bar D_2$, SPS5 &  0.015 & 0.089 & 0.046 \\
        $L_1Q_1\bar D_2$, SPS6 &  0.008 & 0.096 & 0.083 \\
        \hline
        $L_1Q_1\bar D_2$, SPS1a & 0.011 & 0.082 & 0.043\\
        $L_1Q_3\bar D_3$, SPS1a & 0.0013 & 0.03 & 0.048 \\
        $L_3Q_1\bar D_3$, SPS1a & 0.0027 & 0.043 & 0.041 \\
        \hline
        $t\bar t$ & $1.5\cdot 10^{-6}$ & 0.002 & 0.00083 \\
        \hline
    \end{tabular}
  \caption{The fraction of events that passes the various cuts for some $LQ\bar D$ operators,
  as well as $t\bar t$ production. Note that this does not take the different
  cross-sections into account.}\label{tab:LQDcuts} }
\end{TABLE}

The effects of these cuts, on some representative $LQ\bar D $ couplings
as well as on the $t\bar t$ background, are summarized in
Table~\ref{tab:LQDcuts}. All other backgrounds are less important than
$t\bar t$ for these cuts.

We see that for the hard cut we get a satisfactory suppression of the
$t\bar t$ while the loose cut will suffer from a significant
contamination by $t\bar t$ events.

It is also clear that the operators with tau flavour are more
problematic due to fewer leptons.  As expected, of the operators in Table~\ref{tab:LQDcuts},
the most problematic one is $L_1Q_3\bar D_3$ where the signal would be $\nu_ebb$ or
$e^-tb$. With the $t$-quark above the kinematical threshold, only the
$\nu_ebb$-channel is open and we have to rely on the pTmiss cut.  This
last case will appear very similar to an R-parity conserving scenario,
the most significant difference being that the particles now responsible
for the missing $p_T$ are the (massless) neutrinos rather than the
neutralinos. This problem is present for all operators of type $LQ_3\bar D$.

\subsection{Invariant mass distributions}
There are plenty of interesting invariant mass distributions to be
studied with respect to $LQ\bar D$ operators.
Unfortunately, the lack of
knowledge
of the charge of light quark jets implies that the same-sign
subtraction tool is essentially rendered inoperative.
 As a result, most
signals tend to drown in combinatorial and other backgrounds.
This problem would be reduced if the couplings were small enough for
displaced vertices to appear, however, as mentioned before,
we do not make use of such possibilities here.

A further problem is that the boost of the neutralino results in all
the decay products being very close in the detector, thereby raising
the probability of jets merging.  It is, therefore, not always true
that we get two jets and one lepton from each neutralino.  On the
other hand, the closeness of the decay products can be used to choose
which jets and leptons to compute the invariant mass of.

These proximity issues make the choice of jet algorithm very important.
Especially the parameter $R_{\rm jet}$ is very important. To optimize
each analysis, we perform it using jets as defined
by $R_{\rm jet}$ equal to $0.2$, $0.4$ and $0.7$ and show results using
the most favourable value for each analysis. The event selection always uses jets as
defined when $R_{\rm jet}=0.4$.

In Fig.~\ref{fig-LQDInvM1} these points can be seen for $L_1Q_1\bar
D_2$ at SPS1a: if $\ell jj$ where the two jets are within $\Delta R =
1$ from the lepton, are used, one can get a very nice peak at the
neutralino mass (black curve, upper left panel).  If all $\ell j j$
are used, the peak is less pronounced but can be clearly seen both with
$R_{\rm jet}=0.4$ as in the lower left panel, and even better when
$R_{\rm jet}=0.2$, which can be seen in the upper right panel. Since
the two quark jets quite often merge into one, it is also possible to
see a nice peak at the neutralino mass in the distribution of all
$\ell j$ combinations, as seen in the upper left panel. It is also
worth noting that the branching fraction for neutralino decay to $\ell
jj$ as compared to the fraction to $\nu jj$ depends on the SUSY parameter point and e.g.\ SPS1b has a more favourable value, which can be seen in the
lower right panel of Fig.~\ref{fig-LQDInvM1}, where the peak is
clearly more pronounced than in the comparable plot for SPS1a (lower
left panel).

\FIGURE[ht]{
\let\picnaturalsize=N
\def\picsize{15cm}
\ifx\nopictures Y\else{
\let\epsfloaded=Y
\centerline{\hspace{4mm}{\ifx\picnaturalsize N\epsfxsize \picsize\fi
\epsfbox{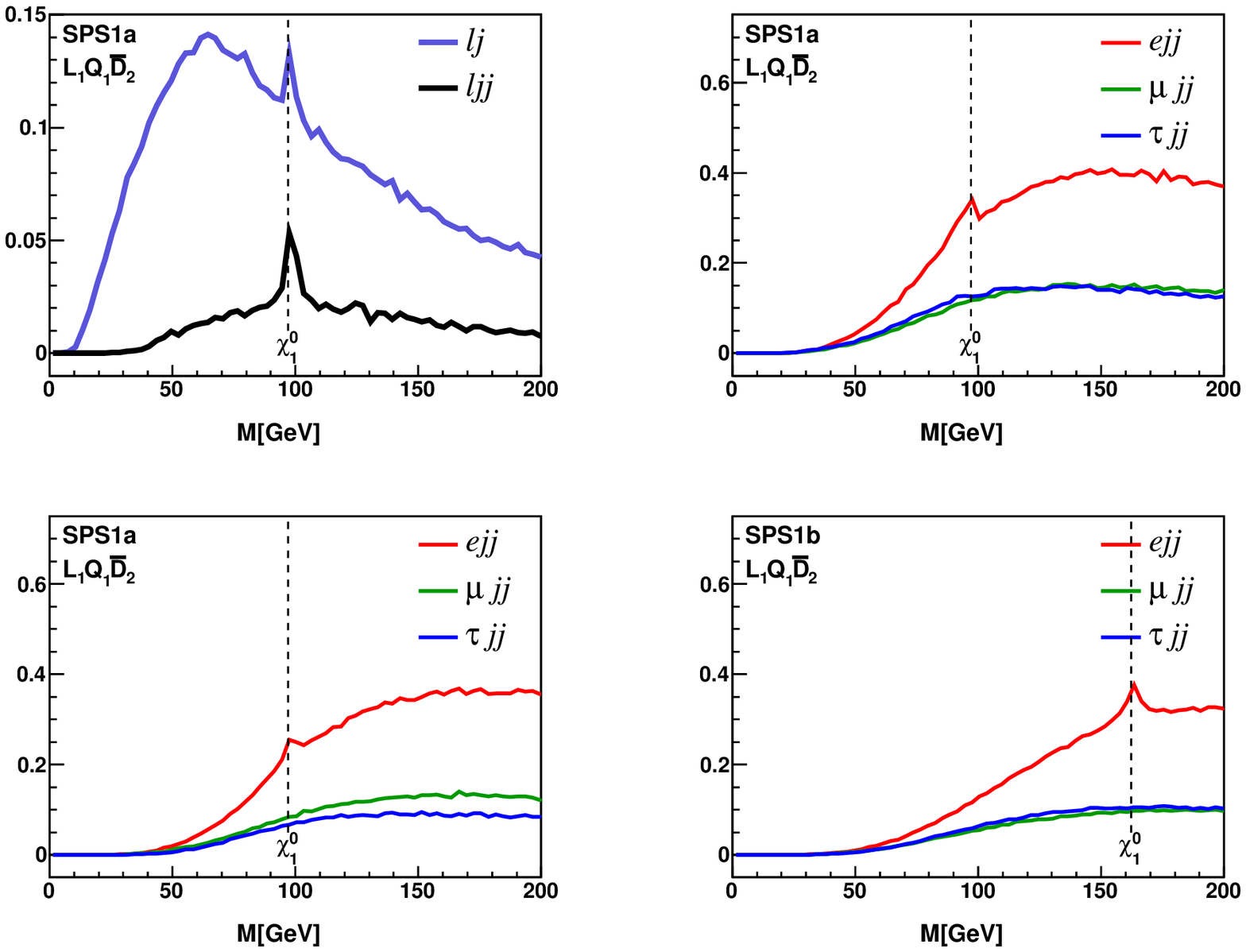}} }
}\fi
\vspace{-6mm}
\caption{Invariant mass distribution for $L_1Q_1\bar D_2$ at SPS1a:
  upper left: violet: $\ell j$ distribution, black: $\ell jj$ where
  the jets are closer than $\Delta R = 1$ to the lepton, this
  distribution has been multiplied by 5 to be visible.  Upper right:
  $\ell jj$ distributions using $R_{\rm jet}=0.2$. Lower left panel:
  same as upper right but with $R_{\rm jet}=0.4$, lower right: same as
  lower left but for SPS1b. In this figure, the hard cut has been
  used.
\label{fig-LQDInvM1} } }

So far we have discussed the case with one light lepton and two light
jets in the decay. Another group of operators have b-quarks in the
decay and that opens up for b-tagging. Assuming that the b-tagging can
also determine the charge of the $\bj$, it becomes possible to use
same-sign subtraction to reduce the combinatorial background.

Fig.~\ref{fig-LQDInvM2} shows $\ell \bj j$ distributions as well as
$\ell \bj $ distributions where same-sign subtraction has been
employed on the $\ell \bj$ pair. In the upper left panel we see that
this gives a good signal for $L_1Q_1\bar D_3$ where we again have a
peak at the neutralino mass. The signal in the $e\bj$ channel looks
like the distribution (\ref{Eq:InvMass}) with a small peak at the high
end (caused by merged jets), which is consistent with our
expectation. Due to the low number of charged leptons in the
neutralino decay at SPS1a, we have used the loose cut to achieve the
upper panels of Fig.~\ref{fig-LQDInvM2}.  For SPS1b we get more
charged leptons and as is clearly seen in the middle panels the hard
cut is fine for extracting these signals.  However, the SUSY
production at SPS1b is also about one order of magnitude smaller than
at SPS1a so using the hard cut is more important to reduce the
backgrounds.

\FIGURE[ht]{
\let\picnaturalsize=N
\def\picsize{15cm}
\ifx\nopictures Y\else{
\let\epsfloaded=Y
\centerline{\hspace{4mm}{\ifx\picnaturalsize N\epsfxsize \picsize\fi
\epsfbox{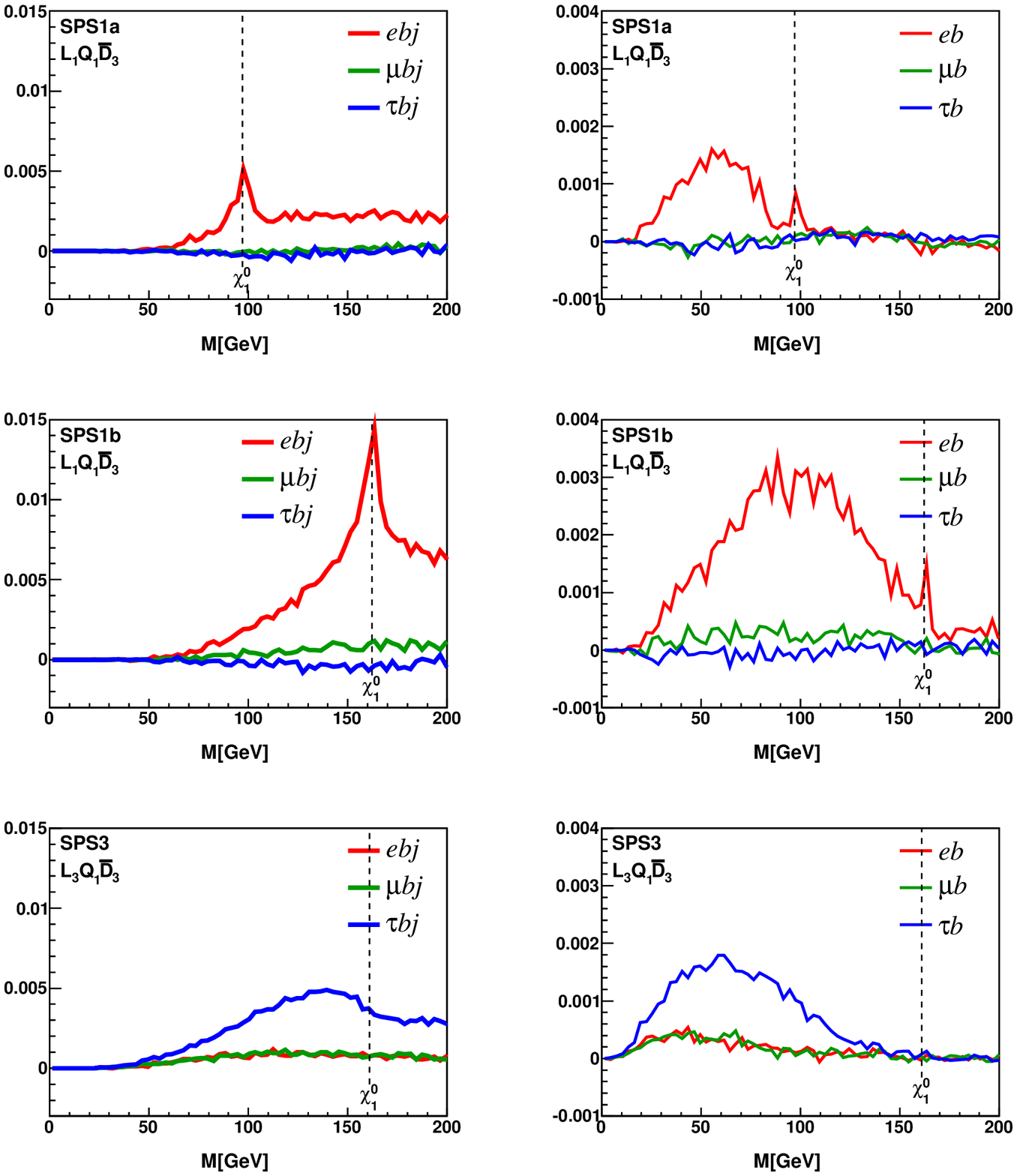}} }
}\fi
\vspace{-6mm}
\caption{Left panels: $\ell\bj j$ distributions with same-sign
  subtraction in the $\ell\bj$ pair.  Right panels: $\ell\bj$
  distributions with same-sign subtraction. Upper and middle panels:
  $L_1Q_1\bar D_3$ at SPS1a (upper) using the loose cut and SPS1b
  (middle) using the hard cut.  Lower panels: $L_3Q_1\bar D_3$ at SPS3
  with $R_{\rm jet}=0.2$ and using the pTmiss cut.
\label{fig-LQDInvM2} } }

If the lepton is a $\tj$, things are much more difficult; first of all
we lose energy to the neutrino, so we do not expect a peak anymore.
However, if we use $R_{\rm jet}=0.2$ together with the pTmiss cut for
SPS3 we see that the expected signals are clearly there.  In the
$\tj\bj j$ distribution (lower left panel) we see a bump below the
neutralino mass.  Perhaps the best and most conclusive signal though,
is in the $\tj\bj$ channel (lower right channel) where we see a
distribution similar to what we discussed in appendix~B for tau decays
(the $\ell\tj$ distribution of Fig.~\ref{fig-Mtau}).

\subsection{Mixing $LQ\bar D$ operators}
The chances to identify two or more simultaneously large $LQ\bar D$
type operators, depend strongly on the specific combination; if the
two operators are identified through different channels, the
identification can be rather straight-forward.

\FIGURE[ht]{
\let\picnaturalsize=N
\def\picsize{15cm}
\ifx\nopictures Y\else{
\let\epsfloaded=Y
\centerline{\hspace{4mm}{\ifx\picnaturalsize N\epsfxsize \picsize\fi
\epsfbox{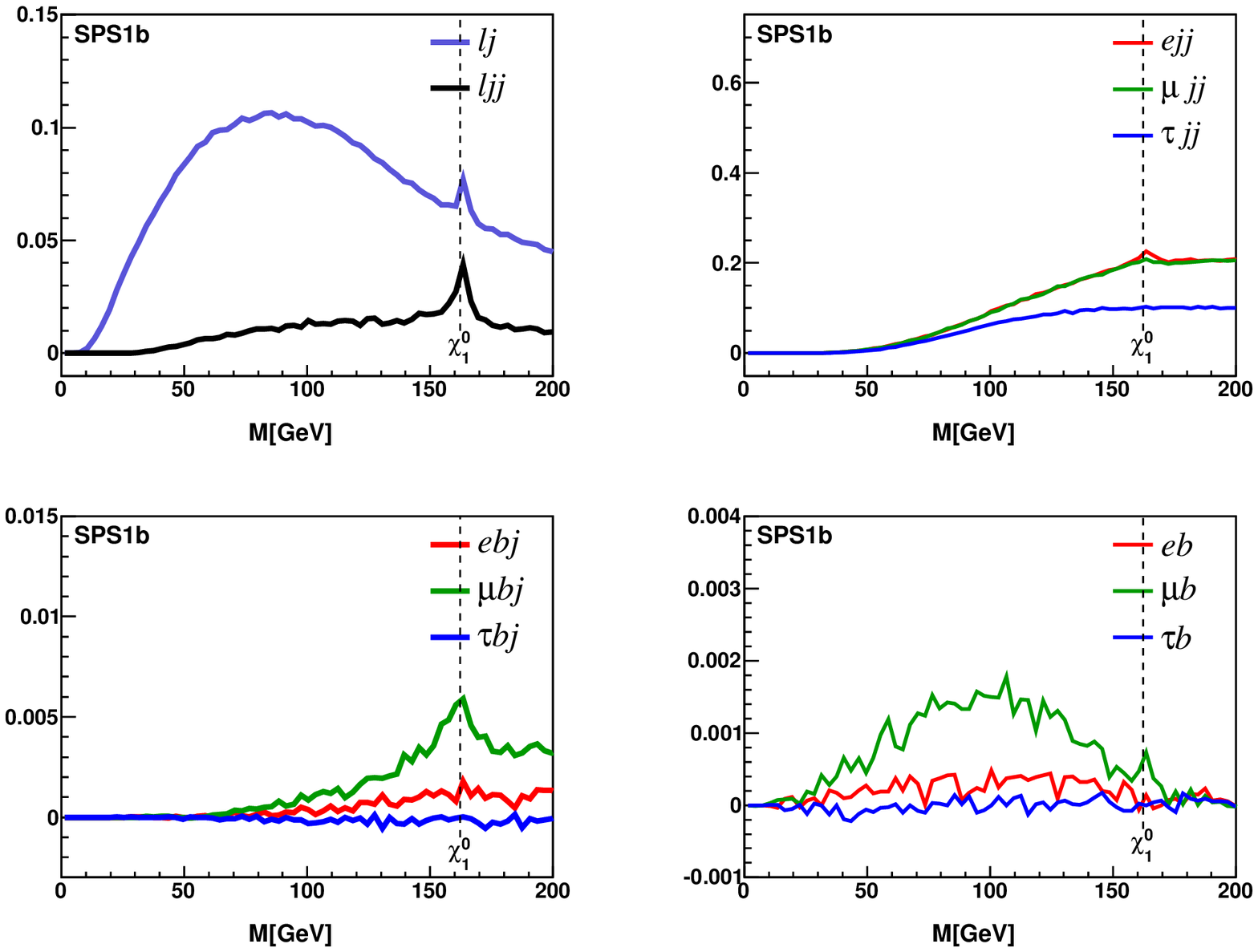}} }
}\fi
\vspace{-6mm}
\caption{Combinations of $LQ\bar D$ operators: Invariant masses for
  $\lambda^{\prime}_{121}= 10^{-4}$ and $\lambda^{\prime}_{223}=
  10^{-4}$ at SPS1b. The two operators differ both in lepton flavours
  and in quark flavours.  Upper left: $\ell j$ and $\ell jj$
  (magnified by a factor 5), in the latter case the two jets are close
  to the lepton (see Fig.~\ref{fig-LQDInvM1}).  Upper right: $\ell
  jj$.  Lower panels show $\ell\bj j$ (left) and $\ell\bj$ (right)
  distributions. All plots use the hard cut.\label{fig-LQDInvMmix2} }
}

One example of this can be seen in Fig.~\ref{fig-LQDInvMmix2} where
one operator gives a signal in $\mu\bj j$ and $\mu\bj$ channels and
the other shows up in the $e jj$ distribution. This case is rather
fortunate since the two operators can be identified independently; if
the lepton flavour of the operators would have been the same, it would
have been much harder to tell that we have one operator with and one
without a $\bar D_3$ component.

In Fig.~\ref{fig-LQDInvMmix5} we show another combination where an
operator ($\lambda^{\prime}_{211}$) giving a peak in the $\mu jj$
distribution, is combined with a more problematic tau-flavour coupling
($\lambda^{\prime}_{313}$) and again we see that both operators can be
identified successfully.

In conclusion, having more than one large $LQ\bar D$ operator at the
same time is no problem as long as the signals from the couplings
appear in different channels and the operators are identifiable on
their own.

\FIGURE[ht]{
\let\picnaturalsize=N
\def\picsize{15cm}
\ifx\nopictures Y\else{
\let\epsfloaded=Y
\centerline{\hspace{4mm}{\ifx\picnaturalsize N\epsfxsize \picsize\fi
\epsfbox{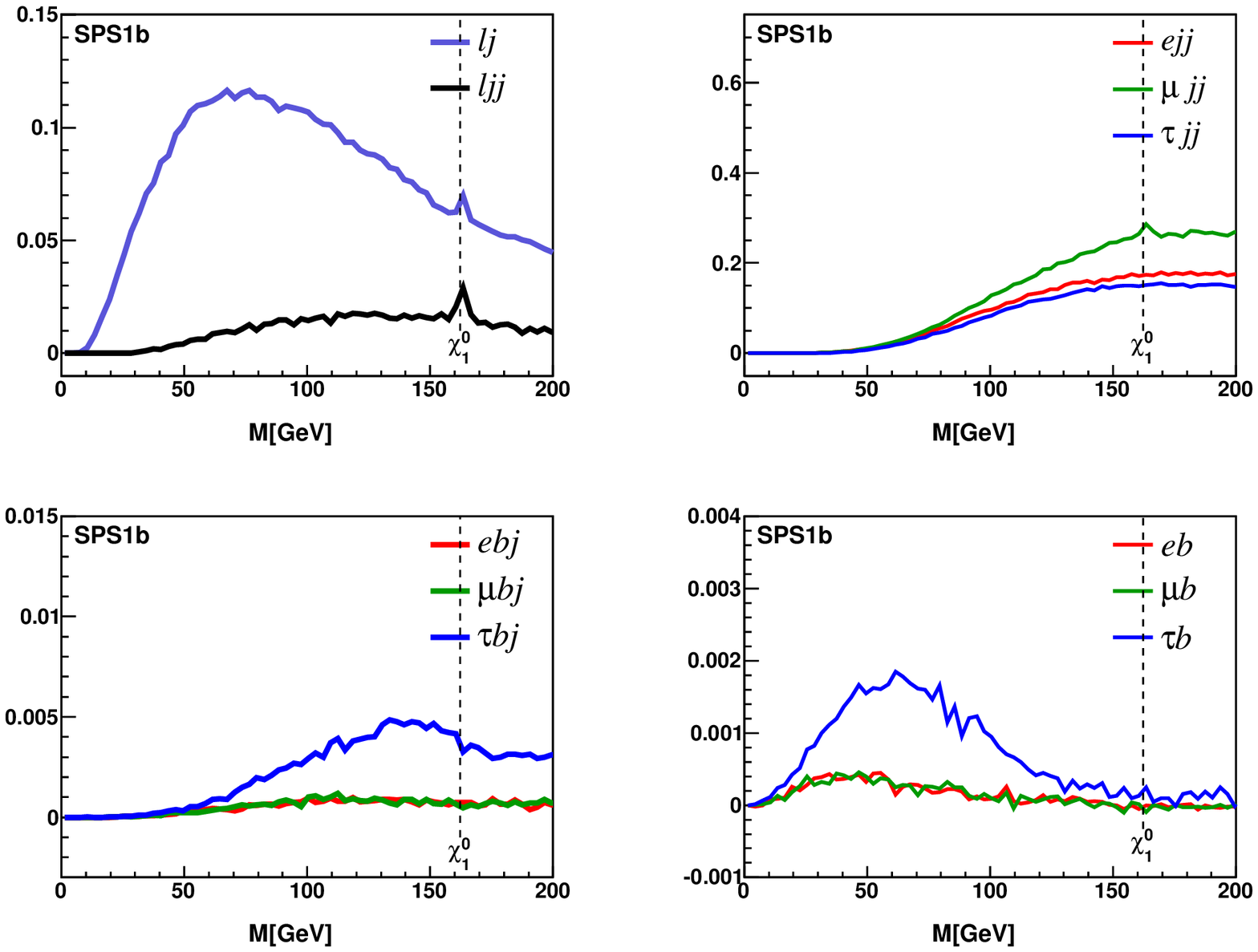}} }
}\fi
\vspace{-6mm}
\caption{Same as Fig.~\ref{fig-LQDInvMmix2} for
  $\lambda^{\prime}_{211}= 10^{-4}$ and $\lambda^{\prime}_{313}=
  10^{-4}$ at SPS1b.  Here the two lower panels use the pTmiss cut as
  well as $R_\text{jet}=0.2$ to get enough
  $\tj$'s. \label{fig-LQDInvMmix5} } }

\subsection{Backgrounds}
As mentioned before, the most important background is $t\bar t$.
Although the hard cut is very effective in suppressing this
background; the need to sometimes use the loose or pTmiss cut makes it
necessary to look at the expected invariant mass distributions from
the $t\bar t$ events.

The interesting distributions here are the invariant masses including
$\bj$ since that is where the weaker signals occur and we expect some
contribution there from $t\bar t$ due to tops decaying to $\ell\bj (\nu)$.

These distributions are shown in Fig.~\ref{fig-ttbarInvM} where we can
see that the only significant distribution is in the $\mu\bj$ and
$e\bj$ channels for the loose cut.  The distribution can be understood
as a triangular distribution ending around 150~GeV from the decay of
the top and a combinatorial background that due to the same-sign
subtraction and the charge correlation between the top and antitop, is
negative.

It is also important to note that we see no significant signal when
the pTmiss cut and $R_\text{jet}=0.2$ are used. Simulations where standard model processes were added, also show that the studied distributions are not significantly changed by these backgrounds.

\FIGURE[ht]{
\let\picnaturalsize=N
\def\picsize{15cm}
\ifx\nopictures Y\else{
\let\epsfloaded=Y
\centerline{\hspace{4mm}{\ifx\picnaturalsize N\epsfxsize \picsize\fi
\epsfbox{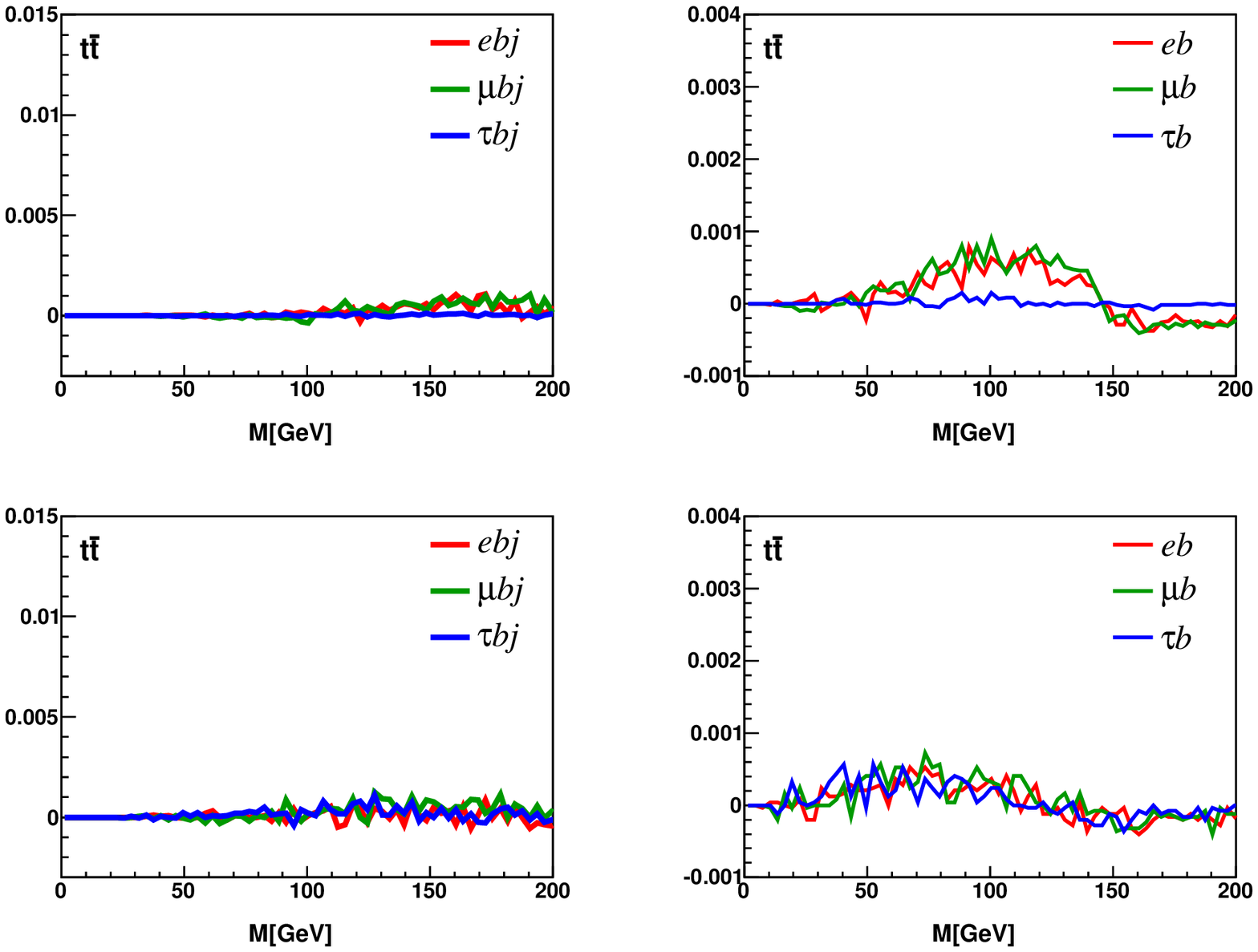}} }
}\fi
\vspace{-6mm}
\caption{Invariant mass distributions for $t\bar t$ events. The left
  panels show $\ell\bj j$ distributions and the right panels show
  $\ell\bj$ distributions.  In the upper panels the loose cut is used
  and in the lower panels the pTmiss cut is used and
  $R_\text{jet}=0.2$. \label{fig-ttbarInvM} } }

\section{$\bar U\bar D\bar D$ coupling}
\label{sect:udd}
\setcounter{equation}{0}
Of the various scenarios with trilinear R-parity violation, the one
that is most difficult to analyse at a hadronic collider is that of
non-zero $\bar U\bar D\bar D$ couplings.  With the lightest neutralino
now decaying into three jets, extracting the signal from the large QCD
background is very challenging~\cite{Baer:1996wa}.
In fact, even for simpler situations like 
resonance production or one-step decays, the reach of 
the Tevatron or the LHC is very limited~\cite{berger_harris_sullivan,Choudhury:2005dg}.

As mentioned earlier, assuming moderately light neutralinos, we expect
them to be highly boosted due to the large center of mass energy of
the collision.  This means that the decay products of the neutralino
will be rather collimated, which has enormous implications for the
identification of $LQ\bar D$ operators as discussed in the previous
section. This would also be the case for $\bar U\bar D\bar D$
operators where the three jets from the decaying neutralino will
sometimes be so collimated that they merge. This very feature can be
turned to an advantage; a $\bar U\bar D\bar D$ signal may be extracted by
studying jet substructure, where one looks for a big jet that looks
like it could consist of three jets that have merged
\cite{Butterworth:2009qa}.

There is, however, an exception to the above picture, namely any of
the three operators $\bar U_3\bar D_i\bar D_j$, for the top flavour of
these couplings ensures a very different behavior. If the neutralino
is lighter than the top, it may appear stable on the detector scale
and we might be facing a fake MSSM scenario.
On the other hand, if the neutralino is heavier than the top, it
decays into a top (or antitop) and two jets that often would tend to
be soft (due to limited phase space).  While a $t \bar t +
n\hbox{--jets}$ final state has a very large QCD background, note that
the latter decreases fast with both increasing $n$ as well as with an
increasing $p_T^{\rm min}({\rm jet})$ (the minimal requirement on the
transverse momenta of these additional putative jets).
And as additional jets
accrue not only from the decay of the neutralinos, but also in earlier
steps in the SUSY cascade, and as at least some of them are likely to
have a sufficiently large $p_T$, this final state is worth
investigating. However, rather than performing this admittedly
difficult task, we recommend concentrating on a relatively cleaner
subsample. Noting that the neutralinos are Majorana particles, the
very last pair of decays is as likely to produce a like-sign top-pair
($tt$ or $\bar t \bar t$) as an opposite-sign one.  The possibility of
same-sign tops makes the extraction of these events rather straight
forward.  One way is to look for same-sign isolated leptons
accompanied by two same-sign b-jets; this approach was taken in a
similar study \cite{Kraml:2005kb}.

However, it turns out that the harder cuts used on $LQ\bar D$
operators are also efficient in this case. This comes as no surprise
since those cuts are based on same-sign leptons. If we, for example,
look at $\bar U_3\bar D_1\bar D_2$ at SPS6 where the neutralino mass
is 189 GeV so that the top decay channel is open, about $1\%$ of the
events pass the hard cut from the $LQ\bar D$ section. This leaves good
hope for detection and if we then look at the $\ell\bj$ invariant
mass, as is done in Fig.~\ref{fig-UDDsps6}, we see the same structure
(although rather weak) as in the corresponding distribution from
$t\bar t$ events as shown in the upper right panel of
Fig.~\ref{fig-ttbarInvM}.  This similarity demonstrates the presence
of top quarks in the events and can therefore be used to confirm the
presence of a $\bar U_3\bar D_j\bar D_k$ operator.

\FIGURE[ht]{
\let\picnaturalsize=N
\def\picsize{10cm}
\ifx\nopictures Y\else{
\let\epsfloaded=Y
\centerline{{\ifx\picnaturalsize N\epsfxsize \picsize\fi
\epsfbox{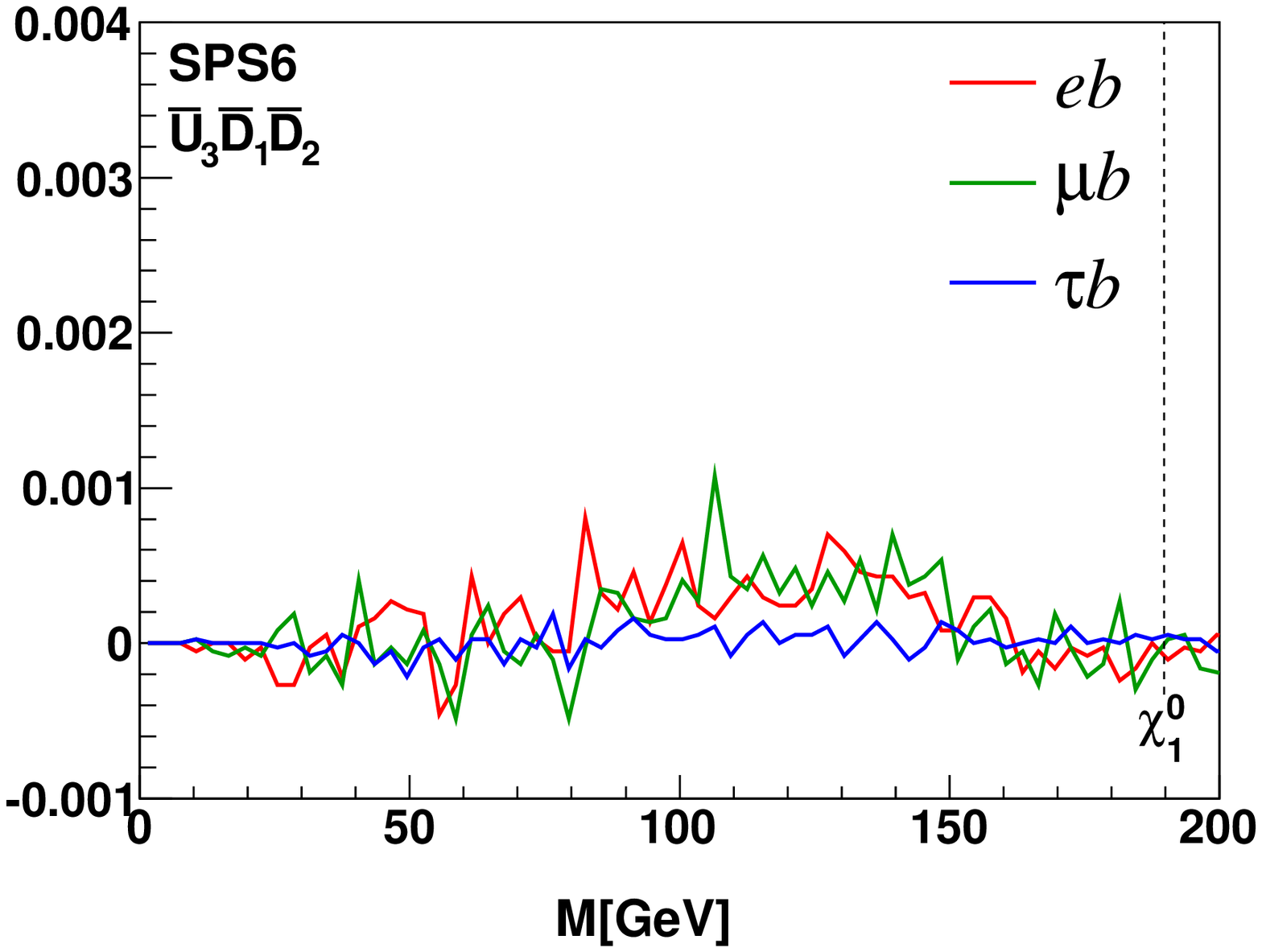}} }
}\fi
\vspace{-6mm}
\caption{Invariant mass distributions for $\lambda^{\prime\prime}_{312}$ at SPS6. The distributions shown are $\ell\bj$ distributions for all lepton flavours.}\label{fig-UDDsps6} }

\section{Mixing and models}
\label{sect:models}
\setcounter{equation}{0}

To date, most phenomenological analyses have assumed the dominance of
a single R-violating operator.  However, once a flavour structure is invoked
(e.g., to explain fermion masses and mixings)~\cite{famsym},
hierarchies in R-violating couplings would be related to the flavour
charges of the respective fields \cite{MODELS}. Even more importantly,
even if only one R-violating operator is to be postulated in the
interaction basis, fermion mixing would, in general, induce non-zero
values for others.  This implies that it is natural to expect a range
of hierarchies for R-violating operators.

\FIGURE[ht]{
\let\picnaturalsize=N
\def\picsize{15cm}
\ifx\nopictures Y\else{
\let\epsfloaded=Y
\centerline{\hspace{4mm}{\ifx\picnaturalsize N\epsfxsize \picsize\fi
\epsfbox{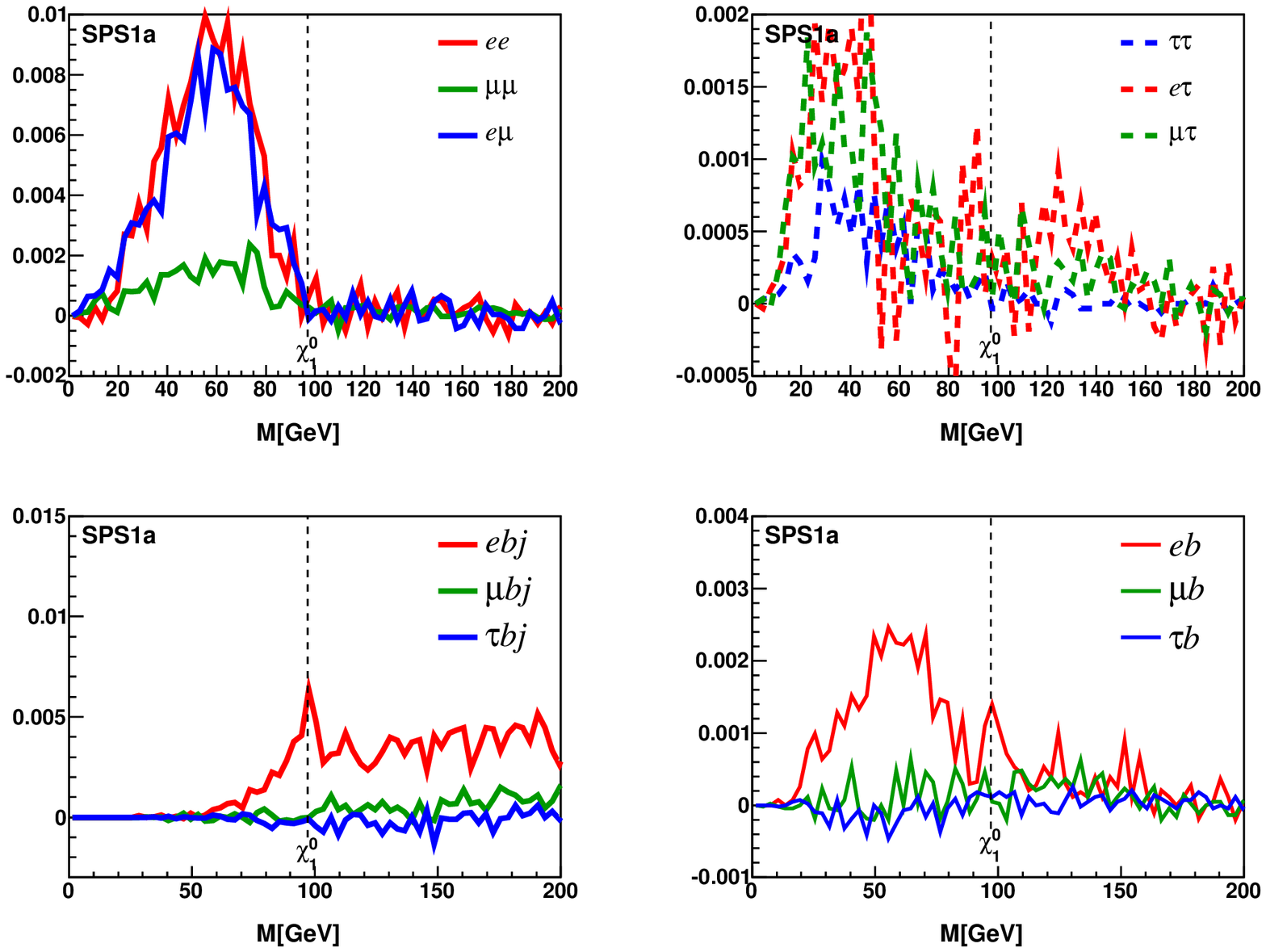}} }
}\fi
\vspace{-6mm}
\caption{Mix of operators: Invariant mass distributions for
  $\lambda_{121}=10^{-5}$ and $\lambda^{\prime}_{123}= 10^{-4}$ at
  SPS1a. Upper panels: di-lepton invariant mass distributions. Lower
  left panel shows $\ell\bj j$ and lower right shows
  $\ell\bj$.}\label{fig-mix5sps1a} }

\subsection{$LL\bar E$ and $LQ\bar D$ operators}
Before entering into a discussion of models that require operators being
related in magnitude, we shall first look at some examples of the
presence of two different operators. Without jeopardizing proton
stability, we can allow for the simultaneous presence of $LL\bar E$
and $LQ\bar D$ operators.

Since both $LQ\bar D$ and $LL\bar E$ operators can be identified due
to the lepton multiplicity, the presence of
one might mask the presence of the other, especially
  if the corresponding decay widths are comparable.  Interestingly, for
  many of the SPS points in question, having one each of $LL\bar E$
  and $LQ\bar D$ operators of the same magnitude, results in the
  decay width associated with the $LL\bar E$ coupling being
  about $10-20$ times larger than that due to the $LQ\bar D$
  operator.\footnote{The reason for this is not far to seek. The
  squarks tend to be much heavier than the sleptons, and, of the
  latter, the stau tends to be the lightest.  Consequently, the purely
  leptonic decay modes of the LSP dominate over the semileptonic
  ones.}
Furthermore, even if the decay widths for the two sets
  of channels were comparable, the $LL\bar E$ ones would be more
  visible due to their higher lepton multiplicity which implies
easier detection. Thus, in order to study scenarios where one might
actually see effects of both operators at the same time, we take the
$LQ\bar D$ operator to be 10 times larger than the $LL\bar E$ one.

While most of the conclusions from the study of single
couplings are still valid when
more than one coupling is relevant, the details
of section \ref{subsect:counting} are modified. In particular, the
differences between the measured and estimated values of the fraction
(\ref{Eq:3l-event-frac}) no longer point to the relevant tau
channels. Therefore, it is more useful to study invariant mass
distributions.

A favourable scenario would be the combination of two operators that
are both relatively easy to identify. One example of this would be
$\lambda_{121}=10^{-5}$ and $\lambda^{\prime}_{123}= 10^{-4}$. The
invariant mass distributions corresponding to this scenario, at SPS1a,
are shown in Fig.~\ref{fig-mix5sps1a}. We clearly see the signals
expected from the $LL\bar E$ coupling in the $e\mu$ and $ee$
distributions (upper left panel) as well as the $\ell b_\text{jet}j$ (lower left panel) 
signals expected from the $LQ\bar D$ coupling. We also note that
the distributions including $\tj$'s (in the upper right plot) show
signals that might suggest a $\tau\tau$ channel, however, those tau
pairs actually come from the cascade chain
and one has to be careful not to interpret them as decay products of
the neutralino.

Whether the potentially confusing background from the cascade
is present or not, depends on the parameter point, in
e.g.\ SPS6 it is not. However, SPS6 comes with another problem; it
turns out that the branching fractions for decay to charged leptons
plus quarks is much smaller (factor $\approx7$) than the branching
fraction to neutrino plus quarks. This, then, results in fewer events
with leptons and, as a consequence, the identification of the $LQ\bar
D$ operator is rendered more difficult.

Similarly, for other experimentally complicated scenarios, such as
operators with a lot of tau flavour, the problems
(in identifying the $LQ\bar D$ operator) noted in
section~\ref{sect:lqd}
remain pretty much unchanged when
combinations are considered. In general, the
identification of the $LL\bar E$ and the $LQ\bar D$ operators can be
handled independently and the only interference between them is that
for the smaller coupling we get less statistics to work with.
This also means that there is an interval where both operators can be
identified while in the periphery we will see only one dominant
operator. This interval is centered roughly where the $LQ\bar D$
operator is about 10 times larger than the $LL\bar E$ coupling. In
general, the $LL\bar E$ operators are easier to identify, which
allows for smaller $LL\bar E$ couplings to be successfully identified.

\subsection{Link with flavour symmetries: Two examples}

For a given flavour model, the relative magnitude of R-violating
operators is, to a large extent, predicted, allowing us to tabulate the
expected signals in neutralino decays. This will be done for two
representative cases, namely Left-Right symmetric models and $SU(5)$.

\subsubsection{Left-Right-symmetric models}

The simplest starting point is a single $U(1)$ family symmetry with
the same charges for the left- and right-handed states (left-right
symmetry) as shown in Table~\ref{table:left-right}, where, e.g., the
choice $a_{i}=(-4,1,0)$ \cite{IR} gives an acceptable pattern for the
mass matrices.

\begin{TABLE}{
\begin{tabular}{|c|ccccccc|}
\hline
& $Q_i$ & $\bar{U}_i$ & $\bar{D}_i$ & $L_i$ & $\bar{E}_i$ & $H_2$ & $H_1$ \\
\hline
$U(1)$ & $a _i$ & $a _i$ & $a _i$ & $b_i$ & $b_i$ & $-2a _3$ & $-2 a _3$
\\ \hline
\end{tabular}
\caption{{\it Assignments of $U(1)$ charges.}} \label{table:left-right}}
\end{TABLE}

With these charge assignments
the quark mass matrices (up to numerical factors and phases,
which, in general, are expected to be of order unity) take the form
\begin{equation}
M^\text{up}\sim\left(
\begin{array}{ccc}
\epsilon ^{8} & \epsilon ^{3} & \epsilon ^{4} \\
\epsilon ^{3} & \epsilon ^{2} & \epsilon \\
\epsilon ^{4} & \epsilon & 1
\end{array}
\right), \quad
M^\text{down}\sim\left(
\begin{array}{ccc}
\ee ^{8} & \ee ^{3} & \ee ^{4} \\
\ee ^{3} & \ee ^{2} & \ee \\
\ee ^{4} & \ee & 1
\end{array}
\right)
\label{equation:LR-mm}
\end{equation}
where $\bar{\epsilon}\approx \sqrt{\epsilon } \approx 0.2$.

We now consider the effect of the $U(1)$ symmetry on the pattern of
allowed $R$-violating interactions~\cite{MODELS}.  In this simple
example, with all fermions of a given family
  having the same charge and with a left-right symmetry, the charges
of the operators
depend only on the combination ($i, j, k$) and are
independent of the type, viz.\ $LL\bar E$, $LQ\bar D$ or $\bar U\bar D\bar D$, (see
Table~\ref{table:3}).

\begin{TABLE}{
\centering
\begin{tabular}{|c |ccccc|}\hline
$ijk$   & { 111} & { 121} &{ 122}
& { 222} & { 131} \\ \hline
$U(1)$ & $-12-w$ &  $-7-w$ &  $-2-w$ &  $3-w$ &
$-8-w$ \\ \hline \hline
$ijk$ & { 133} & { 333} &
{ 223} & { 233} & { 123} \\
\hline
$U(1)$ &  $-4-w$ & $-w$ & $2-w$ &
$1-w$ & $-3-w$ \\
\hline
\end{tabular}
\caption{\it Operator charges in a model (see text) with both
family and Left-Right symmetry.
Here $w$ parametrises flavour-independent contributions
\cite{ELR}.}
\label{table:3} }
\end{TABLE}


The parameter $w$ accounts for the fact that the charge assignment is
not unique in model constructions \cite{ELR} (although it is strongly
constrained by phenomenological and theoretical arguments). For
instance, while the addition of flavour independent contributions
would not modify the fermion mass and mixing ratios, it can affect
other considerations, such as anomaly cancellation
conditions. Flavour-independent contributions could also arise from
additional fields with a non-trivial flavour charge that couple to all
operators. These issues in the context of R-violating hierarchies are
addressed in more detail in Ref.\cite{ELR}.

The above flavour symmetries  cannot ensure, by themselves, that
rapid proton decay is avoided. This is done by imposing a
baryon or a lepton parity
\cite{IR,LR,LOR2}, which imply that  the cases
with $\Delta L \neq 0$ and with $\Delta B \neq 0$
have to be considered separately.
For instance,  lepton-number violating operators can be
eliminated by
imposing a lepton triality \cite{LOR2}, under which the fields transform as
\beq
Z_3: (Q,\bar{U},\bar{D},L,\bar{E},H_1,H_2)
\rightarrow (1,1,1,a,a^2,1,1).
\eeq
This allows only the baryon-number-violating operators and the mass
terms, while forbidding lepton-number-violating ones. On the other hand,
to forbid baryon-number violating operators, we would work instead
with a baryon triality, such as in Ref.\cite{IR}, viz.
\beq
Z_3: (Q,\bar{U},\bar{D},L,\bar{E},H_1,H_2)
\rightarrow (1,a^2,a,a^2,a^2,a^2,a).
\eeq

Along these lines, we can use the hierarchies implied by
Table~\ref{table:3}
to make the link between flavour models
and the expected decay rates in neutralino decays
for two separate cases, namely:
\begin{itemize}
\item
for $LL\bar{E}$ and $LQ\bar{D}$ operators;
\item
for $\bar{U}\bar{D}\bar{D}$ operators.
\end{itemize}

While $w$ can be adjusted in order to ensure that all operators remain
within the experimental bounds, there can be additional sources of
suppression in the couplings (arising due to a small $\tan\beta$ in supersymmetric models,
the form of the K\"ahler potential, or additional, model
dependent, features of the theory that may involve extra fields and
symmetries).
To keep the problem manageable (especially since our purpose
here is to show how specific models can be tested in flavour
neutralino decays),
 we will:
\begin{itemize}
\item
assume that the couplings that correspond to higher flavours are
bigger (have smaller flavour charge). The Yukawa couplings generating
fermion masses are larger for the higher generations, having a smaller
net flavour charge. If this is true for R-violating couplings as well,
then the operators that involve third-generation flavours would
dominate also in this case.  Dominance of the R-violating couplings of
the heavier flavours would also favour radiative over three body
gravitino decays in scenarios with gravitino dark matter
\cite{LOR,LOR2}.
\item
ensure that all couplings remain within the experimental bounds.
\end{itemize}

For $\Delta L \neq 0$ the strictest bounds are on
$L_1Q_1\bar{D}_1$ from nuclear $\beta\beta$ decay and on
$L_1L_3\bar{E}_3$ from bounds on Majorana neutrino masses, constraining
the choice of the charge $w$ to be $|12+w| \geq 2$ and $|4+w| \geq
2$, which are easy to satisfy \cite{ELR}.
We also note that the magnitudes of
the couplings in Table~\ref{table:3} are symmetric in the three
indices $ijk$. This implies, for example, that, at this level, the
$\lambda'_{121}$ and $\lambda'_{112}$ couplings should have similar
magnitudes.  This must be made consistent with the constraint $(L_1
Q_{2} \bar{D}_1)\cdot(L_1 Q_{1} \bar{D}_2) \leq 4 \cdot 10^{-9}$,
which arises from bounds on $\Delta m_K$.  In the present context,
this constraint indicates that the relevant charge $|7+w|$ has to be
large.

The problem gets more complicated by taking into account mixing
effects, which for left-right symmetric models have been studied in
detail in \cite{ELR}, in combination with bounds on both individual
couplings as well as on products. Summarising, it turned out that the
strong correlations were leading to a suppression of all couplings,
and even the 333 flavour, in order to be fully safe, had to be smaller
than 0.006.  Within this framework, therefore, single superparticle
productions are suppressed and the best signal would be pair
productions followed by $R$-violating decays (which favour the
neutralino search channel even further).  Finally, quantum
corrections, expressed through renormalisation group effects, would
only mildly affect the relative hierarchies of R-violating couplings
(although they would modify their absolute values between the GUT and
the low energy scale) \cite{Allanach:1999ic}.

For the $\Delta L \neq 0 $ operators, we may then study the correlated
operators:
\begin{equation} \label{Eq:LLE-LQD}
LL\bar E_\text{eff}+LQ\bar D_\text{eff},
\end{equation}
with
\begin{align}
LL\bar E_\text{eff}
&=\bar\epsilon L_2L_3\bar E_3+\bar\epsilon^2L_1L_2\bar E_2
+{\cal O}(\bar\epsilon^3),
\nonumber \\
LQ\bar D_\text{eff}
&=L_3Q_3\bar D_3+\bar\epsilon(L_2Q_3\bar D_3
+L_3Q_2\bar D_3+L_3Q_3\bar D_2)
+{\cal O}(\bar\epsilon^2),
\end{align}

\FIGURE[ht]{
\let\picnaturalsize=N
\def\picsize{15cm}
\ifx\nopictures Y\else{
\let\epsfloaded=Y
\centerline{\hspace{4mm}{\ifx\picnaturalsize N\epsfxsize \picsize\fi
\epsfbox{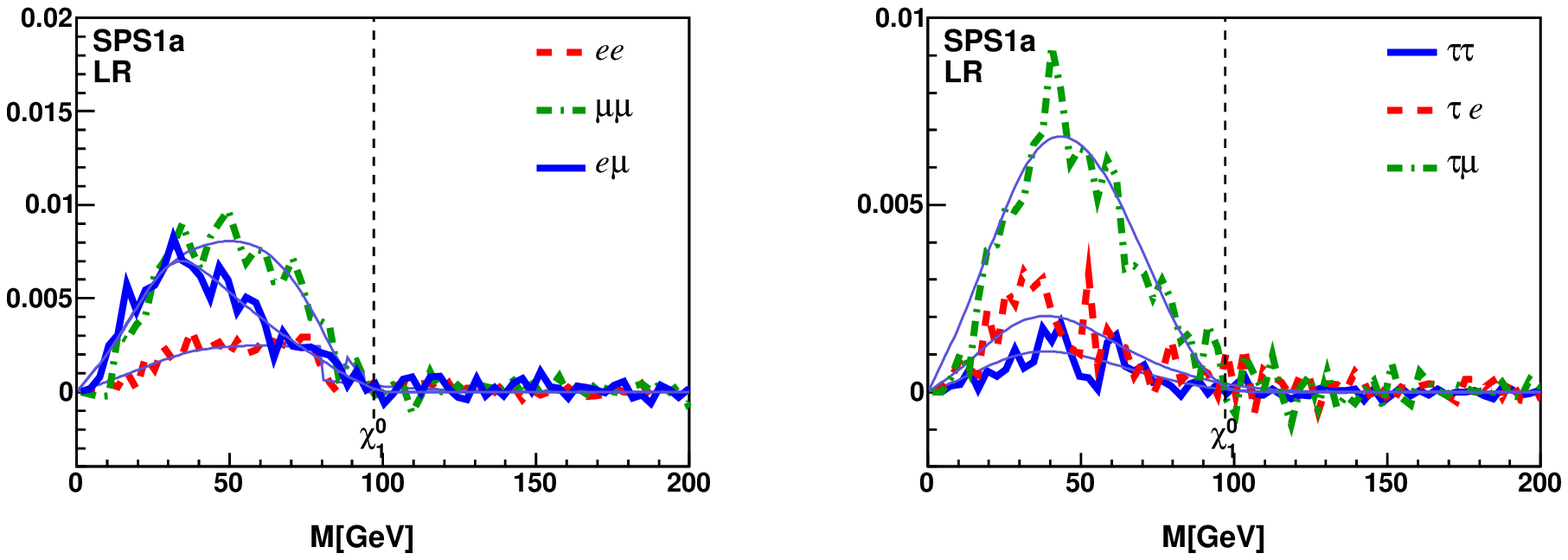}} }
}\fi
\vspace{-6mm}
\caption{The same as Fig.~\ref{fig-InvM-LLE231} for the Left-Right symmetric scenario defined by (\ref{Eq:LLE-LQD}) at SPS1a.
\label{fig-InvM-LR} } }

Due to the larger decay widths associated with the $LL\bar E$
operators as compared to the $LQ\bar D$ operators as well as the lack
of a clear signal associated with the $Q_3$ factor in the dominant
operator; the only thing we can see is the $L_2L_3\bar E_3$ operator
and possibly a hint of the $L_1L_2\bar E_2$ operator. This is
illustrated in Fig.~\ref{fig-InvM-LR} where the leptonic invariant
mass distributions are plotted. The data is consistent with a dominant
contribution from a $\mu\tau$ decay channel giving the $\mu\tj$
distribution, together with contributions of type (\ref{Eq:Mad}) to
the $\mu\mu$ and $e\mu$ distributions. Since the $\mu\mu$ and $e\mu$
distributions have kinematical cutoffs of the same scale as the
$\mu\tj$ distribution, we need an additional contribution of type
(\ref{Eq:InvMass}) to those distributions, consistent with a
subdominant $L_1L_2\bar E_2$ operator.

Due to the dominant $LQ\bar D$ operator, a larger fraction of the
leptons in the final state will come from the upper decay chain rather
than the neutralino decay, as compared to the pure $LL\bar E$ case,
and therefore the numerical observables of
section~\ref{subsect:counting} are too dependent on SPS point and not
very useful. This contamination also causes the invariant mass
distributions to be much less smooth and therefore more difficult to
interpret.

If we now look at the $\bar U \bar D \bar D$ operators, we see that
even though, within this model, the largest coupling is $\bar U_3\bar
D_2\bar D_3$, the phase-space suppression caused by the top quark in
the final state makes the dominant channels the ones with three quark
jets, even at SPS6 where the neutralino is sufficiently heavy to decay
to a top.

\subsubsection{\boldmath{$SU(5)$}}

Another interesting possibility is that the
family symmetry commutes with an $SU(5)$
GUT, where the SM fermions are assigned as follows to the
representations of the group:
\begin{eqnarray}
Q_{(q,u^{c},e^{c})_{i}} &=&Q_{i}^{10}  \label{su5charges} \nonumber\\
Q_{(l,d^{c})_{i}} &=&Q_{i}^{\overline{5}}   \\
Q_{(\nu _{R})_{i}} &=&Q_{i}^{\nu _{R}}  \nonumber
\end{eqnarray}

From the above it immediately follows that :
\begin{enumerate}
\item
The up-quark mass matrix is symmetric (both left- and right-handed
up quarks are in the 10, and thus have the same flavour charge).
\item
the charged lepton mass matrix is the transpose of the down quark mass
matrix.
\end{enumerate}
In this case, a viable choice of charges obeying the
restrictions of the symmetry (see e.g., \cite{Lola:1999un}),
is:
\begin{align}
Q_{1,2,3} &= \bar{E}_{1,2,3} = 3,2,0 \nonumber \\
\bar{D}_{1,2,3} &= L_{1,2,3} = 1,0,0
   \label{su5_qno}
\end{align}
leading to matrices that, apart from other
features, lead to a maximal 2-3 lepton mixing, viz.
\begin{equation}
M^\text{up}\sim\left(
\begin{array}{ccc}
\ee ^{6} & \ee ^{5} & \ee ^{3} \\
\ee ^{5} & \ee ^{4} & \ee^2 \\
\ee ^{3} & \ee^2 & 1
\end{array}
\right), \quad
M^\text{down}\sim\left(
\begin{array}{ccc}
\ee ^{4} & \ee ^{3} & \ee ^{3} \\
\ee ^{3} & \ee ^{2} & \ee^2 \\
\ee  & 1 & 1
\end{array}
\right), \quad
M^{\ell}\sim\left(
\begin{array}{ccc}
\ee ^{4} & \ee ^{3} & \ee\\
\ee ^{3} & \ee ^{2} & 1 \\
\ee^3  & \ee^2 & 1
\end{array}
\right),
\label{Equation:SU5-mm}
\end{equation}
where $\ee \approx 0.2$.

Let us first look at the implications  for the
$LL\bar{E}$ operators.
 Since the
charges of $L_{2,3}$ are  the same, couplings such as
 $L_i L_2 \bar{E}_k$ and $L_i L_3 \bar{E}_k$ would be
 expected to be of similar magnitude.
In short, the $U(1)$ assignments of Eq.~(\ref{su5_qno}) lead to
operator charges as listed in Table~\ref{tab:su5_operators}.

\begin{TABLE}{
\label{tab:su5_operators}
\begin{tabular}{|c |c|c|c|c|c|c|}\hline
$ijk$   &
{ 121,131} & { 231}  &
{ 122,132} & 232 &
{ 123,133} & 233
\\ \hline
$U(1)$ &
$4-w$ &
$3-w$ &
$3-w$ &
$2-w$ &
$1-w$ &
$-w$
\\ \hline \hline
\end{tabular}
\caption{\label{table-su5-1}{\it $LL\bar{E}$ charges in SU(5) enhanced by a U(1) flavour symmetry.}} }
\end{TABLE}

Similarly for $LQ\bar{D}$, where now we have the connection
  $\lambda^\prime_{ijk} = \lambda_{ijk}$ arising directly from the way
  we accommodate the fields in the GUT representation.
For the $LQ\bar D$ operators we also note the following:
\begin{itemize}
\item
Since the $ U(1)$  charges of $L_{2,3}$ are the same, the respective
operators are linked, reducing the number of independent couplings.
\item
Since the $ U(1)$  charges of $\bar{D}_{2,3}$ are the same, the respective
operators are linked, reducing the number of independent couplings.
\end{itemize}
This leaves us with the results of table~\ref{table:SU5-l-viol}.
Finally, for the baryon-number violating operators, we find the results given in table~\ref{table:SU5-b-viol}.

\begin{TABLE}{
\begin{tabular}{|c |c|c|c|c|c|c|}\hline
$ijk^{\prime}$   &
111 & 112,113 & 121 & 122,123 &
131 & 132,133
\\ \hline
$U(1)$ &
$5-w$ &
$4-w$ &
$4-w$ &
$3-w$ &
$2-w$ &
$1-w$ \\ \hline
$ijk^{\prime}$   &
211,311 & 212,213,312 & 221,321 & 222,223,322,323 &
231 & 232,233,332,333
\\ \hline
$U(1)$ &
$4-w$ &
$3-w$ &
$3-w$ &
$2-w$ &
$1-w$ &
$-w$ \\
\hline \hline
\end{tabular}
\caption{{\it $LQ\bar{D}$ charges in SU(5) enhanced by a U(1) flavour symmetry.}}
\label{table:SU5-l-viol}}
\end{TABLE}

\begin{TABLE}{
\begin{tabular}{|c |c|c|c|c|c|c|}\hline
$ijk^{\prime\prime}$   &
{ 112,113} & { 123}  &
{ 212,213} & 223 &
{ 312,313} & 323
\\ \hline
$U(1)$ &
$4-w$ &
$3-w$ &
$3-w$ &
$2-w$ &
$1-w$ &
$-w$
\\ \hline \hline
\end{tabular}
\caption{{\it $\bar{U}\bar{D}\bar{D}$ charges in SU(5) enhanced by a U(1) flavour symmetry.}}
\label{table:SU5-b-viol}}
\end{TABLE}

\FIGURE[ht]{
\let\picnaturalsize=N
\def\picsize{15cm}
\ifx\nopictures Y\else{
\let\epsfloaded=Y
\centerline{\hspace{4mm}{\ifx\picnaturalsize N\epsfxsize \picsize\fi
\epsfbox{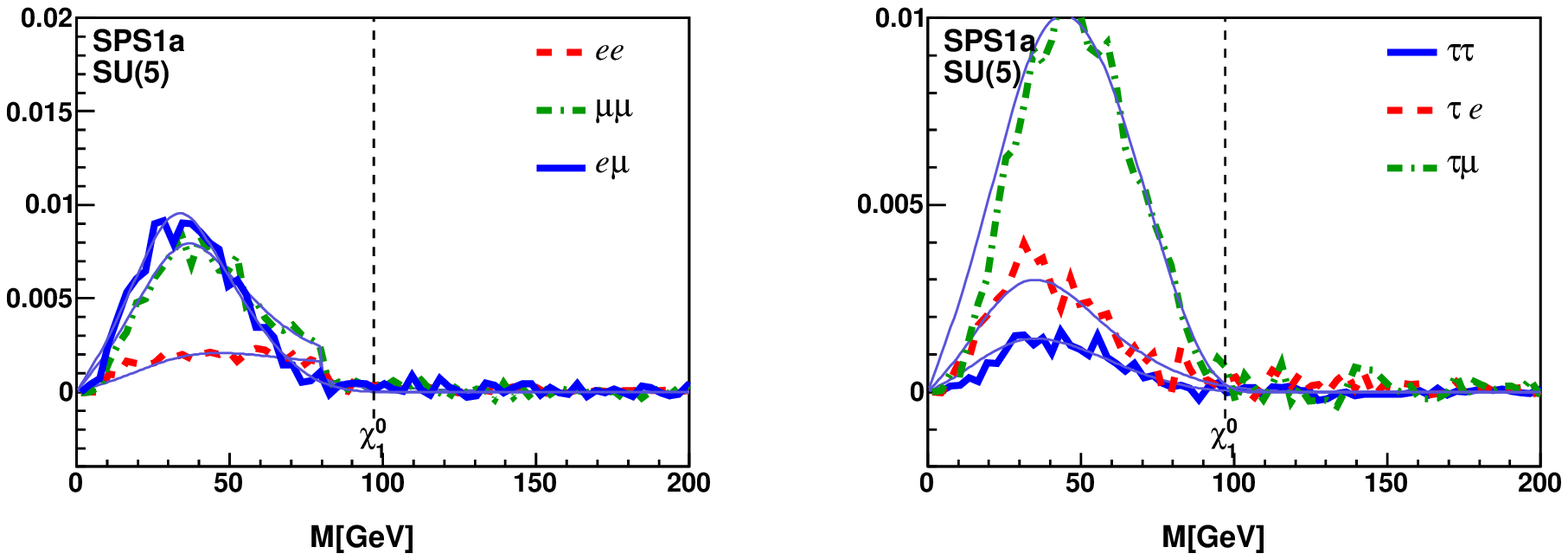}} }
}\fi
\vspace{-6mm}
\caption{The same as Fig.~\ref{fig-InvM-LLE231} for the $SU(5)$ scenario defined by tables \ref{table-su5-1} and \ref{table:SU5-l-viol} at SPS1a.
\label{fig-InvM-SU5} } }

As for the $L$-violating part, for $w = 0$, the $L_2L_3\bar
E_3$ coupling is as large as the largest of the $LQ\bar D$ couplings and
therefore the prospects of accurately identifying this coupling are
improved. Moreover, the leading
subdominant $LL\bar E$ operators, namely
$L_1L_3\bar E_3$ and $L_1L_2\bar E_3$, also carry a lot
of tau flavour.   This
ensures an even larger dominance of the
$L_2L_3\bar E_3$ operator after cuts are applied (in the Left-Right
symmetric case the cuts partially compensate for the difference in
width between the $L_2L_3\bar E_3$ and the $L_1L_2\bar E_2$
operator). All in all, the prospects for identifying the $L_2L_3\bar
E_3$ operator are good but it does not appear possible to extract any
of the subdominant couplings.

This can be seen in the invariant mass distributions shown in
Fig.~\ref{fig-InvM-SU5} where all distributions look much like we
expect for a pure $L_2L_3\bar E_3$ operator. Note that all the light
lepton distributions of the left panel of Fig.~\ref{fig-InvM-SU5}
vanish between $\approx 80$ GeV and the kinematical end-point of the
$\mu\tj$ distribution ($\approx M_{\chi_1^0}$) and we can therefore
exclude any significant neutralino decay to light leptons.

Note that this strong statement is linked to the simplifying
assumption that like for the Yukawa couplings, R-violating couplings
of the third generation are in general larger. If this assumption is
dropped, the picture would change. For instance, for the $LL\bar{E}$
operators:
\begin{itemize}
\item
$w = 1$ would imply a simultaneous
dominance of the $L_1L_2\bar{E}_3$  and $L_1L_3\bar{E}_3$ couplings,
\item
$w = 2$ leads to  a dominant $L_2L_3\bar{E}_2$ coupling, and
\item
$w = 3$ results in a simultaneous dominance  of the $L_2L_3\bar{E}_1$,
$L_1L_2\bar{E}_2$
and $L_1L_3\bar{E}_2$ couplings.
\end{itemize}
A similar situation
(but for different respective flavours) would occur for the
rest of the operators.

For the baryon-violating case, $w=0$ leads to a
stronger dominance of the $\bar U_3\bar D_j\bar D_k$ operators as compared to
the Left-Right symmetric case. Consequently, the decay to top plus 3 jets
will dominate if the neutralino is heavy enough, e.g.\ in SPS6.

\section{Summary}
\label{sect:summary}
Within a supersymmetric theory with trilinear R-parity violation and a
neutralino LSP,\footnote{The lightest neutralino could even be the
NLSP if, for example, we invoke a gravitino LSP for dark
matter; such an assumption does not affect this study.} one would
expect the usual pair production of squarks and gluinos, followed by
cascade decays down to the neutralino, to be the dominant scenario at
the LHC. The novel feature introduced by the R-parity violation would
then be the three-body decay of the neutralino.

Within such a model, the neutralino can decay through any of the 45
trilinear R-parity violating operators and therefore this channel
allows us to study all those operators and their internal hierarchies
simultaneously. On account of the typically
  large number of detectable particles (especially leptons) produced
  in such neutralino decays, the inferences drawn from such a study are
not as dependent on the underlying SUSY scenario as R-parity
conserving SUSY scenarios can be.

In general, bilinear and trilinear terms are generated in different ways
(for instance an underlying string theory would favour trilinear couplings; on the other
hand, flavour symmetries could in principle favour either possibility). 
See, however, Ref.~\cite{AristizabalSierra:2004cy}.
The decay modes
$\tilde\chi_1^0\to W^\pm\ell^\mp$ and $\tilde\chi_1^0\to Z\nu$
exist in bilinear schemes, leading to interesting signatures
\cite{Mukhopadhyaya:1998xj,Porod:2000hv}.  For these terms
to dominate, one would require non-negligible neutrino-neutralino mixing.

We have shown that the prospects of identifying trilinear operators of
the $LL\bar E$ type are rather good, mostly due to the abundance of
invariant mass distributions one can explore.
While one can, in principle, also identify
such operators from pure counting measures, the corresponding
conclusions would not only be more sensitive to backgrounds,
but also to potential competing operators, as well as to the SUSY model.
In addition to a significant reduction in such
dependences, the invariant
mass distributions are also promising for measuring the neutralino
mass; this has been shown to be feasible thanks to the knowledge of
the theoretically expected distributions, despite the lack of clear
peaks or edges.

As compared to $LL\bar E$ operators, $LQ\bar D$ operators have both
advantages and disadvantages. One disadvantage is that background in
the form of $t\bar t$ events is more challenging, but the large lepton
multiplicity is still sufficient to get a clean event
sample. Operators that lead to charged leptons and jets also have the
advantage of producing peaks in the relevant invariant mass
distributions, which not only allow good determination of the
operator, but also accurate measurements of the neutralino mass.
Particularly promising are the operators leading to $b$-quarks in the
decay since, in this case, $b$-tagging can also be used to extract the
signal.

With a tau as the lepton of the operator, things are not equally
promising, due to the loss of energy to neutrinos in the tau
decay. However, if the tau is accompanied by a $b$-quark, then $b$ and
tau tagging together might provide the information we need. For the
case with a tau and two light jets, and for $LQ_3\bar D$ operators
(which only gives neutrino and two jets) we do not have an obvious
method of identifying the operator.

If more than one lepton number violating operators are large at the
same time, we have shown that the prospects of identifying them are
good (in most cases, the signals we are looking for appear in
different channels for different operators and these signals are
mostly independent of each other). It is also worth mentioning that if
both $LL\bar E$ and $LQ\bar D$ operators are large at the same time we
expect to see only the $LL\bar E$ operator since it gets a larger
branching fraction and is easier to detect.

From flavour models one may expect large R-violating hierarchies,
similar to those of the Yukawa couplings that generate fermion
masses. Fermion mass terms are dominated by the heavier flavours.  If
this feature persists for the R-violating couplings as well, it would
make the identification of the $LQ\bar D$ operators very difficult
since the dominant ones will either contain tau flavour or a $Q_3$
operator, with either case implying a low chance of identification. In
addition, accompanying $LL\bar E$ operators have the advantage of
larger branching fraction as well as easier detection. The conclusion
is that in this subclass of flavour models we would expect to see
dominant $LL\bar E$ operators with heavy flavours, i.e.\ lots of tau flavour.

Finally we have looked at the case where a heavy neutralino
($M_{\chi_1^0}>M_\text{top}$) decays via a $\bar U_3\bar D_j\bar D_k$
operator to a top quark and two jets, and have shown that this
scenario can be identified.  This is important for flavour physics
since in many models the R-parity violating couplings with heavy
flavours will be the dominant ones.
\bigskip

{\bf Acknowledgements.}  We thank the NORDITA program {\it ``TeV
scale physics and dark matter''}, for hospitality while this work was
initiated. SL thanks the CERN Theory Division, for
kind hospitality.
The research of PO has been supported by the Research
Council of Norway.

\section*{Appendix~A. Energy spectra from three-body decays}
\setcounter{equation}{0}
\renewcommand{\thesection}{A}
For any decay into a final state comprising more than two particles,
the energy distribution of
a given final state particle is determined
not only by kinematics, but has a dependence on the dynamics as well.
In particular, let us consider the decay
of a neutralino to three fermions, namely
\begin{equation}
\tilde\chi_1^0\to f _1f_2 f_3.
\end{equation}
There are three contributions,
characterized by the identity of the sfermion (corresponding
to the fermion coupling to the neutralino).
Adopting the notation of Fig.~\ref{fig-LSP-decay},
the corresponding matrix element will be of the form
\begin{equation}
{\cal M}\sim [\bar u(p_1) u(p)] \; [\bar u(p_2) v(p_3)].
\end{equation}
Here, we disregard the energy dependence of the propagator (valid for
a heavy sfermion) and the pseudoscalar part of the coupling,
since its presence will not modify any differential decay rate. Then,
\begin{equation}
\sum_\text{spins} {\cal M}^\dagger {\cal M}
\propto (p\cdot p_1)(p_2\cdot p_3)
\end{equation}
\FIGURE[ht]{
\let\picnaturalsize=N
\def\picsize{13cm}
\ifx\nopictures Y\else{
\let\epsfloaded=Y
\centerline{\hspace{10mm}{\ifx\picnaturalsize N\epsfxsize \picsize\fi
\epsfbox{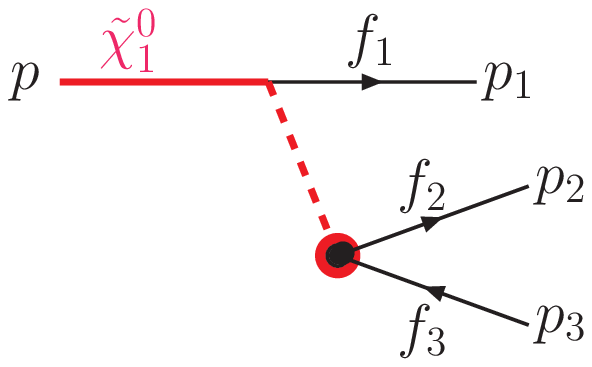}} }
}\fi
\caption{Kinematics of three-body neutralino decay.
\label{fig-LSP-decay} } }

In the decay, two of the final-state particles will be charged, and
potentially observable. For a particular final-state particle, the
energy distribution depends on whether it comes from the upper or
lower vertex in Fig.~\ref{fig-LSP-decay}. With the notations
$p=\{M,\boldsymbol{0}\}$, $p_i=\{E_i,\boldsymbol{p}_i\}$, and
$\epsilon_i=E_i/E_i^\text{max}\equiv 2E_i/M$, we find
\begin{equation} \label{Eq:dGamma-depsilon}
\frac{1}{\Gamma}\frac{d\Gamma}{d\epsilon_i}
=
\begin{cases}
12\epsilon_i^2(1-\epsilon_i) &i=1,\\
2\epsilon_i^2(3-2\epsilon_i) &i=2,3.
\end{cases}
\end{equation}
For $i=1$, the distribution vanishes at the upper end due to
angular momentum conservation. For comparison,
if the distribution had been determined by kinematics alone,
it would have read
\begin{equation}\label{Eq:epsilonPS}
\frac{1}{\Gamma}\frac{d\Gamma}{d\epsilon_i}
=
2\epsilon_i.
\end{equation}
To see the implications of this, consider the operator
$L_1L_2\bar{E}_3$. The last factor represents a coupling to a tau or a
stau (both charged), whereas the first two (by SU(2) invariance)
represent couplings to a charged field and a neutral one, one of which
will be of the first family, whereas the other will be of the second
family. We thus have two distinct decay channels:
\begin{equation}
\tilde\chi_1^0\to \tau^+e^-\nu_\mu,\qquad \tilde\chi_1^0\to \tau^+\mu^-\nu_e,
\end{equation}
and charge conjugates.
Each of these channels
is governed by three diagrams of the type shown in Fig.~\ref{fig-LSP-decay}.
Assuming that all the sfermions have comparable masses, and neglecting
interferences, the energy distribution for each of the decay products
would be averaging the
distributions (\ref{Eq:dGamma-depsilon}) over all three values of $i$,
resulting in
\begin{equation}\label{Eq:E-distr}
\frac{1}{\Gamma}\frac{d\Gamma}{d\epsilon}
=\epsilon^2\left(8-\frac{20}{3}\epsilon\right).
\end{equation}
The above assumptions are not entirely justified, e.g.\ we do expect
differences in sparticle masses to introduce differences between the
three diagrams due to propagator effects and we have no reason to
ignore interference altogether. However, Eq.~(\ref{Eq:E-distr}) should
still be a good approximation and the assumptions involved provide a
useful formula.

The distributions~(\ref{Eq:dGamma-depsilon}), (\ref{Eq:E-distr}) as well as (\ref{Eq:epsilonPS}) are shown in Fig.~\ref{fig-epsilonTheory}.
\FIGURE[ht]{
\let\picnaturalsize=N
\def\picsize{10cm}
\ifx\nopictures Y\else{
\let\epsfloaded=Y
\centerline{\hspace{10mm}{\ifx\picnaturalsize N\epsfxsize \picsize\fi
\epsfbox{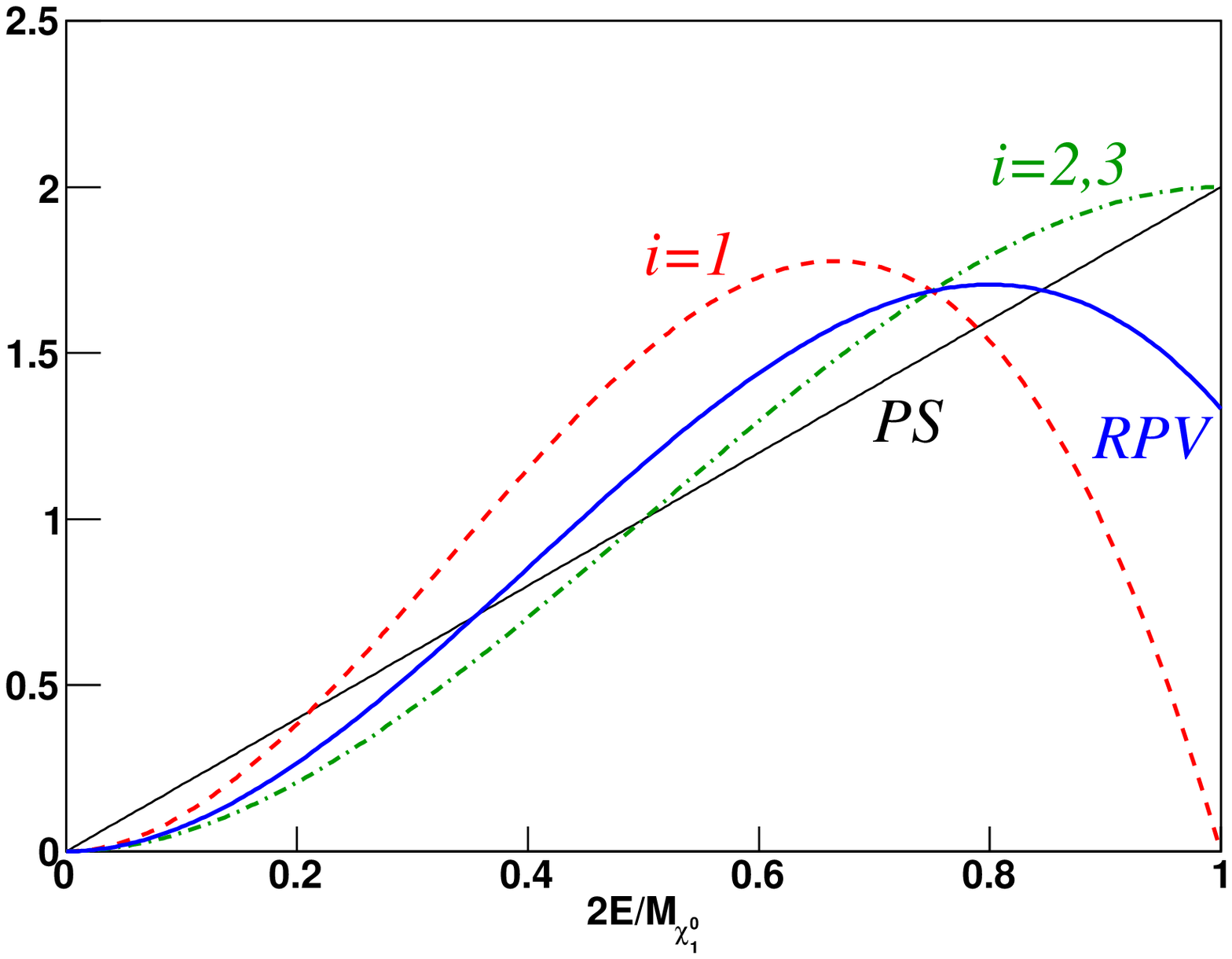}} }
}\fi
\vspace{-6mm}
\caption{Energy distributions for the decay products in a three body decay, Eq.~(\ref{Eq:dGamma-depsilon}). The straight black line shows the distribution achieved when only phase space (PS) is taken into account. The average, Eq.~(\ref{Eq:E-distr}), is denoted `RPV'.
\label{fig-epsilonTheory} } }

\section*{Appendix~B. Theoretical  invariant mass distributions}
\setcounter{equation}{0}
\renewcommand{\thesection}{B}
\label{App:InvMass}

In order to calculate the theoretically expected invariant mass distributions \cite{Baer:1995va,Gjelsten:2004ki} from the leptonic neutralino decays, we start with the energy distribution of the decay products, as given by Eq.~(\ref{Eq:E-distr}).

Let us now denote the three decay products $a$, $b$ and $c$, with corresponding four-momenta $p_a$, $p_b$ and $p_c$. The invariant mass $M_{ab}$ of particles $a$ and $b$ is given by
\begin{equation}
    M^2_{ab}= (p_a+p_b)^2=M^2-2ME_c+m_c^2,
\end{equation}
where $E_c$ denotes the energy in the neutralino rest frame and $M$ is the neutralino mass.
Assuming that particle $c$ is massless and follows the distribution given by Eq.~(\ref{Eq:E-distr}) we get:
\begin{equation}\label{Eq:InvMass}
    f_{M_{ab}}(M_{ab})=\frac{8M_{ab}}{3M^8}(M^2-M_{ab}^2)^2(M^2+5M_{ab}^2).
\end{equation}
This distribution is shown in Fig.~\ref{fig-MabTheory} (solid curves).

\FIGURE[ht]{
\let\picnaturalsize=N
\def\picsize{16cm}
\ifx\nopictures Y\else{
\let\epsfloaded=Y
\centerline{\hspace{10mm}{\ifx\picnaturalsize N\epsfxsize \picsize\fi
\epsfbox{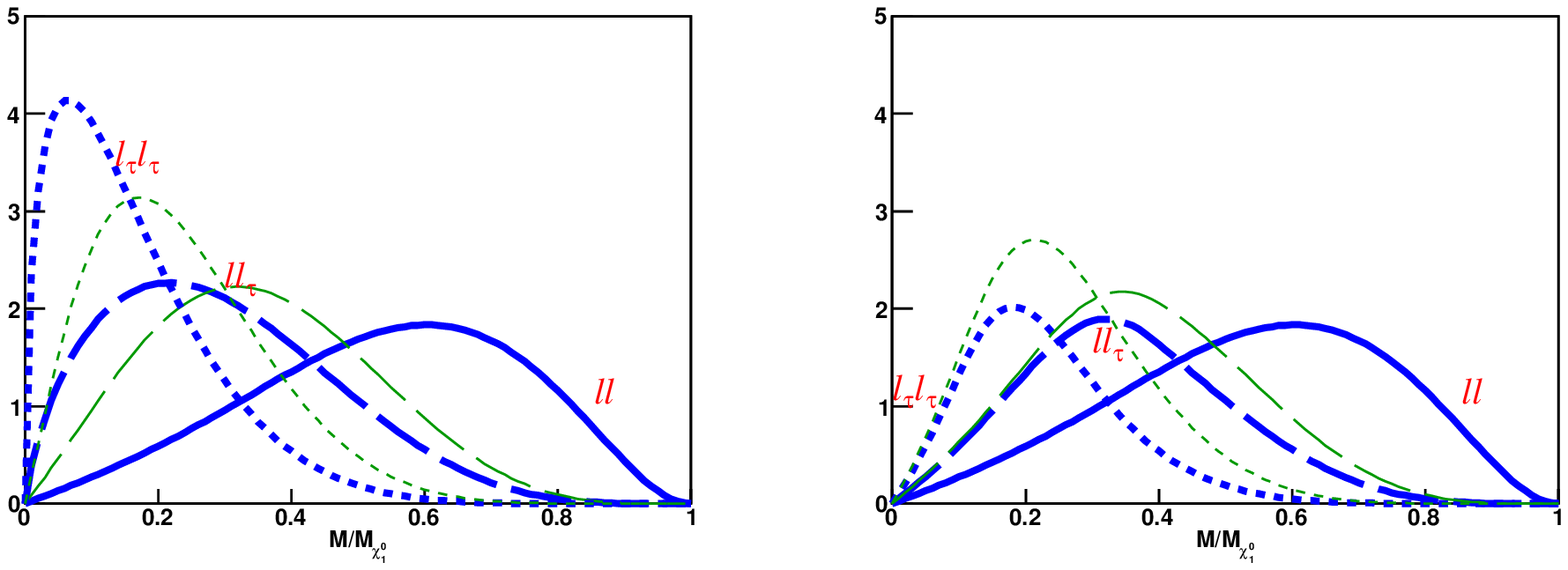}} }
}\fi
\vspace{-6mm}
\caption{Theoretical invariant di-lepton mass distributions for massless particles from $\tilde\chi_1^0$ decay.
Left: No cuts. The $\ell\ell$ curve is given by Eq.~(\ref{Eq:InvMass}). Right: Cuts as indicated by Eq.~(\ref{Eq:Mad}).
Solid curves: directly produced leptons; long-dashed curves: one lepton comes from a decaying tau;
short-dashed curves: both leptons come from decaying taus. Blue heavy curves do not take tau polarization into account while the thin green curves do. The labels refer to the heavy curves (blue).
\label{fig-MabTheory} } }

If one of the charged particles is a $\tau$, we get a
contribution to the di-lepton invariant mass only if the $\tau$ decays
leptonically. To calculate the distribution from this, we need to
estimate a distribution for the fraction of the energy of the tau that
is transferred to the charged lepton.

To do this, we assume that the tau decays into three massless
particles (one charged lepton and two neutrinos).  The kinematics of a
tau decay will be the same as discussed in Appendix~A for a neutralino
decay. However, with a $V-A$ coupling, $i=1$ in
Eq.~(\ref{Eq:dGamma-depsilon}) represents the antiparticle among the
decay products.  In the rest frame of the tau, the charged lepton will
then follow the distribution for $i=2,3$ with $M=M_\tau$.

The next step is to perform a Lorentz boost to the rest frame of the
neutralino. If we denote the momentum four-vector of the charged
lepton in the rest frame of the tau, $(E_l,\boldsymbol{P}_l)$ and that
of the tau in the rest frame of the neutralino, $(E,\boldsymbol{P})$,
we get the energy of the lepton in the rest frame of the neutralino
as:
\begin{equation}\label{Eq:El1}
    \frac{E_lE}{M_\tau}+\frac{\boldsymbol{P}_l\cdot\boldsymbol{P}}{M_\tau}.
\end{equation}
Since we assume that the lepton is massless, i.e., $E_l =
|\boldsymbol{P}_l|$, and that the lepton carries a fraction $x$ of
$M_\tau/2$ in the rest frame of the tau, i.e. $E_l = xM_\tau/2$,
Eq.~(\ref{Eq:El1}) can be written:
\begin{equation}\label{El2}
    \frac{x}{2}\left(E+|\boldsymbol{P}| \cos\theta\right),
\end{equation}
where $\theta$ is the angle between $\boldsymbol{P}$ and
$\boldsymbol{P}_l$. Let us now take the limit of a massless tau, i.e.,
$E = |\boldsymbol{P}|$, then we get:
\begin{equation}\label{El3}
    \frac{xE}{2}\left(1+\cos\theta\right).
\end{equation}

Assuming that the direction of the lepton is chosen
isotropically\footnote{Since the tau will in general be left-handed,
  this is not a very accurate assumption. For a discussion of the
  effects of tau polarization, see below.} in the rest frame of the
tau, implies that the distribution of $\cos\theta$ is flat. From that
we can calculate the distribution of $y=1+\cos\theta$ as:
\begin{equation}\label{Eq:ydistr}
    f_y(y)= \frac{1}{2},
\end{equation}
where $y\in[0,2]$. We are interested in the fractional lepton energy with respect to that of the tau, in the $\tilde\chi_1^0$ rest frame, i.e., the  product $z=\frac{x}{2}y$ where $x$ by construction follows the distribution (\ref{Eq:dGamma-depsilon}) with $i=2,3$. The distribution $f_z(z)$ is now given by:
\begin{equation}\label{zdistr}
    f_z(z)=\frac{5-9z^2+4z^3}{3}.
\end{equation}
With all decay products from the neutralino massless, we can simplify the invariant mass formula:
\begin{equation}
    M^2_{ab}= (p_a+p_b)^2 = 2E_aE_b(1-\cos\phi),
\end{equation}
where $\phi$ is the angle between $\boldsymbol{P}_a$ and $\boldsymbol{P}_b$. If we then assume that particle $b$ is a tau and decays to a charged lepton $d$, the resulting invariant mass $M_{ad}$ is given by:
\begin{equation}
    M_{ad}= M_{ab}q,
\end{equation}
where $q=\sqrt{z}$ and has a distribution $f_q(q)=\frac{2q}{3}(5-9q^4+4q^6)$. $M_{ad}$ then has a distribution:
\begin{equation}\label{Eq:Maduncut}
    f_{M_{ad}}(M_{ad})=\int_{M_{ad}/M}^1 \frac{f_q(q)f_{M_{ab}}(M_{ad}/q)}{q}dq.
\end{equation}
 We see from Eq.~(\ref{zdistr}) that there are many low-energy leptons coming from the tau decay. These low-energy leptons will, however, not pass the $p_T$ cuts we impose and therefore we need to take the cuts into account when calculating the distribution (this is not required for the direct decay to leptons since the amount of low energy leptons there is small). The simplest thing to do to include some cuts, is to constrain the lower limit in the integral of Eq.~(\ref{Eq:Maduncut}):
\begin{equation}\label{Eq:Mad}
    f_{M_{ad}}(M_{ad})=\int_{\max(M_{ad}/M,0.4)}^1 \frac{f_q(q)f_{M_{ab}}(M_{ad}/q)}{q}dq.
\end{equation}

The introduction of this cut does not preserve the normalization of
the distribution; up to this point all distributions have been
normalized to unity as one would expect for probability
distributions. However, this is no longer true for Eq.~(\ref{Eq:Mad}),
moreover, the new normalization will depend on the distribution to
which this kind of cut is applied, due to the larger impact on the
lower end of the distribution.

Since we are more concerned about the shape rather than the exact
values from the distributions, there is no need to normalize
Eq.~(\ref{Eq:Mad}). Note also that introducing the cuts this way is
simple and gives a reasonably good result, but it is very ad hoc, the
specific value 0.4 is adopted for no other reason than the fact that
it gives a good approximation to the Monte Carlo results.

\FIGURE[ht]{
\let\picnaturalsize=N
\def\picsize{16cm}
\ifx\nopictures Y\else{
\let\epsfloaded=Y
\centerline{\hspace{4mm}{\ifx\picnaturalsize N\epsfxsize \picsize\fi
\epsfbox{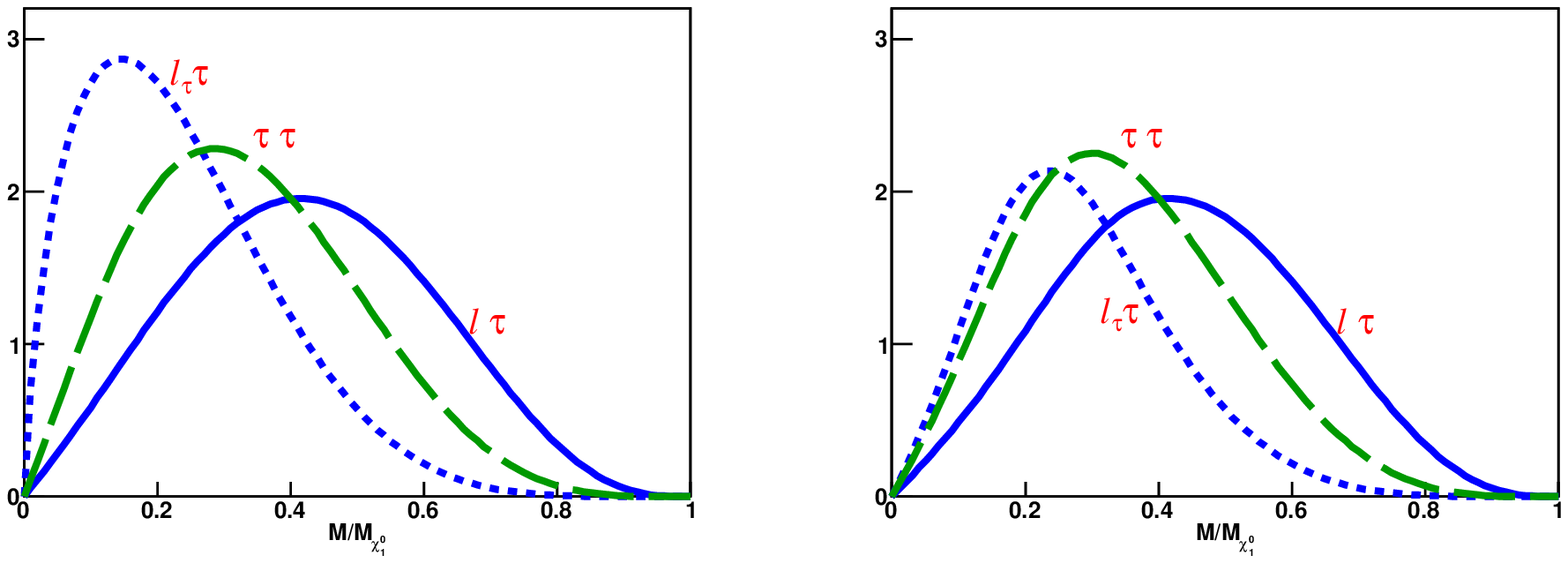}} }
}\fi
\vspace{-6mm}
\caption{Effects of cuts on invariant mass distributions involving taus.
Left: No cuts. Right: Cuts as indicated by Eq.~(\ref{Eq:Mad}).
Long-dashed (green) curve: two tau jets are combined; solid (blue): one directly produced lepton combined with a tau jet; short-dashed (blue): a tau together with a lepton from a tau decay.
\label{fig-Mtau} } }

\FIGURE[ht]{
\let\picnaturalsize=N
\def\picsize{16cm}
\ifx\nopictures Y\else{
\let\epsfloaded=Y
\centerline{\hspace{4mm}{\ifx\picnaturalsize N\epsfxsize \picsize\fi
\epsfbox{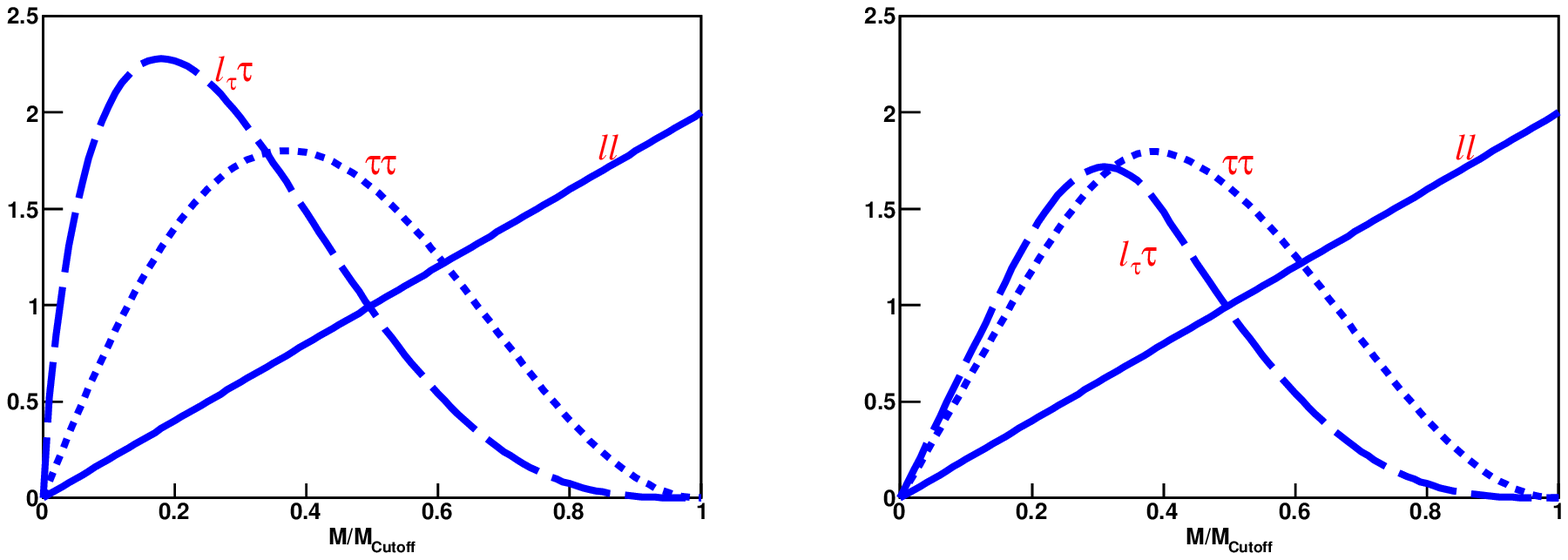}} }
}\fi
\vspace{-6mm}
\caption{Invariant mass distributions expected from the higher RPC chain.
Left: No cuts. Right: Cuts as indicated by Eq.~(\ref{Eq:Mad}).
Solid: the familiar  triangular di-lepton distribution; short-dashed: the di-tau-jet invariant mass; long-dashed: lepton-tau-jet mass distribution, expected when one of the taus decays leptonically.
\label{fig-MRPC} } }

We can also extend the analysis to the case when both leptons come from decaying taus by multiplying $M_{ad}$ with $q=\sqrt{z}$ once again.
The result of this is shown as the long-dashed curves of Fig.~\ref{fig-MabTheory} and the distributions of Eq.~(\ref{Eq:Maduncut}) and Eq.~(\ref{Eq:Mad}) are shown as the short-dashed curves.

So far we have ignored any effects of tau polarization. However, since
the neutralino will in general decay to left-handed taus, the spin of
the tau will affect the mass distribution. More precisely, we get an
extra factor $\half (1+\cos\theta)$ which amounts to replacing
Eq.~(\ref{Eq:ydistr}) with $f_y(y)= \frac{y}{2}$. The result of this
can be seen in the thin green curves of Fig.~\ref{fig-MabTheory}. The
tau polarization pushes the distributions to higher energy but the
effect is smaller than the uncertainties regarding the effects of the
cuts. Given that PYTHIA also does not take polarization into account,
we use the distributions without polarization effects for the
comparison with Monte Carlo.

In the right panel of Fig.~\ref{fig-MabTheory} we see the result of
simulating the cuts with a constraint on the integral as done in
Eq.~(\ref{Eq:Mad}). As mentioned before, this suppresses the lower end
of the distributions and changes the normalizations.  We can see that
the change in normalization is different without (blue curves) and
with tau polarization effects (green curves). Including tau
polarization effects, the distributions are pushed towards higher
masses and are therefore less affected by the cuts.

If the tau instead decays hadronically, we need to estimate the energy
of the resulting tau-jet ($\tj$).  The simplest thing to do is to
assume a three-body decay of the tau where the neutrino gets a
fraction $c$ of the tau mass and the rest of the energy goes into the
$\tj$. The distribution of $c$ will then follow
(\ref{Eq:dGamma-depsilon}) for $i=2,3$. We get the invariant mass of a
$\tj$ combined with a lepton, $M_{\tj l}$ as:
\begin{equation}\label{Eq:Mtaul}
    M_{\tj l}=M_{ab}\sqrt{1-c},
\end{equation}
where $M_{ab}$ and $c$ are as defined above. In line with the above
procedure we introduce the cuts by requiring $\sqrt{1-c}>0.4$.

This procedure can be repeated to produce the distribution of two
$\tj$'s as well as lepton-$\tj$ distributions where the lepton comes
from the decay of a tau. The result of this is shown in Fig.~\ref{fig-Mtau}.

For comparison, in Fig.~\ref{fig-MRPC} we show the expected signals
from the RPC chain. One clearly sees the similarities between the RPV
and RPC signals in channels including $\tj$'s. This makes the study of
these channels much more difficult.


\end{document}